\documentclass{article}
\pdfoutput=1 

\usepackage[utf8]{inputenc}
\usepackage[margin=1in]{geometry}
\usepackage{amsfonts,amsmath,amssymb,amsthm,mathtools,stmaryrd}

\usepackage{thmtools}
\usepackage{thm-restate}

\usepackage{caption}
\usepackage[noend,noline,ruled]{algorithm2e}

\usepackage{tocloft}
\usepackage[nottoc]{tocbibind} 

\usepackage{enumitem}
\setlist{noitemsep}
\setlist[itemize]{label=$-$}

\usepackage{hyperref}
\usepackage[dvipsnames]{xcolor}
\hypersetup{
	colorlinks=true,
	urlcolor=MidnightBlue,
	linkcolor=MidnightBlue,
	citecolor=MidnightBlue
}

\usepackage[capitalise,nameinlink,noabbrev]{cleveref}
\crefname{equation}{}{}
\crefrangelabelformat{section}{#3#1#4--#5\crefstripprefix{#1}{#2}#6}

\usepackage{macros} 

\begin{document}
\newgeometry{margin=1.3in,top=1.7in,bottom=1in}

\begin{center}
{\huge Randomised Composition and Small-Bias Minimax}
\\[1.3cm] \large

\setlength\tabcolsep{1.1em}
\begin{tabular}{cccc}
Shalev Ben-David&
Eric Blais&
Mika G\"o\"os&
Gilbert Maystre\\[-1mm]
\small\slshape University of Waterloo &
\small\slshape University of Waterloo &
\small\slshape EPFL &
\small\slshape EPFL
\end{tabular}

\large

\vspace{10mm}

{\today}
	
\vspace{10mm}

\normalsize
\bf Abstract
\end{center}

\noindent
We prove two results about randomised query complexity $\R(f)$. First, we introduce a \emph{linearised} complexity measure $\LR$ and show that it satisfies an \emph{inner-optimal} composition theorem: $\R(f\circ g) \geq \Omega(\R(f) \LR(g))$ for all partial $f$ and $g$, and moreover, $\LR$ is the largest possible measure with this property. In particular, $\LR$ can be polynomially larger than previous measures that satisfy an inner composition theorem, such as the max-conflict complexity of Gavinsky, Lee, Santha, and Sanyal~({\small ICALP 2019}).

Our second result addresses a question of Yao~({\small FOCS 1977}). He asked if $\epsilon$-error \emph{expected} query complexity $\bR_\epsilon(f)$ admits a distributional characterisation relative to some hard input distribution. Vereshchagin~({\small TCS 1998}) answered this question affirmatively in the bounded-error case. We show that an analogous theorem \emph{fails} in the small-bias case $\epsilon=1/2-o(1)$.

\vspace{10mm}

\setlength{\cftbeforesecskip}{2pt}
\setlength{\cftbeforepartskip}{15pt}
\renewcommand\cftsecfont{\mdseries}
\renewcommand{\cftsecpagefont}{\normalfont}
\renewcommand{\cftsecleader}{\cftdotfill{\cftdotsep}}
\setcounter{tocdepth}{1}
\tableofcontents

\thispagestyle{empty}
\setcounter{page}{0}
\newpage
\restoregeometry

\section{Introduction}

This paper is motivated by the following basic open problem in boolean function complexity theory.
\begin{conjecture} \label{prob}
$\R(f\circ g) \geq \Omega(\R(f)\R(g))$ for all total boolean functions $f,g$.
\end{conjecture}
Let us unpack what this conjecture is claiming. The randomised $\epsilon$-error query complexity $\R_\epsilon(f)$ of a boolean function~$f\colon\{0,1\}^n\to\{0,1\}$ is defined (see~\cite{Buhrman2002} for the classic reference) as the least number of queries a randomised algorithm (decision tree) needs to make, on the worst-case input, to the bits $x_i$ of~$x\in\{0,1\}^n$ in order to compute $f(x)$ correctly with error at most $\epsilon$. We write $\R\coloneqq\R_{1/3}$ for the bounded-error case. For functions $f$ and~$g$ over $n$ and $m$ bits, their composition $f\circ g$ is defined over $nm$ bits by
\[
(f\circ g)(x) ~\coloneqq~ f(g(x^1),\ldots,g(x^n)) \qquad \text{where}\qquad x=(x^1,\ldots,x^n)\in(\{0,1\}^m)^n.
\]
In particular, we have $\R(f\circ g)\le O(\R(f)\R(g)\log\R(f))$ for all $f,g$. This holds since we can run an algorithm for $f$ with query cost $\R(f)$ and whenever it queries an input bit, we can run, as a subroutine, an $\epsilon$-error algorithm for $g$ of cost $\R_\epsilon(g)$. Setting $\epsilon\ll 1/\R(f)$ makes sure that the errors made by the subroutines do not add up. Moreover, we have $\R_\epsilon(g)\leq O(\R(g)\log(1/\epsilon))=O(\R(g)\log\R(f))$ by standard error reduction techniques. \Cref{prob} thus postulates that a converse inequality always holds (without the log factor).

The analogue of \cref{prob} has been long resolved for many other well-studied complexity measures: deterministic query complexity satisfies a perfect multiplicative composition theorem, $\D(f\circ g)=\D(f)\D(g)$~\cite{Savicky2002}, quantum query complexity satisfies $\Q(f\circ g)=\Theta(\Q(f)\Q(g))$~\cite{Reichardt2011,Lee2011}, and yet more examples (degree, certificate complexity, sensitivity, rank) are discussed in~\cite{Tal2013,Gilmer2016,Dahiya2021}. In the randomised case, however, the conjecture has proved more delicate, exhibiting a far richer, and more surprising, structure.

\paragraph{Partial counterexamples.}
\Cref{prob} is known to be false if we relax the requirement that $f,g$ are total and instead consider \emph{partial} functions (promise problems), which are undefined on some inputs~$x$, $f(x)=*$. Indeed, works by Gavinsky, Lee, Santha, and Sanyal~\cite{Gavinsky2019} and Ben-David and Blais~\cite{BenDavid2020comp} have culminated in examples of partial functions $f$, $g$ such that $\R(f\circ g)\leq o(\R(f)\R(g))$.
Motivated by these counterexamples, we ask: \emph{What is the best possible composition theorem one can prove for partial functions?}

\subsection{A new composition theorem}
Our first result is an \emph{inner-optimal} composition theorem for partial functions. To state this result, we start by introducing a new \emph{linearised} complexity measure defined for a partial function $f\colon\{0,1\}^n\to\{0,1,*\}$ by
\[
\LR(f) ~\coloneqq~ \min_R\max_x \frac{\cost(R,x)}{\bias_f(R,x)},
\]
\begin{itemize}
\item where $R$ ranges over randomised decision trees;
\item $x$ ranges over the domain of $f$, namely, $\Dom(f)\coloneqq f^{-1}(\{0,1\})$;
\item $\cost(R,x)$ denotes the \emph{expected} number of queries $R$ makes on input $x$; and
\item $\bias_f(R,x)$ denotes the bias $R$ has of guessing the value $f(x)$ correctly; formally, $\bias_f(R,x)\coloneqq \max\{1-2\err_f(R,x),0\}$ where $\err_f(R,x)\coloneqq \Pr_R[R(x)\neq f(x)]$. We often omit the subscript $f$ for brevity.
\end{itemize}
This definition might seem mysterious at first sight. To get better acquainted with it, let us first note that
\begin{equation} \label{eq:lr-def}
\forall f\colon\qquad\Omega(\sqrt{\R(f)}) ~\leq~ \LR(f) ~\leq~ O(\R(f)).
\end{equation}
Indeed, the second inequality follows by considering a bounded-error decision tree $R$, with $\cost(R,x)\leq \R(f)$ and $\bias(R,x)\geq 1/3$. For the first inequality, if we let $R$ be a randomised tree that achieves the minimum in the definition of $\LR(f)$, we can amplify the bias of $R$, which is possibly tiny, as follows. On input $x$ we run~$R(x)$ repeatedly until we have made a total of $\LR(f)^2$ queries, and then output the majority answer over all runs. We expect this simulation to run $R(x)$ for $\LR(f)^2/\cost(R,x)\geq 1/\bias(R,x)^2$ many times, which, by standard Chernoff bounds, is enough to amplify the bias to a constant. This shows $\R(f)\leq O(\LR(f)^2)$.

Both extremes in \cref{eq:lr-def} can be realised. First, consider the $n$-bit \emph{parity} function $\Xor_n$. It is not hard to see that any randomised tree that achieves bias $\delta$ for $\Xor_n$ needs to query all the $n$ bits with probability at least $\delta$, resulting in expected query cost at least $\delta n$. This shows $\LR(\Xor_n)=\R(\Xor_n)=n$. Second, consider the partial $n$-bit \emph{gap-majority} function (here $|x|$ denotes the Hamming weight)
\[
\GapMaj_n(x) ~\coloneqq~
\begin{cases}
1 & \text{if}\enspace |x| \geq n/2+\sqrt{n},\\
0 & \text{if}\enspace |x| \leq n/2-\sqrt{n},\\
* & \text{otherwise}.
\end{cases}
\]
It is well known that $\R(\GapMaj_n)=\Theta(n)$. By contrast, the algorithm $R$ that queries and outputs a uniform random bit of $x$ has $\cost(R,x)=1$ and $\bias(R,x)\geq \Omega(1/\sqrt{n})$, which shows $\LR(\GapMaj_n)\leq O(\sqrt{n})$.

\bigskip\noindent
Our first main result shows that a multiplicative composition theorem holds when the inner function is measured according to $\LR$, and moreover, our choice of $\LR$ is optimal among all inner complexity measures. Ultimately, these theorems are what lends naturalness to our definition of $\LR$.
\begin{restatable}{theorem}{Comp}
\label{thm:comp}
$\R(f \circ g)\geq \Omega(\R(f)\LR(g))$ for all partial boolean functions $f,g$.
\end{restatable}
\vspace{-3mm}
\begin{restatable}{theorem}{Optimal}
\label{thm:optimal}
\Cref{thm:comp} is optimal: If $\M$ is any complexity measure such that $\R(f \circ g)\geq \Omega(\R(f)\M(g))$ for all partial $f,g$, then $\LR(g)\geq \Omega(\M(g))$ for all partial $g$.
\end{restatable}

Additionally, $\LR$ itself satisfies a composition theorem as well.

\begin{restatable}{theorem}{LRComp}
\label{thm:LRcomp}
$\LR(f\circ g)\ge \Omega(\LR(f)\LR(g))$
for all partial boolean functions $f,g$.
\end{restatable}

\subsection{Comparison with previous work} \label{sec:previous-work}

The randomised composition conjecture for general boolean functions was first explicitly raised in~\cite{Ben-David2016}. Several complexity measures have since been shown to satisfy an inner composition theorem, including:
\begin{enumerate}
\item (block-)sensitivity $\s$, $\bs$~\cite{Aaronson2016},
\item randomised sabotage complexity $\RS$~\cite{Ben-David2016},
\item randomised complexity $\R_\delta$ with small-bias error $\delta \coloneqq {1/2-1/n^4}$~\cite{Anshu2018},
\item max-conflict complexity $\chibar$~\cite{Gavinsky2019} (also studied in~\cite{Li2021}).
\end{enumerate}
By our optimality theorem, we have $\LR(f)\geq \Omega(\M(f))$ for all $\M\in\{\s,\bs,\RS,\R_\delta,\chibar\}$ and all $f$. In fact, we can show that the largest of the above measures, namely $\chibar$, can sometimes be polynomially smaller than $\LR$.\footnote{Technically,
it does not seem to be known in the literature whether $\R_\delta$
is always at most $\chibar$; this doesn't matter much for our
purposes, as $\LR$ is larger than both
and it is easy to separate $\LR$ from $\R_\delta$
(for example with the $\textsc{Or}$ function).}
\begin{restatable}{lemma}{ChibarGap} \label{lem:chibar-gap}
There exists a partial $f$ such that $\LR(f)\geq \Omega(\chibar(f)^{1.5})$.
\end{restatable}

Previous work has also investigated complexity measures $\M$ that admit an \emph{outer} composition theorem, that is, $\R(f\circ g)\geq\Omega(\M(f)\R(g))$ for all partial $f,g$. These measures include:
\begin{enumerate}
\item sensitivity $\s$~\cite{Goos2018} (which was applied in~\cite{Ambainis2016}),
\item fractional block sensitivity $\fbs$~\cite{Bassilakis2020},
\item noisy randomised complexity~$\noisyR$~\cite{BenDavid2020comp} (also studied in~\cite{Girish2021}).
\end{enumerate}
In particular, $\noisyR$ is known to be \emph{outer-optimal}: if we have $\R(f \circ g)\geq \Omega(\M(f)\R(g))$ for all partial $f,g$, then~$\noisyR(f)\geq \Omega(\M(f))$ for all partial $f$. Our result can be viewed as an inner analogue of this.

Finally, we mention that randomised composition has also been studied in the \emph{super-multiplicative} regime, where we have examples of functions $f,g$ with $\R(f\circ g)\geq \omega(\R(f)\R(g))$. Tight bounds exist when the outer function is identity~\cite{Blais2019} (building on~\cite{Jain2010,Ben-David2016}), parity~\cite{Brody2020}, or majority~\cite{BenDavid2020amp,Goos2021}.

\subsection{On small-bias minimax}\label{sec:small_bias_minimax}

Our second result addresses a question of Yao~\cite{Yao1977}. Yao-style minimax theorems are routinely used to construct and analyse hard input distributions (including in our proof of the  new composition theorem). For example, $\R_\epsilon$ admits a distributional characterisation as
\begin{equation}\label{eq:worst-minimax}
\R_\epsilon(f) ~=~ \max_\mu \min_{R\in\R(f,\epsilon,\mu)} \depth(R),
\end{equation}
where $\mu$ ranges over distributions on $\Dom(f)$; the set $\R(f,\epsilon,\mu)$ consists of trees $R$ with $\E_{x\sim\mu}[\err(R,x)]\leq \epsilon$; and $\depth(R)$ is the worst-case cost of $R$, that is, maximum number of queries over all inputs (and internal randomness if $R$ is randomised). While the worst-case cost setting is perhaps what is most widely studied up to this day, Yao's original paper discussed, in fact, exclusively the expected cost setting. It is the expected cost setting that is currently undergoing a renaissance as it has proven important in the randomised composition literature surveyed above (\cref{sec:previous-work}).

\paragraph{Minimax for expected cost.}
We define the \emph{$\epsilon$-error expected query complexity} and the \emph{$\epsilon$-error distributional expected query complexity} by
\begin{align*}
\bR_{\epsilon}(f) &~\coloneqq~~~ \min_{\mathclap{R\in\R(f,\epsilon)}}~~~ \max_x~~ \cost(R,x),\\
\bD_\epsilon(f) &~\coloneqq~~~ \max_\mu~~~ \min_{\mathclap{R\in\R(f,\epsilon,\mu)}}~~ \cost(R,\mu),
\end{align*}
where $\R(f,\epsilon)$ is the set of randomised trees $R$ such that $\err(R,x)\leq \epsilon$ for all inputs $x$; and $\cost(R,\mu)\coloneqq \E_{x\sim\mu}[\cost(R,x)]$ is the expected cost over $\mu$ (and internal randomness of $R$). We note that the set $\R(f,\epsilon,\mu)$ is sometimes restricted to contain only deterministic algorithms wlog (as can be done in \cref{eq:worst-minimax}), but in the expected cost setting this may not necessarily be t†the case (see~\cref{open:small_bias}); hence we allow $\R(f,\epsilon,\mu)$ to contain randomised trees.

Yao showed an exact distributional characterisation for zero-error algorithms, namely, $\bR_0(f)=\bD_0(f)$, and moreover, the optimal distributional algorithm is deterministic. He asked if a similar characterisation holds in the case $\epsilon>0$. He observed that the ``easy'' direction of minimax, $\bD_\epsilon(f) \leq \bR_\epsilon(f)$, certainly holds (although Yao's version of this inequality had some loss in parameters as he was restricted to deterministic algorithms). Vereshchagin~\cite{Vereshchagin1998} proved the ``hard'' direction with a modest loss in parameters; in summary,
\[\textstyle
\bD_\epsilon(f) ~\leq~ \bR_\epsilon(f) ~\leq~ 2\bD_{\epsilon/2}(f).
\] 
These bounds give a satisfying distributional characterisation in the bounded-error case. What happens in the small-bias case $\epsilon=1/2-o(1)$? Our second result shows that, surprisingly, the distributional characterisation fails in a particularly strong sense. We write $\dot{\delta} = (1-\delta)/2$ for short.
\begin{theorem} \label{thm:failure-minimax}
There is an $n$-bit partial function $f$ and a bias $\delta(n)=o(1)$ such that
$\bR_{\dot{\delta}}(f) \geq \bD_{\dot{\delta}}(f)^{1+\Omega(1)}.$
\end{theorem}
This theorem says that there is no way to capture $\bR_\epsilon(f)$ relative to a \emph{single} hard distribution. However, there does exist a distributional characterisation using a pair of distributions, as we explore next.

\subsection{Discussion: How are our two results related?}

Suppose we want to prove an inner composition theorem. All the previous proofs~\cite{Ben-David2016,Anshu2018,Gavinsky2019} revolve around the following high-level idea.
Let $R$ be a randomised tree that on input $x$ seeks to compute~$f(g(x^1),\ldots,g(x^n))$. The tree can invest different numbers of queries~$q_i$ to different components~$x^i$, making $q=\sum_i q_i$ queries in total. If we had a complexity measure $\M(g)$ that allowed us to bound the bias the tree has for the $i$-th component $g(x^i)$ as \emph{a linear function} of $q_i$---say, the bias for~$g(x^i)$ is at most~$q_i/\M(g)$---then, by \emph{linearity of expectation}, the expected total sum of the biases for all components~$g(x^1),\ldots,g(x^n)$ is at most $q/\M(g)$. This would allow us to track the total progress $R$ is making across all the inner functions.

What is the largest such ``linearised'' measure $\M$? The most natural attempt at a definition (which the authors of this paper studied for a long time before finding the correct definition of $\LR$) runs as follows. The measure should be such that with $q \coloneqq \bR_{\dot{\delta}}(f)$ queries one gets bias at most $\delta\leq q/\M(f)$. Optimising for~$\M(f)$ this suggest the following definition (a competitor for $\LR$)
\[
\ULR(f)
~\coloneqq~ \min_{\delta>0} \frac{\bR_{\dot{\delta}}(f)}{\delta}
~=~ \min_R\max_{x,y} \frac{\cost(R,x)}{\bias(R,y)}.
\]
We call it \emph{uniform}-$\LR$, since the tree $R$ that achieves the minimum has an upper bound on $\cost(R,x)$ that is uniformly the same for all $x$, and similarly there is a uniform lower bound on $\bias(R,x)$ for all $x$. By contrast, the definition of $\LR(f)$ is \emph{non-uniform}: a tree $R$ that achieves the minimum for $\LR(f)$ has only a bound on the cost/bias \emph{ratio}, but the individual cost and bias functions can vary wildly as a function of $x$.

We clearly have $\LR(f)\leq \ULR(f)$ by definition. How about the converse? It is enlightening to compare the distributional characterisations of these two measures, which can be derived using the recent minimax theorem for ratios of bilinear functions~\cite{BenDavid2020minimax}:
\begin{align}
\LR(f) ~\coloneqq~ \min_R\max_x \frac{\cost(R,x)}{\bias(R,x)}
&~=~ \max_\mu \min_R \frac{\cost(R,\mu)}{\bias(R,\mu)} ,\\[1mm]
\ULR(f) ~\coloneqq~ \min_R\max_{x,y} \frac{\cost(R,x)}{\bias(R,y)}
&~=~ \max_{\mu,\nu} \min_R \frac{\cost(R,\mu)}{\bias(R,\nu)}. \label{eq:minimax-ulr}
\end{align}
Here, $\LR$ is captured using a single hard distribution $\mu$ such that both cost and bias are measured against it. By contrast, $\ULR$ needs a pair of distributions $\mu,\nu$, one to measure the cost, one to measure the bias. The upshot is that we are able to show that the two measures are polynomially separated.
\begin{restatable}{theorem}{LrUlr}
\label{thm:lr-ulr}
There is an $n$-bit partial function $f$ such that $\ULR(f)\geq \Omega(\LR(f)^{5/4})\geq n^{\Omega(1)}$.
\end{restatable}

Our optimality theorem thus implies that $\ULR$ \emph{cannot} satisfy an inner composition theorem. This means that our attempt at finding a ``linearised'' measure at the start of this section missed a subtlety, namely, Yao's question: can we capture our measure relative to a single hard distribution? Our proof of the composition theorem will rely heavily on the fact that $\LR$ admits a single hard distribution. Our separation of $\LR$ and~$\ULR$ is what allows us to prove the impossibility of capturing~$\bR_\epsilon(f)$ relative to a single distribution. Indeed, \cref{thm:failure-minimax} can be derived from \cref{thm:lr-ulr} simply as follows.

\begin{proof}[Proof of \cref{thm:failure-minimax}]
Let $f$ be as in \cref{thm:lr-ulr} and let $R$ be a randomised tree witnessing $\LR(f)$. We may assume wlog that $\cost(R,x)\geq 1$ for all $x$. (If $R$ places a lot of weight on a $0$-cost tree, we may re-weight~$R$ without affecting the cost/bias ratio; see~\cref{lem:LR-always-query} for details.) Thus $\bias(R,x)\geq 1/n\eqqcolon\delta$ for all $x$.
We show the following inequalities, which would prove \Cref{thm:failure-minimax}.
\begin{align}
\bR_{\dot{\delta}}(f) &~\geq~ \delta\cdot\ULR(f), \label{eq:bR} \\
\bD_{\dot{\delta}}(f) &~\leq~ \delta\cdot\LR(f), \label{eq:bD}
\end{align}
 Indeed, \cref{eq:bR} holds since $\ULR(f)\leq \bR_{\dot{\delta}}(f)/\delta$ by the definition of $\ULR$. For \cref{eq:bD} consider any input distribution $\mu$. Define $R'$ as the randomised tree that with probability $\lambda\coloneqq\delta/\bias(R,\mu)$ runs $R$, and with probability $1-\lambda$ makes no queries and outputs a random 0/1 answer. Then $\bias(R',\mu)=\lambda\bias(R,\mu)=\delta$ and $\cost(R',\mu)= \lambda\cost(R,\mu) =\delta\cost(R,\mu)/\bias(R,\mu)\leq \delta\LR(f)$, as desired.
\end{proof}

\subsection{Techniques}

\paragraph{Composition theorem.}
Our first result, the inner-optimal composition theorem, is proved in \cref{part:one}.
As in other composition theorems for randomised algorithms, we start with
a randomised algorithm $R$ for the composition $f\circ g$ as well as hard distributions
$\mu_0$ and $\mu_1$ for $g$ (corresponding to distributions on $g^{-1}(0)$ and $g^{-1}(1)$),
and we construct a randomised algorithm $R'$ for $f$ whose cost is significantly
lower than that of $R$ (we need the cost to decrease by a factor of $\LR(g)$).
The algorithm $R'$ will simulate $R$, but not every query that $R$ makes to the
large, $mn$-sized input to $f\circ g$ will turn into a query to the smaller,
$n$-sized input to $f$ that $R'$ has access to. Instead, $R'$ will attempt
to delay making a true query as long as possible, and instead when $R$ makes
a query $(i,j)$ (querying position $j$ inside copy $i$ of an input to $g$),
$R'$ will return an answer that is generated according to $\mu_0$ and $\mu_1$,
so long as these two distributions approximately agree on the answer to that query.

So far, this is the same strategy employed by several other composition theorems,
including in particular that of \cite{Gavinsky2019}. Our innovation comes
from the precise way we choose when to query the bit $i$ versus when to return
an artificially-generated query answer to the query $(i,j)$. Specifically,
in \sec{DTsim}, we prove the following simulation theorem for decision trees.
Suppose we are given two distributions $\mu_0$ and $\mu_1$, we are asked
to answer online queries to the bits of a string sampled from $\mu_b$
without knowing the value of $b$; moreover, suppose we have access to a big red button
that, when pressed, provides the value of $b\in\B$. Then there is a strategy to answer
these online queries with perfect soundness (i.e.\ with distribution
identical to sampling a string from $\mu_b$)
with the following guarantee: if the decision tree that is making the online queries is $D$,
then the probability we press the button is at most $\TV(\tran(D,\mu_0),\tran(D,\mu_1))$
(the total variation distance between the query outputs $D$ receives when run on $\mu_0$
and the query outputs $D$ receives when run on $\mu_1$).

This simulation theorem, though somewhat technical, ends up being stronger
than the simulation guarantee used by Gavinsky, Lee, Santha, and Sanyal~\cite{Gavinsky2019}
to provide their composition result for max-conflict complexity.
To get a composition theorem, we need to convert this total variation distance
between transcripts into a more natural measure; this can be done via some minimax
arguments, and the resulting measure is $\LR$. We note that
a similarly structured argument occurred in \cite{BenDavid2020comp},
but the squared-Hellinger distance between the transcripts appeared instead
of the total variation distance; in that result, the authors showed that
this squared-Hellinger distance between transcripts characterized $\R(g)$,
but they failed to construct a randomised algorithm $R'$ for $f$,
instead constructing only a ``noisy'' randomised algorithm. This gave them the
result $\R(f\circ g)=\Omega(\noisyR(f)\R(g))$. In contrast, the total variation
distance allows us to get $\R(f)$ on the outside, at the cost of getting only
$\LR(g)$ on the inside.

The measure $\LR$ is arguably more natural than max-conflict complexity,
but the real advantage is that our composition theorem turns out to be
the best possible of its type: if $\R(f\circ g)=\Omega(\R(f)\M(g))$ for all
partial functions $f$ and $g$, then $\LR(g)=\Omega(\M(g))$. To show this, we give
a characterization of $\LR(g)$ in terms of randomised
query complexity: there is a family of partial
functions $f_m$ such that for all partial functions~$g$,
we have
\[\LR(g)=\Theta\left(\frac{\R(f_m\circ g)}{\R(f_m)}\right),\]
where $m$ is the input size of $g$. Once we have this,
it clearly follows that $\R(f\circ g)=\Omega(\R(f)\M(g))$
implies $\LR(g)=\Omega(\M(g))$. The function family $f_m$
turns out to be the same as the one introduced in
\cite{BenDavid2020comp} (based on a family of relations
introduced in \cite{Gavinsky2019}); the randomised
query complexity $\R(f_m)$ was already established in that
paper, so all we need is an upper bound on $\R(f_m\circ g)$
which uses the existence of an $\LR$-style algorithm for~$g$.
The linear dependence on the bias which is built into the definition
of $\LR(g)$ turns out to be precisely what is needed to
upper bound $\R(f_m\circ g)$ (see \sec{optimality} for details).

\paragraph{Failure of small-bias minimax.}
Our second result, separation of $\LR$ and $\ULR$, is proved in \cref{part:two}. The function~$f$ that witnesses the separation $\ULR(f)\geq\Omega(\LR(f)^{5/4})$ is not hard to define. For simplicity, we denote its input length by $N\coloneqq Bn$ and think of the input as being composed of $B=n^c$ blocks (for some large constant $c$) of $n$ bits each. We define $f$ as a composition of $\Maj_B$ as an outer function, and~$\Xor_n$ as an inner function, where we are able to switch individual $\Xor$-blocks to be easy (requiring~$O(1)$ queries) or hard (requiring $n$ queries). Moreover, we make the following promises about the input. Either
\begin{enumerate}[label=(\arabic*)]
\item all blocks are easy, and a random block has a value with bias $1/n$ towards the majority value; or
\item $b\coloneqq n^{-3/4}$ fraction of the blocks are hard, and a random block has bias $\Omega(b)$ towards the majority.
\end{enumerate}
We claim that this function is easy for $\LR$, namely, $\LR(f)=O(n)$. To see this, consider the algorithm~$R$ that chooses a block at random, computes it, and outputs its value. For inputs $x$ of type (1) we have $\cost(R,x)=O(1)$ and $\bias(R,x)\geq 1/n$ so that cost/bias ratio is $O(n)$. For inputs $x$ of type (2) we have $\cost(R,x)= bn +(1-b)O(1)\leq O(bn)$ and $\bias(R,x)=b$ so that cost/bias ratio is $O(n)$ again. 

The difficult part is to show that $\ULR(f)\geq \Omega(n^{5/4})$. For example, the above algorithm $R$ has $\ULR$-style measure $\max_{x,y}\cost(R,x)/\bias(R,y)=O(bn)/(1/n)=O(n^{5/4})$, and we would like to show that this is optimal. Intuitively, it is hard to get large bias for inputs of type (1) (although query cost is small here) and it is hard to get low query cost for inputs of type (2) (although bias is relatively high here). We first argue that an algorithm that wants to keep $\cost(R,x)$ small uniformly for all $x$ (even those $x$ with high $\bias(R,x)$) cannot afford to solve hard blocks very often. This is formalised by picking an appropriate pair of hard distributions for $f$ according to the minimax formulation~\cref{eq:minimax-ulr}. What remains is the following task: Show that any algorithm that does not solve hard blocks, has large cost/bias ratio relative to a \emph{single} hard distribution, that is, show an $\LR$-style lower bound.

To this end, we develop a suite of techniques to prove lower bounds on the cost/bias trade-off achievable by decision trees in the small-bias expected cost setting, which has not really been studied in the literature before. Consequently, we end up having to re-establish some basic facts in the expected-cost setting that have been long known in the worst-case setting. For example, we show any algorithm for $\GapMaj_n$ (with~$\sqrt{n}$ gap promise) can achieve bias at most~$O(\sqrt{\cost/n})$ (see \cref{sec:two-hardness}). The proof here exploits the ``$\And$-trick'' used by Sherstov~\cite{Sherstov2012} to prove a lower bound on the (worst-case) randomised communication complexity of the gap-Hamming problem. These techniques also come in handy when we separate $\LR$ from~$\chibar$ for the proof of \cref{lem:chibar-gap}.

\subsection{Open questions}

The foremost open question is to resolve \cref{prob}. We can equivalently formulate it as follows.
\begin{open}[\cref{prob} rephrased]
Does $\LR(f)=\Theta(\R(f))$ for all total functions $f$?
\end{open}

One intriguing open problem regarding our new-found measure $\LR$ is to show that it is lower-bounded by quantum query complexity $\Q$. Indeed, the bias of a quantum algorithm can be amplified linearly in the query cost, so it seems sensible to conjecture this is so. However, quantum query complexity has mostly been studied in the worst-case setting, and it is unclear how one should even define quantum query complexity in expectation (in such a way that it supports linear bias amplification).
\begin{open}
Does it hold that $\LR(f)\ge \Q(f)$?
\end{open}
There is a second reason to care about this question, having
to do with the \emph{composition limit} of randomised algorithms.
Define $\R^*(f)\coloneqq \lim_{k\to\infty}\R(f^{\circ k})^{1/k}$;
this is the limit of the $k$-th root of the randomised query complexity
of the $k$-fold composition of $f$. Our results here imply that
$\R^*(f)\ge\Omega(\LR(f))$ for all (possibly partial) functions $f$.
Due to the composition theorem for quantum query complexity,
it is also known that $\R^*(f)\ge\Omega(\Q(f))$. The above
open problem asks whether one of these results dominates the other.
More generally, it would be nice to characterize $\R^*(f)$
in terms of a simpler measure (for instance, one which is efficiently
computable given the truth table of the function).

Our inner-optimal composition theorem for $\LR$, together with the outer-optimal composition theorem for~$\noisyR$~\cite{BenDavid2020comp} give a relatively satisfying picture of composition in the case of partial functions. However, we can still ask whether there remain other \emph{incomparable} composition theorems.
\begin{open}
Are there multiplicative composition theorems, stating that $\R(f\circ g)\geq\Omega(\M_1(f)\M_2(g))$ for all partial $f,g$, that can sometimes prove better lower bounds than $\Omega(\max\{\R(f)\LR(g),\noisyR(f)\R(g)\})$?
\end{open}

Regarding the failure of the distributational characterization of $\overline{\R}_\epsilon$ in the low bias regime (\cref{thm:failure-minimax}), one may wonder whether the definition of $\overline{\D}_\epsilon$ should really involve randomized decision trees instead of deterministic ones. As hinted in \cref{sec:small_bias_minimax}, while considering deterministic trees is the natural choice in the bounded error regime, we feel it might not be in the regime where $\epsilon \approx 1/2$. Indeed, while a randomised decision tree can get cost arbitrarily close to zero for $\epsilon$ approaching $1/2$ (by taking an appropriate mixture with the zero-query tree), a deterministic one will get stuck at making one query and thus cost 1. Deciding whether the two versions are equivalent (up to constant factors and additive terms) is our last open question.

\begin{open}\label{open:small_bias}
Let $\overline{\D}_\epsilon^\star(f) \coloneqq \max_{\mu} \min_{D \in \D(f, \epsilon, \mu)} \cost(D, \mu)$ where $\D(f, \epsilon, \mu)$ is the set of all deterministic decision trees solving $f$ with error at most $\epsilon$ relative to inputs sampled from $\mu$. For any partial $f$ and $\epsilon$, do we have $\overline{\D}_\epsilon^\star(f) \leq O(\overline{\D}_\epsilon(f)+ 1)$?
\end{open}

\section{Preliminaries}

\subsection{Query complexity notation}

Fix a natural number $n\in\bN$.
A total boolean function is a function $f\colon\B^n\to\B$.
We will consider several generalizations of total boolean functions:
first, there are \emph{partial} boolean functions, which
are defined on a domain which is a subset of $\B^n$.
We use $\Dom(f)\subseteq\B^n$ to denote the domain of such a function.
A further way to generalize boolean functions is to expand the
input and output alphabets; that is, for finite sets
$\Sigma_I$ and $\Sigma_O$, we can consider functions
$f\colon\Dom(f)\to\Sigma_O$ with $\Dom(f)\subseteq\Sigma_I^n$,
which take in input strings over the alphabet $\Sigma_I$
and output a symbol in $\Sigma_O$.

A still further way to generalize such functions is to consider
\emph{relations} instead of partial functions. A relation
is a subset of $\Sigma_I^n\times\Sigma_O$, or alternatively,
it is a function that maps $\Sigma_I^n$ to a subset of $\Sigma_O$.
Any partial function can be viewed as a (total) relation,
where on an input $x$ which is not in the domain of the partial function,
the corresponding relation relates all output symbols to $x$
(meaning that if $x$ is the input, any output symbol is considered
valid).

Given a boolean function $f$ (or, more generally, a relation),
we will denote its \emph{deterministic query complexity}
by $\D(f)$. This is the minimum height of a \emph{decision tree}
$D$ which correctly computes $f(x)$ on any $x\in\Dom(f)$;
in other words, it is the minimum number of worst-case adaptive queries
required by a deterministic algorithm computing $f$.
For a formal definition, see \cite{Buhrman2002}.

In this work we will mostly be dealing with randomised algorithms
rather than deterministic ones, so let us more carefully define those.
A \emph{randomised query algorithm} or \emph{randomised decision tree}
will be a probability distribution over deterministic decision trees.
Such deterministic decision trees will have internal nodes labeled
by $[n]\coloneqq\{1,2,\dots,n\}$
(representing the index of the input to query),
arcs labeled by $\Sigma_I$ (representing the symbol we might
see after querying an index),
and leaves labeled by $\Sigma_O$
(representing output symbols to return at the end of the algorithm).
We will assume that no internal node shares a label with an
ancestor, meaning that a deterministic algorithm does not query
the same index twice.

For such a randomised algorithm $R$
and for an input $x\in\Sigma_I^n$,
we denote by $R(x)$ the random variable we get by sampling
a deterministic tree $D$ from $R$, and returning $D(x)$
(the label of the leaf of $D$ reached after starting
from the root and taking the path determined by $x$).
For a function $f$, we write $\err_f(R,x)\coloneqq\Pr_R[R(x)\ne f(x)]$
(or $\Pr_R[R(x)\notin f(x)]$ if $f$ is a relation),
and we write $\bias^{\pm}_f(R,x)\coloneqq 1-2\err_f(R,x)$,
$\bias_f(R,x)\coloneqq \max\{\bias_f^{\pm}(R,x),0\}$;
we omit the subscript $f$ when it is clear from context.

For a deterministic tree $D$, let $\cost(D,x)$ be the number
of queries $D$ makes on input $x$; this is the height
of the leaf of $D$ that is reached when $D$ is run on $x$.
For a randomised algorithm $R$, we then define
$\cost(R,x)\coloneqq\E_{D\sim R}[\cost(D,x)]$
(this is the expected number of queries $R$ makes when run on $x$).

We extend both of the above to distributions $\mu$ over $\Sigma_I^n$
instead of just inputs $x$; that is, define
\[\bias^{\pm}_f(R,\mu)\coloneqq \E_{x\sim\mu}[\bias^{\pm}_f(R,x)]=\E_{x\sim\mu}\E_{D\sim R}[\bias^{\pm}_f(D,x)],\]
\[\cost(R,\mu)\coloneqq \E_{x\sim\mu}[\cost(R,x)]=\E_{x\sim\mu}\E_{D\sim R}[\cost(D,x)],\]
with $\bias_f(R,\mu)\coloneqq\max\{\bias_f^{\pm}(R,\mu),0\}$.
We also define $\tran(R,\mu)$ to be the random
variable we get by sampling a decision tree $D$ from $R$,
a string $x$ from $\mu$, and returning the pair $(D,\ell)$,
where $\ell$ is the leaf of $D$ reached when $D$ is
run on $x$. Intuitively, $\tran(R,\mu)$ is the ``transcript''
when $R$ is run on an input sampled from $\mu$,
and such a transcript records all information
that an agent running $R$ knows about the
input $x$ at the end of the algorithm.
We will use $\TV(\mu, \nu) \coloneqq \frac{1}{2} \sum_{x \in \mathcal{X}} |\mu[x] - \nu[x]|$ to denote the total variation distance between distributions $\mu$ and $\nu$ over set $\mathcal{X}$. Most often, we will employ it with respect to the transcript of $R$ on two
different distributions as a way to quantify
the extent to which $R$ can tell these distributions
apart.

We say that a randomised algorithm $R$
computes $f$ to error $\epsilon$ if
$\err_f(R,x)\le \epsilon$ for all $x\in\Dom(f)$.
We then let $\bR_\epsilon(f)$ the minimum
possible value of $\max_x\cost(R,x)$
over randomised algorithms $R$ satisfying
$\err_f(R,x)\le\epsilon$ for all $x\in\Dom(f)$.
We also use $\R_\epsilon(f)$
to denote the minimum number $T$ such that
there is a randomised algorithm $R$ with
$\err_f(R,x)\le \epsilon$ for all $x\in\Dom(f)$
such that all decision
trees in the support of $R$ have height at most $T$.
The difference between $\R_\epsilon(f)$
and $\bR_\epsilon(f)$ is that the former
measures the worst-case cost of an algorithm
computing $f$ to error $\epsilon$
(maximizing over both the input string and the
internal randomness), while the latter measures
the expected worst-case cost of the algorithm
computing $f$ to error $\epsilon$
(this still maximizes over the input strings $x$,
but takes an expectation over the internal randomness
of the algorithm $R$).

It is easy to see that 
$\bR_\epsilon(f)\le\R_\epsilon(f)$
for all $f$. The other direction also holds if
we tolerate a constant-factor loss, as well as
an additive constant loss in $\epsilon$;
to see this, note that if we cut off the 
$\bR_\epsilon(f)$ algorithm after it makes
$10$ times more queries than it is expected to,
then the probability of reaching such a cutoff is at most
$1/10$ by Markov's inequality, and hence the error
probability of the algorithm increases by at most
$1/10$; this converts an $\bR_\epsilon(f)$
algorithm into a $\R_\epsilon(f)$ algorithm.

Standard error reduction techniques imply that
for a boolean function $f$,
$\R_\epsilon(f)$ is related to $\R_{\epsilon'}(f)$
by a constant factor that depends only on $\epsilon$
and $\epsilon'$, so long as both are in $(0,1/2)$.
For this reason, the value of $\epsilon$ does
not matter when $\epsilon$ is a constant in $(0,1/2)$
(so long as we ignore constant factors and so long
as the function is boolean), so we omit
$\epsilon$ when $\epsilon=1/3$. The same error
reduction property holds for $\bR_\epsilon(f)$.
Combined with the Markov inequality argument
above, both $\R(f)$ and $\bR(f)$ are the same
measure (up to constant factors) for a boolean function
and for constant values of $\epsilon$.

We warn that these equivalences break if $f$
is not boolean (especially if $f$ is a relation)
or if the value of $\epsilon$ is not constant;
in particular, when $\epsilon=1/n$ or when
$\epsilon=1/2-1/n$, the values of
$\R_\epsilon(f)$ and $\bR_\epsilon(f)$
may differ by more than a constant factor.

\subsection{Linearised \texorpdfstring{$\R$}{R}} \label{sec:lr-characterisation}

For a (possibly partial) boolean function $f$ on $n$ bits,
we define
\[\LR(f)\coloneqq \min_R\max_x \frac{\cost(R,x)}{\bias(R,x)}.\]
Here $R$ ranges over randomised decision trees and $x$
ranges over the domain of $f$, and 
we treat $0/0$ as $\infty$.

We call this measure \emph{linearised randomised query complexity}.
The name comes from the linear dependence on the bias achieved
by the algorithm. Note that if we wanted to amplify bias $\gamma$
to constant bias, we would, in general, have to repeat the algorithm
$\Theta(1/\gamma^2)$ times to do so. In some sense, then,
the measure $\R(f)$ charges $1/\gamma^2$ for an algorithm
that achieves bias $\gamma$ instead of achieving constant bias.
The measure $\LR(f)$, in contrast, charges only $1/\gamma$
for such an algorithm, so it can be up to quadratically smaller
than $\R(f)$.

A minimax theorem for ratios such as
\cite{BenDavid2020minimax} (Theorem~2.18) can show that
\begin{equation}
\LR(f)=\max_{\mu} \min_D \frac{\cost(D,\mu)}{\bias(D,\mu)},
\label{eq:LRminimax}
\end{equation}
where $D$ ranges over deterministic decision trees and $\mu$
ranges over probability distributions over $\Dom(f)$.

It is not hard to see that the maximizing distribution $\mu$
above will place equal weight on $0$ and $1$ inputs.
This is because otherwise, we could take $D$ to be a decision
tree that makes $0$ queries, and then $\cost(D,\mu)$ would be $0$
while $\bias(D,\mu)$ would be positive.

If $\mu$ is balanced over $0$ and $1$ inputs, we may express it as $\mu \coloneqq \mu_0/2 + \mu_1/2$ and it is not hard to show that for the best possible
choice of leaf labels for an unlabeled decision tree $D$,
we have
\begin{equation}\label{eq:tv-bias}
\bias(D,\mu)^\pm=\bias(D,\mu)=\TV(\tran(D,\mu_0),\tran(D,\mu_1)).
\end{equation}
This follows, for example, from \cite{BenDavid2020minimax} (Lemma~3.9);
to see this intuitively, recall that $\tran(D,\mu)$ is the
random variable for the leaf of $D$ reached when $D$ is run on $\mu$,
and note that the best choice of leaf label if $D$ reaches
a leaf $\ell$ is $0$ if the probability of $D$ reaching $\ell$
is higher when run on $\mu_0$ than on $\mu_1$, and it is $1$ otherwise.
Therefore, the bias for the best choice
of leaf labels is the sum, over leaves $\ell$ of $D$, of
$2\max\{\Pr_{\mu_0}[\ell],\Pr_{\mu_1}[\ell]\}-1$, which is easily
seen to be the total variation distance between the
two distributions over leaves.

Given \eq{tv-bias}, we can also write
\[\LR(f)=\max_{\mu_0,\mu_1}\min_D
\frac{\cost(D,\frac{\mu_0+\mu_1}{2})}
{\TV(\tran(D,\mu_0),\tran(D,\mu_1))},\]
where $\mu_0$ ranges over probability distributions
with support $f^{-1}(0)$ and $\mu_1$ ranges over probability
distributions with support $f^{-1}(1)$. Observe that neither
the top nor the bottom depend on the leaf labels of $D$,
so we can now assume $D$ is an unlabeled decision tree if we wish.
Note also that $\cost(D,\mu)$ is linear in the second argument,
so we can write
\[\LR(f)=\max_{\mu_0,\mu_1}\min_D
\frac{\cost(D,\mu_0)+\cost(D,\mu_1)}
{2\TV(\tran(D,\mu_0),\tran(D,\mu_1))}.\]

We clearly have
\[\LR(f)\ge \max_{\mu_0,\mu_1}\min_D
\frac{\min\{\cost(D,\mu_0),\cost(D,\mu_1)\}}
{\TV(\tran(D,\mu_0),\tran(D,\mu_1))}.\]

\begin{lemma}
For any fixed $\mu_0$ and $\mu_1$, we have
\[\min_D \frac{\cost(D,\mu_1)}{\TV(\tran(D,\mu_0),\tran(D,\mu_1))}
\le 6
\min_D \frac{\cost(D,\mu_0)}{\TV(\tran(D,\mu_0),\tran(D,\mu_1))}.\]
\end{lemma}

\begin{proof}
Let $D$ minimize the right-hand side. Suppose by contradiction
that $D$ makes a lot more queries against $\mu_1$ than it does
against $\mu_0$. The idea is to cut off paths in $D$ that
are too long, since in those paths we essentially already
know we are running on $\mu_1$ instead of $\mu_0$. The
new truncated tree $D'$ will still have roughly
the same total variation distance between the transcripts
when run on $\mu_0$ and $\mu_1$, but it won't make too many
more queries on $\mu_1$ than $D$ made on $\mu_0$.

Formally, we define $D'$ to be the same tree $D$ except
that we cut off an internal node $u$ of $D$ (making it a leaf in $D'$)
if $\Pr_{\mu_1}[u]/\Pr_{\mu_0}[u]>3$.
The new tree $D'$ makes fewer
queries on every input than $D$ did, so we clearly have
$\cost(D',\mu_b)\le \cost(D,\mu_b)$ for $b=0,1$.
Moreover, note that $\cost(D,\mu)$ is the sum, over all
internal nodes of $D$, of the probability that $D$ reaches
that node when run on $\mu$. Now, for all internal nodes of $D'$,
we know that $\Pr_{\mu_1}[u]\le 3\Pr_{\mu_0}[u]$, and hence we have
$\cost(D',\mu_1)\le 3\cost(D',\mu_0)\le 3\cost(D,\mu_0)$.

We next want to show that the distance
$\TV(\tran(D',\mu_0),\tran(D',\mu_1))$
is not much smaller than the distance
$\TV(\tran(D,\mu_0),\tran(D,\mu_1))$.
To see this, let $V$ be the set of all leaves in $D'$
that are also leaves in $D$, and let $*$ denote the event
that a cutoff occurred in $D'$. Then
\begin{align*}
\TV(\tran(D,\mu_0),\tran(D,\mu_1))
&=\frac{1}{2}\sum_v |\Pr_{D,\mu_0}[v]-\Pr_{D,\mu_1}[v]|
\\&\le \frac{1}{2}\sum_{v\in V} |\Pr_{D,\mu_0}[v]-\Pr_{D,\mu_1}[v]|
+\frac{1}{2}\sum_{v\notin V} \max\{\Pr_{D,\mu_0}[v],\Pr_{D,\mu_1}[v]\}
\\&\le \frac{1}{2}\sum_{v\in V} |\Pr_{D,\mu_0}[v]-\Pr_{D,\mu_1}[v]|
+\frac{\Pr_{D,\mu_0}[*]+\Pr_{D,\mu_1}[*]}{2},
\end{align*}
while
\begin{align*}
\TV(\tran(D',\mu_0),\tran(D',\mu_1))
&=\frac{1}{2}\sum_v |\Pr_{D',\mu_0}[v]-\Pr_{D',\mu_1}[v]|
\\&= \frac{1}{2}\sum_{v\in V} |\Pr_{D,\mu_0}[v]-\Pr_{D,\mu_1}[v]|
+\frac{1}{2}\sum_{v\notin V}
    \frac{\Pr_{D,\mu_1}[v]-\Pr_{D,\mu_0}[v]}
    {\Pr_{D,\mu_1}[v]+\Pr_{D,\mu_0}[v]}
    \cdot (\Pr_{D,\mu_1}[v]+\Pr_{D,\mu_0}[v])
\\&\ge \frac{1}{2}\sum_{v\in V} |\Pr_{D,\mu_0}[v]-\Pr_{D,\mu_1}[v]|
+\frac{1}{4}\sum_{v\notin V}
\Pr_{D,\mu_1}[v]+\Pr_{D,\mu_0}[v]
\\&= \frac{1}{2}\sum_{v\in V} |\Pr_{D,\mu_0}[v]-\Pr_{D,\mu_1}[v]|
+\frac{\Pr_{D,\mu_0}[*]+\Pr_{D,\mu_1}[*]}{4}.
\end{align*}
Hence the worst possible ratio between them is $1/2$,
and the greatest possible ratio between the left-hand side
and right-hand side of the original lemma is 
$6$, completing the proof.
\end{proof}

\begin{corollary}\label{cor:LR}
\[\max_{\mu_0,\mu_1}\min_D
\frac{\min\{\cost(D,\mu_0),\cost(D,\mu_1)\}}
{\TV(\tran(D,\mu_0),\tran(D,\mu_1))}\le\LR(f)\le 6\max_{\mu_0,\mu_1}\min_D
\frac{\min\{\cost(D,\mu_0),\cost(D,\mu_1)\}}
{\TV(\tran(D,\mu_0),\tran(D,\mu_1))}.\]
\end{corollary}

One useful property of $\LR$ complexity is that up to a multiplicative factor of 2, we can consider only randomised decision trees that always query at least one bit of their input.

\begin{lemma}
\label{lem:LR-always-query}
For every non-constant partial function $g$, there is a randomised decision tree $A$ that always queries at least one bit of $g$'s input and satisfies, for every $x$,
\[
\frac{\cost(A,x)}{\bias(A,x)} \le 2 \cdot \LR(g).
\]
\end{lemma}

\begin{proof}
Let $R$ be a randomised tree that achieves the minimum for $\LR(g)$. Write $R=\lambda_0 T_0 +\lambda_1 T_1+ \lambda_2 R_{\geq 1}$, $\lambda_0+\lambda_1+\lambda_2=1$, $\lambda_i\geq 0$, where $T_i$ is the cost-$0$ deterministic tree that makes no queries and outputs $i$, and where $R_{\geq 1}$ is a randomised tree that always makes at least $1$ query. Note that $\lambda_0,\lambda_1< 1/2$ as otherwise the tree would answer incorrectly with probability $\geq 1/2$ on some input. Let us assume wlog that $\lambda_0\leq \lambda_1$. Re-weight $R$ by defining a new randomised tree $R'\coloneqq \lambda'_1 T_1 + \lambda'_2 R_{\geq 1}$ where $\lambda'_1\coloneqq (\lambda_1-\lambda_0)/(1-2\lambda_0)$ and~$\lambda'_2\coloneqq \lambda_2/(1-2\lambda_0)$. Then, for all $x$, using $\cost((T_0+T_1)/2,x)=\bias((T_0+T_1)/2,x)=0$,
\[
\frac{\cost(R,x)}{\bias(R,x)}
=
\frac{2\lambda_0\cost((T_0+T_1)/2,x)+(1-2\lambda_0)\cost(\lambda_1'T_1+\lambda'_2 R_{\geq1},x)}%
{2\lambda_0\bias((T_0+T_1)/2,x)+(1-2\lambda_0)\bias(\lambda_1'T_1+\lambda'_2 R_{\geq1},x)}
=
\frac{\cost(R',x)}{\bias(R',x)}.
\]
Note that $\lambda'_1<1/2$ and thus $\cost(R',x)\geq 1/2$. Consider finally the tree $A\coloneqq \lambda'_1 T'_1 + \lambda'_2 R_{\geq 1}$ where $T'_1$ is the cost-$1$ tree that makes one (arbitrary) query and then outputs $1$. We have $\cost(A,x)/\bias(A,x)\leq (\cost(R',x)+1/2)/\bias(R',x)\leq 2\cost(R',x)/\bias(R',x)=2\cdot\LR(g)$.
\end{proof}

As a corollary, we obtain a universal lower bound on the $\LR$ complexity of every non-constant function.

\begin{corollary}
\label{cor:LR-universal-lb}
For every non-constant partial function $g$, $\LR(g) \ge \frac12$.
\end{corollary}

\begin{proof}
By \Cref{lem:LR-always-query}, there exists a randomised decision tree $A$ that always queries at least one bit of its input and satisfies $\cost(A,x)/\bias(A,x) \le 2 \cdot \LR(g)$ for all $x$ in the domain of $g$. But since $A$ always makes at least one query, $\cost(A,x) \ge 1$. And by definition, $\bias(A,x) \le 1$, so the cost-bias ratio of $A$ is always bounded below by $1$.
\end{proof}

\part{Composition Theorem}\label{part:one}

In this \cref{part:one}, we prove our inner-optimal composition theorem, \cref{thm:comp,thm:optimal} restated below, along with related results.

\Comp*
\Optimal*

The heart of the proof of \cref{thm:comp} is a simulation theorem showing that for any two distributions $\mu_0$ and $\mu_1$ and any decision tree $T$, it is possible to simulate $T$ on inputs drawn from $\mu_b$ for some initially unknown $b \in \{0,1\}$ while querying the actual value of $b$ with probability bounded by the total variation distance between the two distributions $\mu_0$ and $\mu_1$. This result,  \cref{thm:simulation}, is established in \cref{sec:DTsim}. 

In \cref{sec:composition}, we use the simulation theorem  to complete the proof of the main composition theorem, \cref{thm:composition}, a slightly more general version of \cref{thm:comp}. We also use the simulation theorem to establish the perfect composition for $\LR$ complexity, \cref{thm:LRcomp}, in this section.

The proof of \Cref{thm:optimal} is completed in \cref{sec:optimality}. Finally, in \cref{sec:chibar-sep}, we establish the separation between LR complexity and max-conflict complexity of \cref{lem:chibar-gap}.

\section{Decision tree simulation theorem}
\label{sec:DTsim}

An \emph{online decision tree simulator} is a randomised algorithm
that is given two distributions $\mu_0$ and $\mu_1$ on inputs
$\{0,1\}^n$, oracle access to a bit $b \in \{0,1\}$, and a stream of
queries $i_1,\ldots,i_k \in [n]$ that represent the queries made by a
decision tree $T$ that is not known to the algorithm. The goal of an
online decision tree simulator is to answer the queries according to
the distribution $\mu_b$ while querying the value of $b$ itself with
as small probability as possible. We think of this protocol
as having a big red button that gives $b$, and it tries
to pretend to have a sample from $\mu_b$ without pressing
the button for as long as possible.

\begin{theorem}
\label{thm:simulation}
There exists an online decision tree simulator that simulates the queries of $T$ on $\mu_b$ while querying the value of $b$ with probability 
$\TV\big(\tran(T,\mu_0),\tran(T,\mu_1)\big)$.
\end{theorem}

The algorithm that satisfies the theorem is stated below. In the algorithm, $x \in \{0,*,1\}^n$ is a partially defined boolean string: the coordinates labelled with $*$ are undefined. Given a string $x \in \{0,*,1\}^n$, an index $i \in [n]$, and a value $a \in \{0,1\}$, the notation $x^{(i \gets a)}$ denotes the string $y$ which equals $x$ on all coordinates except $i$, where it takes the value $y_i = a$.

\begin{algorithm}[ht]
\caption{\textsc{OnlineQuerySimulator}($\mu_0, \mu_1$)}
\For{all $x \in \{0,*,1\}^n$}{
	$\mu_{\mathrm{min}}(x) \gets \min\{\mu_0(x), \mu_1(x)\}$\;
}
$x \gets *^n$\;
$b \gets *$\;

\BlankLine
\While{more queries remain}{
	$i \gets \textsc{NextQuery}$\;
	$u \gets \mu_{\mathrm{min}}(x^{(i \gets 0)}) + \mu_{\mathrm{min}}(x^{(i \gets 1)})$\;

	\BlankLine
    \If{ $b = *$}{
    	With probability $1 - u/\mu_{\mathrm{min}}(x)$, query the value of $b$\;
    }

    \BlankLine
    \If{ $b = *$}{
    	$x_i \gets \mathrm{Ber}\big( \mu_{\mathrm{min}}(x^{(i \gets 1)})/u\big)$\;
    } \Else {
    	$x_i \gets \mathrm{Ber}\left( \frac{\mu_b(x^{(i \gets 1)}) - \mu_{\mathrm{min}}(x^{(i \gets 1)})}{\mu_b(x) - u} \right)$\;
    }
}
\end{algorithm}

Note that each vertex in a decision tree $T$ corresponds to the partial string $x \in \{0,1,*\}^n$ of the values revealed on the path to that vertex in $T$. Our main task is to show that each vertex in $T$ (including each leaf) is reached with probability $\mu_b(x)$ in the algorithm and that the probability that we reach $x$ \emph{and} don't reveal $b$ along the way is $\mu_{\mathrm{min}}(x)$.

\begin{lemma}
\label{lem:main}
For every $x \in \{0,1,*\}^n$, when we run the \textsc{OnlineQuerySimulator}, then
\begin{enumerate}
\item We reach the vertex $x$ with probability $\mu_b(x)$, and
\item We reach the vertex $x$ \emph{and} don't query the value $b$ on the way to $x$ with probability $\mu_{\mathrm{min}}(x)$.
\end{enumerate}
\end{lemma}

\begin{proof}
We prove the claim by induction on the number of defined coordinates on $x$. The base case corresponds to $x = *^n$, which trivially satisfies both conditions of the claim.

Consider now any $x \neq *^n$. Let $z$ be the parent of $x$ in the decision tree $T$, and let $i$ denote the coordinate where $z_i = *$ and $x_i \neq *$. Define also $y$ to be $x$'s sibling in $T$. Let us assume that $x_i = 1$. (The case where $x_i = 0$ is essentially identical.)

By the induction hypothesis, the probability that we reach $z$ and don't query the value $b$ is $\mu_{\mathrm{min}}(z)$. With probability $\big(\mu_{\mathrm{min}}(x) + \mu_{\mathrm{min}}(y)\big)/\mu_{\mathrm{min}}(z)$, we don't query the value of $b$ while processing the query $i$ either. And when this occurs the algorithm next reaches $x$ with probability $\mu_{\mathrm{min}}(x)/\big(\mu_{\mathrm{min}}(x) + \mu_{\mathrm{min}}(y)\big)$. So the overall probability that we reach $x$ without querying $b$ along the way is
\[
\mu_{\mathrm{min}}(z) \cdot \frac{\mu_{\mathrm{min}}(x) + \mu_{\mathrm{min}}(y)}{\mu_{\mathrm{min}}(z)} \cdot \frac{\mu_{\mathrm{min}}(x)}{\mu_{\mathrm{min}}(x) + \mu_{\mathrm{min}}(y)} = \mu_{\mathrm{min}}(x).
\]

Next, by the induction hypothesis again the probability that we query the value of $b$ either on the way to $z$ or while processing the query $i$ is
\[
(\mu_b(z) - \mu_{\mathrm{min}}(z)) + \mu_{\mathrm{min}}(z) \cdot \left( 1 - \frac{\mu_{\mathrm{min}}(x) + \mu_{\mathrm{min}}(y)}{\mu_{\mathrm{min}}(z)} \right) = \mu_b(z) - \big( \mu_{\mathrm{min}}(x) + \mu_{\mathrm{min}}(y) \big).
\]
Then the probability we output $x$ conditioned on having revealed $b$ is 
\[
\frac{\mu_b(x) - \mu_{\mathrm{min}}(x)}{\mu_b(z) - (\mu_{\mathrm{min}}(x) + \mu_{\mathrm{min}}(y))},
\]
so that the overall probability that we reach $x$ and reveal $b$ along the way is $\mu_b(x) - \mu_{\mathrm{min}}(x)$. Therefore, the overall probability that we reach $x$ is $\mu_b(x)$.
\end{proof}

The proof of \thm{simulation} is now essentially complete, as it just requires combining the lemma with a simple identity on total variation distance.

\begin{proof}[Proof of \cref{thm:simulation}.]
\lem{main} implies that the OracleQuerySimulator indeed reaches each leaf with the correct probability $\mu_b(x)$. And the probability that it queries the value of $b$ is $1 - \sum_{\ell \in T} \min\{\mu_0(\ell), \mu_1(\ell)\}$, which is the total variation distance between $\tran(T,\mu_0)$ and $\tran(T,\mu_1)$.
\end{proof}

\section{Composition theorems}
\label{sec:composition}

The inner-optimal composition theorem, \cref{thm:comp}, is established in \cref{sec:composition-inner-optimal}. In fact, we establish a slight generalization of that theorem, stated below in \cref{thm:composition}. Then the perfect composition theorem for LR complexity, \cref{thm:LRcomp}, is established in \cref{sec:composition-LR}.

\subsection{Composition for randomised query complexity}
\label{sec:composition-inner-optimal}

For a boolean string $y\in\B^n$ and a pair
of distributions $\mu_0,\mu_1$, we define $y\circ (\mu_0,\mu_1)$
to be the product distribution $\bigotimes_{i=1}^n \mu_{y_i}$.
In particular, if $\mu_0$ and $\mu_1$ are hard distributions
for the $0$- and $1$-inputs of $g$ respectively, and if $y$
is an input to $f$, then $y\circ (\mu_0,\mu_1)$ will give
a distribution over the inputs to the composed $f\circ g$
(all of which correspond to the same $f$-input $y$).

We prove the following composition theorem, which is a slightly more general version of \cref{thm:comp}.

\begin{theorem}
\label{thm:composition}
Let $\Sigma_I$ and $\Sigma_O$ be finite alphabets, and let
$n,m\in\mathbb{N}$.
Let $f\subseteq\B^n\times \Sigma_O$ be a (possibly partial)
relation on $n$ bits,
and let $g\colon \Dom(g)\to \B$ be a (possibly partial) boolean function,
with $\Dom(g)\subseteq \Sigma_I^m$. Let $\epsilon\in[0,1/2)$.
Then
\[\bR_{\epsilon}(f\circ g)\ge \bR_{\epsilon}(f)\LR(g)/6.\]
\end{theorem}

\begin{proof}
Let $\mu_0$ and $\mu_1$ be distributions over
the $0$-inputs and $1$-inputs to $g$, respectively,
that maximize the expression in the right-hand side of \cor{LR}.
Let $\Pi$ be the online decision tree simulator from
\thm{simulation}.
Let $R$ be a randomised algorithm that computes $f\circ g$
to error $\epsilon$ using $\bR_{\epsilon}(f\circ g)$
expected queries.
We describe
a randomised algorithm $R'$ for computing $f$ on worst-case
inputs.

Given input $y\in\B^n$, the algorithm $R'$ will instantiate
$n$ copies of $\Pi$, which we denote $\Pi_1,\Pi_2,\dots,\Pi_n$,
one for each bit of the input;
if protocol $\Pi_i$ presses the button, it gets $y_i$
(and this causes $R'$ to make a real query to the real input).
Each of these copies of $\Pi$ will assume the distributions
to be simulated are $\mu_0$ and $\mu_1$.
Then $R'$ will run $R$,
and whenever $R$ makes a query $(i,j)$ (corresponding
to querying bit $j$ inside of the $i$-th copy of $g$),
the algorithm $R'$ will ask $\Pi_i$ to give an answer to query
$j$, and it will use that answer to determine the next
query of $R$.

Note that since the protocols $\Pi_i$ are guaranteed to
be sound, the outcome of the simulation of $R$ made
by $R'$ is precisely the same (in distribution) as the outcome of
running $R$ on an input sampled from $y\circ(\mu_0,\mu_1)$.
Therefore, by the correctness guarantee of $R$, the output will be
a valid output for $f(y)$ except with error probability $\epsilon$.
It remains to show that for each $y\in\Dom(f)$, the expected
number of real queries $R'$ makes when run on $y$ is at most
$6\bR_\epsilon(f\circ g)/\LR(g)$.

Fix any $y\in\Dom(f)$.
Now, when $R'$ is run on $y$, let $T$ be the expected number
of fake queries it makes; in other words, let
$T=\cost(R,y\circ(\mu_0,\mu_1))\le \bR_{\epsilon}(f\circ g)$.
For each $i$,
let $T_i$ be the expected number of queries to $\Pi_i$
that $R'$ makes when run on $y$, so that $T_1+T_2+\dots+T_n=T$.
Let $p_i$ the overall probability that $\Pi_i$ presses
the button when $R'$ runs on $y$;
the sum $q=p_1+p_2+\dots+p_n$ is therefore the expected number
of real queries made by $R'$ on $y$. We would like to show that
$q\le 6T/\LR(g)$, or equivalently, $T/q\ge \LR(g)/6$.

Since $T/q=(T_1+\dots+T_n)/(p_1+\dots+p_n)$,
there must be some $i$ such that $T/q\ge T_i/p_i$.
It will therefore suffice to show that $T_i/p_i\ge \LR(g)/6$
for all $i\in[n]$. Fix such $i$, and recall that $T_i$
is the number of (fake) queries $R'$ makes to $\Pi_i$
when run on $y$, and $p_i$ is the probability that $\Pi_i$
presses the button when $R'$ is run on $y$.
Consider the algorithm $R_{y,i}$ which takes in an input
$x$ in $\Dom(g)$, generates $n-1$ additional fake inputs
to $g$ from the distributions $\mu_{y_\ell}$ for $\ell\ne i$,
places the real input $x$ as the $i$-th input among the
$n$ inputs to $g$, and runs $R$ on this tuple (treating
it as an input to $f\circ g$). Note that when $R_{y,i}$
is run on an input from $\mu_{y_i}$, its behavior is exactly
the same as the behavior of $R$ when run on $y\circ (\mu_0,\mu_1)$;
therefore, it makes $T_i$ expected queries.
Consider running $R_{y,i}$ with query answers generated by $\Pi$
instead of by making real queries; then when $\Pi$ uses the hidden
bit $y_i$ and simulates the distributions $\mu_0,\mu_1$,
the behavior of $R_{y,i}$ is the same as when we run it on
$\mu_{y_i}$, and hence the expected number of queries it makes to $\Pi$
is $T_i$ and the probability that $\Pi$ presses the button is exactly
$p_i$.

Now, by \thm{simulation}, we know that $\Pi$ presses the button
with probability $\TV(\tran(D,\mu_0),\tran(D,\mu_1))$
when simulating a deterministic decision tree $D$. For a random
decision tree such as the one given by $R_{y,i}$, the probability
$p_i$ of the button being pressed will be the mixture of the
values $\TV(\tran(D,\mu_0),\tran(D,\mu_1))$ for the deterministic
decision trees $D$ in the support of $R_{y,i}$. Also,
the expected number of queries $T_i$ that $R_{y,i}$ makes
is a matching mixture of the expected number of queries
made by the decision trees $D$ in the support of $R_{y,i}$;
the latter is $\cost(D,\mu_{y_i})$. Hence to lower bound
$T_i/p_i$, it will suffice to lower bound
$\frac{\cost(D,\mu_{y_i})}{\TV(\tran(D,\mu_0),\tran(D,\mu_1))}$
for all deterministic decision trees $D$ acting on inputs in $\Dom(g)$.
We now write
\[\frac{\cost(D,\mu_{y_i})}{\TV(\tran(D,\mu_0),\tran(D,\mu_1))}
\ge \frac{\min\{\cost(D,\mu_0),\cost(D,\mu_1)\}}
{\TV(\tran(D,\mu_0),\tran(D,\mu_1))}\ge\LR(g)/6\]
(using \cor{LR}). The desired result follows.
\end{proof}

\subsection{Composition for \texorpdfstring{$\LR$}{LR} complexity}
\label{sec:composition-LR}

\LRComp*

We actually prove the more explicit result
$\LR(f\circ g)\ge \LR(f)\LR(g)/6$.

\begin{proof}
The proof is similar to that of \thm{composition}.
We fix hard distributions $\mu_0$ and $\mu_1$ for $\LR(g)$,
and we fix a randomised algorithm $R$ for $f\circ g$
such that $\max_z \cost(R,z)/\bias(R,z)\le \LR(f\circ g)$.
We then define a randomised algorithm $R'$ for $f$;
this time, unlike in the proof of \thm{composition}, we want $R'$
to solve $f$ in the $\LR(f)$ sense instead of being
a randomised algorithm that solves $f$ to error $\epsilon$.
We define $R'$ as before: on input $y\in\Dom(f)$,
$R'$ instantiates $n$ protocols $\Pi_i$, one for each bit of $y$;
it instantiates each with the distributions $(\mu_0,\mu_1)$,
and gives $\Pi_i$ the hidden bit $y_i$ if it presses the button.
Then $R'$ will run $R$, and whenever $R$ makes a query $(i,j)$
(to the bit $j$ inside the $i$-th input to $g$), $R'$ will
ask $\Pi_i$ for bit $j$.

Note that by the soundness of the protocols $\Pi_i$,
we have $\bias(R',y)=\bias(R,y\circ(\mu_0,\mu_1))$.
We will next show that
$\cost(R',y)\le 6\cost(R,y\circ(\mu_0,\mu_1))/\LR(g)$;
This way, we will have
\[\LR(f)=\max_y\frac{\cost(R',y)}{\bias(R',y)}
\le \frac{6}{\LR(g)}\max_y\frac{\cost(R,y\circ(\mu_0,\mu_1))}
    {\bias(R,y\circ(\mu_0,\mu_1))}
\le \frac{6}{\LR(g)}\max_z\frac{\cost(R,z)}{\bias(R,z)}
\le \frac{6\LR(f\circ g)}{\LR(g)}.\]

Fix any $y\in\Dom(f)$; it remains to show that
$\cost(R,y\circ(\mu_0,\mu_1))/\cost(R',y)\ge \LR(g)/6$.
For every $i\in[n]$, let $T_i$ be the expected number of queries
$R$ makes to the $i$-th input
on $y\circ(\mu_0,\mu_1)$, and let $p_i$ be the probability that
$R'$ queries the $i$-th bit when run on input $y$. Let
$T=T_1+\dots +T_n$, and let $q=p_1+\dots +p_n$. We wish to show
$T/q\ge \LR(g)/6$. This precise statement was shown in the proof
of \thm{composition}, which completes this proof as well.
%
\end{proof}


\section{Optimality of the composition theorem}
\label{sec:optimality}

We complete the proof of \cref{thm:optimal} in this section.

\Optimal*

The proof of \cref{thm:optimal} is obtained by a characterization of $\LR$ complexity in terms of the complexity of functions composed with the \emph{approximate index} partial function $\textsc{ApproxIndex}_k : \{0,1\}^k \times \{0,1,2\}^{2^k} \to \{0,1,*\}$ defined by
\[
\textsc{ApproxIndex}_k(a,y) = 
\begin{cases}
y_a & \mbox{if $y_a \in \{0,1\}$,} \\
& \quad \mbox{$y_b = y_a$ for all $|b-a| \le \frac k2 - 2\sqrt{k \log k}$, and} \\
& \quad \mbox{$y_b = 2$ for all other $b$} \\
* & \mbox{otherwise.}
\end{cases}
\]
The randomised query complexity of the approximate index function is as follows.

\begin{lemma}[{\cite[Lemma 27]{BenDavid2020comp}}]
\label{lem:ApproxIndex}
$\R(\textsc{ApproxIndex}_k) = \Theta\left( \sqrt{k \log k}\right)$.
\end{lemma} 

The key to the proof of \cref{thm:optimal} is the following characterization of LR complexity in terms of composition with the approximate index function.

\begin{lemma}
\label{lem:LR-characterization}
For every partial boolean function $g : \Sigma^m \to \{0,1,*\}$, when $k \in \mathbb{N}$ satisfies $\frac{k}{\log k} \ge (36m)^2$ then
\[
\LR(g) = \Theta \left( \frac{\R(\textsc{ApproxIndex}_k \circ g)}{ \R(\textsc{ApproxIndex}_k)} \right).
\]
\end{lemma}

\begin{proof}
The lemma trivially holds when $g$ is a constant function. For the rest of the proof, fix $g$ to be any non-constant partial function.
\thm{comp} implies the upper bound 
\[
\LR(g) = O \left( \frac{\R(\textsc{ApproxIndex}_k \circ g)}{ \R(\textsc{ApproxIndex}_k)} \right).
\]
The goal of the remainder of the proof is to establish a matching lower bound by showing that
\[
\bR(\textsc{ApproxIndex}_k \circ g) = O \left( \sqrt{k \log k} \cdot \LR(g) \right).
\]
This bound suffices to complete the proof because $\bR(f) = \Theta(\R(f))$ for every partial function $f$.

Let $R$ denote a randomised algorithm that satisfies 
\[
\cost(R,x) \le 2\cdot \LR(g) \cdot \mathrm{bias}(R,x)
\]
for all $x$ in the domain of $g$ and always queries at least one bit of its input. Such an algorithm is guaranteed to exist by \Cref{lem:LR-always-query}. 
We define a new randomised algorithm $A$ that proceeds as follows: it runs the algorithm $R$ sequentially on the first instances $x_1,x_2,\ldots,x_\ell$ of $g$ which correspond to the initial address bits of the input to $\textsc{ApproxIndex}_k$. It continues this process until the total number of queries made to the underlying inputs exceeds $36\sqrt{k \log k} \cdot \LR(g)$. By the choice of $k$ and the trivial bound $\LR(g) \le m$, this process terminates when $R$ has computed the first $\ell$ instances of $g$ with some biases $b_1,\ldots,b_\ell$ for some $\ell \le k$. The algorithm $A$ then guesses the value of the remaining $k - \ell$ bits of the address. It finally computes the value of $g$ on the instance corresponding to the address obtained with error probability at most $\frac19$ and returns that value.

Let $c_1,\ldots,c_\ell$ denote the query cost incurred by $R$ when running on the $\ell$ computed instances of $g$. The random variables $(X_i)_{i \le k}$ defined by $X_i = \sum_{j \le i} c_j - \cost(R,x_j)$ form a discrete-time martingale and $\ell$ is the stopping time of this martingale. By the optional stopping theorem, $\mathrm{E}[X_\ell] = 0$. 
So $\mathrm{E}[\sum_{i \le \ell} c_i] = \sum_{i \le \ell} \cost(R,x_i)$. By Markov's inequality, the probability that the total cost exceeds $6$ times the expected cost on the same inputs is at most $1/6$; let us consider from now on only the case when this does not occur. In this case,
\[
\sum_{i=1}^\ell \cost(R,x_i) \ge \frac16 \sum_{i=1}^\ell c_i \ge 6\sqrt{k \log k}\cdot \LR(g).
\]
By our choice of $R$, the biases $\beta_1,\ldots,\beta_\ell$ on the values $g(x_1),\ldots,g(x_\ell)$ satisfy 
$\sum_{i=1}^\ell \beta_i \ge 3 \sqrt{k \log k}$
and so if we let $b \in \{0,1\}^k$ denote the address computed by the algorithm, we observe that
\[
\mathrm{E}\big[|b - a|\big] = \sum_{i=1}^k \Pr[ b_i \neq g(x_i)] \le \frac{k}2 - 3\sqrt{k \log k}.
\]
Furthermore, each of the $k$ events $b_i \neq g(x_i)$ are independent. So by Hoeffding's bound the probability that more than $\frac{k}2 - 2\sqrt{k \log k}$ of these events occur is at most $e^{-2\log^2 k}$, which is less than $\frac19$ when $k \ge 3$. 
When this event does not occur, the address $b$ computed by the algorithm satisfies $x_b = x_a$. Since $A$ lastly computes $g(x_b)$ with error at most $\frac19$, in total it computes $\textsc{ApproxIndex}_k \circ g$ with error at most $\frac13$.

It remains to show that the expected query cost of the algorithm $A$ satisfies the desired bound. The first round of the algorithm uses at most $36\sqrt{k \log k} \cdot \LR(g)$ queries plus the number of queries of the instance of $R$ run on $x_\ell$. In expectation, this additional number of queries is at most $\cost(R,x_\ell) \le \LR(g)$. And then computing $g(x_b)$ requires another $\R(g) \le m < \sqrt{k}$ queries. So the overall expected query complexity of $A$ is at most $(36\sqrt{k \log k} + 1) \cdot \LR(g) + \sqrt{k}$. By \Cref{cor:LR-universal-lb}, $\LR(g) \ge \frac12$ for every non-constant function $g$ so this query complexity is bounded above by $O\big(\sqrt{k \log k} \cdot \LR(g)\big)$, as required.
\end{proof}

The proof of \cref{thm:optimal} now follows easily from \cref{lem:LR-characterization}.

\begin{proof}[Proof of \cref{thm:optimal}]
Let $M$ be a measure that satisfies the condition of the theorem. Then, choosing $f$ to be the $\textsc{ApproxIndex}_{k}$ function for a large enough value of $k$ and applying \cref{lem:LR-characterization}, we obtain
\[
M(g) = O\left( \frac{\R(\textsc{ApproxIndex}_{k} \circ g)}{\R(\textsc{ApproxIndex}_{k})}\right) 
= O\left(\LR(g)\right). \qedhere
\]
\end{proof}

\section{Separation from \texorpdfstring{$\chibar$}{chi}} \label{sec:chibar-sep}
In this section, we exhibit a polynomial separation between $\LR$ and $\chibar$, the max conflict complexity introduced by Gavinsky, Lee, Santha
and Sanyal in~\cite{Gavinsky2019} (see \cref{sec:max_conflict} for a formal definition of $\chibar$).%
\ChibarGap*
\begin{proof}
The function $f$ we build takes input of size $n^2 + \sqrt{n}$ with format $(x_1,\, \dots,\, x_{n^2},\, a_1,\, \dots,\,  a_{\sqrt{n}})$. The function value is given as the parity of $\GapMaj(x)$  and $\Xor(a)$, i.e.:
$$f(x_1,\, \dots,\, x_{n^2},\, a_1,\, \dots,\,  a_{\sqrt{n}}) = \GapMaj^{n^2}_{n^{-1/2}}(x_1,\, \dots\, , \, x_{n^2}) \oplus \Xor_{\sqrt{n}}(a_1, \, \dots, \, a_{\sqrt{n}})$$
$\GapMaj^{n^2}_{n^{-1/2}}(x)$ is the majority function on $n^2$ bits with promise that $|x| \notin [n^2/2 - n^{3/2}, n^2/2 + n^{3/2}]$ so that returning the value of a random index holds bias at least $n^{-1/2}$. Thus, $f$ is a partial function whose domain is constrained by the gap majority instance. \cref{lemma:lr_F} shows that $\LR(f) \geq \Omega\left(n^{3/4}\right)$ and \cref{lemma:chibar_F} that $\chibar(f) \leq O(n^{1/2})$, as desired.
\end{proof}

\begin{lemma}\label{lemma:lr_F}
$\LR(F) \geq \Omega\left(n^{3/4}\right)$
\end{lemma}
\begin{proof}
To obtain the lower-bound, we define a pair of distributions $P^0$, $P^1$ over $f^{-1}(0), f^{-1}(1)$ and use the minimax theorem for $\LR$ (see \eq{LRminimax}):
$$\LR(f) = \max_{\mu} \min_{D} \frac{\cost(D, \mu)}{\bias_f(D, \mu)} \geq \min_D \frac{\cost(D, P)}{\TV(\tran(D, P^0), \tran(D, P^1))} \quad\text{where}\quad P \coloneqq (P_1 + P_2)/2$$
Let $\mu^0$ be the hard distribution for no-instances of $\GapMaj$, i.e $\mu^0$ is uniform over all strings of Hamming weight $n^2 - n^{3/2}$. Similarly, we let $\mu^1$ be the uniform distribution of all strings of Hamming weight $n^2 + n^{3/2}$. Define further $\nu^0$, respectively $\nu^1$ to be the uniform distribution over even-parity, respectively odd-parity strings of size $n^{1/2}$. With those base distribution in hand, we define $P^0$ and $P^1$:
$$P^0 \coloneqq \frac{\mu^0 \times \nu^0}{2} + \frac{\mu^1 \times \nu^1}{2} \quad\text{and}\quad P^1 \coloneqq \frac{\mu^1 \times \nu^0}{2} + \frac{\mu^0 \times \nu^1}{2}$$
Fix now any decision tree $D$ and let us argue that $\cost(D, P)/\TV(\tran(D, P^0), \tran(D, P^1)) \geq \Omega\left(n^{3/4}\right)$. Without loss of generality, we may assume that $D$ has depth bounded by $n^{3/4}$ (see \cref{lemma:simplifying_depth}). Observe that any leaf $\ell \in \L(D)$ can be written as $\ell = \ell_x \circ \ell_a$ where $\ell_x$ is exclusively over variables $x_1, \, \dots,\, x_{n^2}$ and $\ell_a$ over $a_1,\, \dots, \, a_{\sqrt{n}}$. We further let $\L^s \coloneqq \left\lbrace \ell_x \circ \ell_a \in \L(D): |\ell_a| = \sqrt{n}\right\rbrace$ be the set of leaves that solve the $\Xor$ instance. 
Observe that for any $\ell \notin \L^s$, $\left\vert P^0[\ell] - P^1[\ell] \right\vert = 0$, indeed $\nu^0[\ell_a] = \nu^1[\ell_a] = 1/2$ so that:
$$\left\vert P^0[\ell] - P^1[\ell] \right\vert = \frac{1}{2} \cdot \left\vert \mu^0[\ell_x]\nu^0[\ell_a] + \mu^1[\ell_x]\nu^1[\ell_a] - \mu^1[\ell_x]\nu^0[\ell_a] - \mu^0[\ell_x]\nu^1[\ell_a] \right\vert = 0$$
We can therefore focus on bounding the bias contribution of leaves in $\L^s$. To that end, fix $\Gamma$ to be the set of all conjunction of size $n^{1/2}$ over $a_1, \, \dots, \, a_{\sqrt{n}}$ and for $\gamma \in \Gamma$, fix $\mathcal{L}_\gamma$ to be the set of leaves $\ell$ over $x_1, \, \dots, \, x_n$ such that $\ell \circ \gamma \in \L(D)$ (note that it is possible that $\L_\gamma = \emptyset$ for some $\gamma$). The bias of $D$ on $P$ can now be re-expressed as:
\begin{equation}\label{equation:bias_chibar}
    \TV(\tran(D, P^0), \tran(D, P^1)) = \frac{1}{2}\sum_{\gamma \in \Gamma} \sum_{\ell \in \L_\gamma} \left\vert P^0[\ell \circ \gamma] - P^1[\ell \circ \gamma]\right\vert = \sum_{\gamma \in \Gamma} \nu[\gamma] \sum_{\ell \in \L_\gamma} \left\vert \mu^0[\ell] - \mu^1[\ell]\right\vert
\end{equation} 
Observe that $\L_\gamma$ can be seen as the leaves of $D_\gamma$, the tree which is obtained from \emph{compressing} $D$ with $\gamma$. For instance if $\gamma_1 = 1$, we swap any node querying $a_1$ with its children sub-tree corresponding to $a_1 = 1$. The inner sums can thus be interpreted as the bias $D_\gamma$ holds in distinguishing $\mu^0$ from $\mu^1$. To bound those bias, it is convenient to replace $\mu$ (which is hyper-geometric) with a binomial variant $\tilde{\mu}$. We let $\tilde{\nu} \coloneqq (\tilde{\nu}^0 + \tilde{\nu}^1)/2$, where $\tilde{\mu}^0$, $\tilde{\mu}^1$ yield $n^2$ iid $\Bern(1/2 - n^{-1/2})$, respectively $\Bern(1/2 + n^{-1/2})$ random variables. Because all $D_\gamma$ have depth $\leq n^{3/4}$, we may relate $\mu$ to $\tilde{\mu}$ and use the hardness of \cref{sec:ber-opposite} to get:
\begin{align*}
\sum\nolimits_{\ell \in \L(D_\gamma)} \left\vert \mu^0[\ell] - \mu^1[\ell] \right\vert &= \sum\nolimits_{\ell \in \L(D_\gamma)} \left\vert \tilde{\mu}^0[\ell] - \mu^0[\ell] \right\vert + \left\vert \tilde{\mu}^0[\ell] - \tilde{\mu}^1[\ell] \right\vert + \left\vert \tilde{\mu}^1[\ell] - \mu^1[\ell] \right\vert\\
&\leq \sum\nolimits_{\ell \in \L(D_\gamma)} \left\vert \tilde{\mu}^0[\ell] - \tilde{\mu}^1[\ell] \right\vert  + 24n^{-1/2} \left(\tilde{\mu}^0[\ell] + \tilde{\mu}^1[\ell] \right) \tag{by \cref{lemma:hypergeometric_vs_binomial}}\\
&= \TV(\tran(D_\gamma, \tilde{\mu}^0), \tran(D_\gamma, \tilde{\mu}^1)) + 48n^{-1/2}\\
&\leq O(1) \cdot n^{-1/2}\sqrt{\cost(D_\gamma, \tilde{\mu})} \tag{by \cref{theorem:pi_vs_pis} with $b \coloneqq n^{-1/2}$}\\
&\leq O(1) \cdot n^{-1/2}\sqrt{\cost(D_\gamma, \mu)} \tag{by \cref{lemma:hypergeometric_vs_binomial}}
\end{align*}
Note that by \cref{lemma:hypergeometric_vs_binomial}, we can extend \cref{equation:bias_chibar} by using this bound and Cauchy-Schwarz inequality:
\begin{align*}
\TV(\tran(D, P^0), \tran(D, P^1)) &\leq O(1) \cdot n^{-1/2} \sum_{\gamma \in \Gamma} \nu[\gamma]\sqrt{\cost(D_\gamma, \mu)}\\
&\leq O(1) \cdot n^{-1/2} \sqrt{\sum\nolimits_{\gamma \in \Gamma} \nu[\gamma]} \sqrt{\sum\nolimits_{\gamma \in \Gamma} \nu[\gamma]\cost(D_\gamma, \mu)}
\end{align*}
The cost of $D$ on $P$ is easily lower-bounded by only taking leaves of $\L^s$ into account:
$$\cost(D, P) \geq \sum\nolimits_{\gamma \in \Gamma} \sum\nolimits_{\ell \in \L_\gamma} \left(\sqrt{n} + |\ell|\right) \nu[\gamma]\mu[\ell] = \sqrt{n} \sum\nolimits_{\gamma \in \Gamma} \nu[\gamma] + \sum\nolimits_{\gamma \in \Gamma} \nu[\gamma] \cost(D_\gamma, \mu)$$
Combining both, we get the desired bound on the $\LR$ ratio of $D$:
\begin{align*}
\frac{\cost(D, P)}{\TV(\tran(D, P^0), \tran(D, P^1))} &\geq \Omega(1) \cdot \max \left\lbrace \frac{ n \cdot \sqrt{\sum_{\gamma \in \Gamma} \nu[\gamma]}}{\sqrt{\sum_{\gamma \in \Gamma} \nu[\gamma] \cost(D_\gamma, \mu)}}, \,  \frac{n^{1/2} \cdot \sqrt{\sum_{\gamma \in \Gamma} \nu[\gamma] \cost(D_\gamma, \mu)}}{\sqrt{\sum_{\gamma \in \Gamma} \nu[\gamma]}} \right\rbrace\\
&\geq \Omega(n^{3/4}) \qedhere
\end{align*}
\end{proof}

\subsection{An upper bound for $\chibar$}\label{sec:max_conflict}

We recall here the definition of max conflict complexity (but see \cite{Gavinsky2019} for an in-depth treatment of the measure). Let $f$ be a fixed boolean function, $\mu^0, \mu^1$ a pair of distribution over $f^{-1}(0)$ and $f^{-1}(1)$ respectively and $D$ a deterministic decision tree solving $f$. For each node $v$ in $D$, we let $\mu^0|_v, \mu^1|v$ be the distributions conditioned on reaching $v$ and $q(v)$ be the index queried at node $v$. Furthermore, we associate to each $v \in \mathcal{N}(D)$ a number $R^D_\mu(v)$ inductively. If $v$ is the root of $D$, we let $R^D_\mu(v) = 1$ and if $v$ is the child of $w$ which is reached when the query answer to $q(w)$ is $b \in \B$:
$$R^D_\mu(v) = R^D_\mu(w) \cdot \min\left\lbrace\Pr\nolimits_{x \sim \mu^0|_w}[x_{q(w)} = b], \Pr\nolimits_{x \sim \mu^1|_w}[x_{q(w)} = b]\right\rbrace$$
Finally, we define $\Delta^D_\mu(v)$ for each $v \in \mathcal{N}(D)$ with:
$$\Delta^D_\mu(v) \coloneqq \left\vert \Pr\nolimits_{x \sim \mu^0|_w}[x_{q(v)} = 0] - \Pr\nolimits_{x \sim \mu^1|_w}[x_{q(v)} = 0] \right\vert$$
$R^D_\mu(v)$ can be interpreted as the probability of reaching node $v$ in a random walk that starts at the root and with probability $\min\{\Pr_{x \sim \mu^0|_v}[x_i = 0], \Pr_{x \sim \mu^1|_v}[x_i = 0]\}$ moves left, with probability $\min\{\Pr_{x \sim \mu^0|_v}[x_i = 1], \Pr_{x \sim \mu^1|_v}[x_i = 1]\}$ moves right and with remaining probability $\Delta^D_\mu(v)$ stops. As such, it holds that $\sum_{v \in \mathcal{N}(D)} \Delta^D_\mu(v) R^D_\mu(v) = 1$ and that for any partition $\Gamma$ of $\B^n$ we have $\sum_{\gamma \in \Gamma} R^D_\mu(\gamma) \leq 1$. The max conflict complexity $\chibar(f)$ is defined as:
$$\chibar(f) \coloneqq \max_{\mathcal{Q}} \min_{D \in \D(f)} \EXSUB{\mu \sim \mathcal{Q}}{\sum\nolimits_{v \in \mathcal{N}(D)} |v| \Delta^D_\mu(v) R^D_\mu(v)}$$
Where $\mathcal{Q}$ ranges over distributions of pairs of distributions over $f^{-1}(0)$ and $f^{-1}(1)$ and $\D(f)$ is the set of all decision tree solving $f$ correctly.

\begin{lemma}\label{lemma:chibar_F}
$\chibar(F) \leq O(n^{1/2})$
\end{lemma}
\begin{proof}
Let $\mathcal{Q}$ be the witness distribution over pairs of distribution for $\chibar(F)$ so that:
$$\chibar(F) = \min_{D \in \D(f)} \EXSUB{\mu \sim \mathcal{Q}}{\sum\nolimits_{v \in \mathcal{N}(D)} |v| \Delta^D_\mu(v) R^D_\mu(v)}$$
We build a decision tree $D \in \D(F)$ and show that it witnesses $\chibar(F) \leq O(n^{1/2})$. Let $\Gamma$ be the set of all conjunction of size $n^{1/2}$ over $a_1, \, \dots, \, a_{\sqrt{n}}$. $D$ starts by querying all the $\Xor$ variables $a_1,\, \dots, \, a_{\sqrt{n}}$. We then append to each branch $\gamma \in \Gamma$ a decision tree $D_\gamma$ on variables $x_1, \dots, x_{n^2}$ depending on $\mathcal{Q}$. For each $\gamma \in \Gamma$, let $f|_\gamma$ be the function $f$ with $a$ set to $\gamma$ and let us define a distribution $\mathcal{Q}_\gamma$ over distributions on $f|_\gamma^{-1}(0)$ and $f|_\gamma^{-1}(1)$. For each $\nu \in \support(\mathcal{Q})$, we put the conditioned distribution $\nu|_\gamma$ in $\mathcal{Q}_\gamma$ and set its probability mass as:
$$\mathcal{Q}_\gamma[\nu|_\gamma] = \frac{\mathcal{Q}[\nu]R^D_\nu(\gamma)}{z_\gamma} \quad\text{where}\quad z_\gamma = \sum\nolimits_{\nu \in \support(\mathcal{Q})} \mathcal{Q}[\nu] R^D_\nu(\gamma)$$
While $D$ refers to the whole tree, $z_\gamma$ does not actually depend on the choice of $D_\gamma$ so that we are still free to choose it as
$$D_\gamma \coloneqq \argmin_{D' \in \D(f|_\gamma)} \EXSUB{\mu \sim \mathcal{Q}_\gamma}{\sum\nolimits_{v \in \mathcal{N}(D')} |v| \Delta^{D'}_\mu(v) R^{D'}_\mu(v)}$$
We now show that $D$ witnesses $\chibar(F) \leq \sqrt{n}$. Indeed, we have:
\allowdisplaybreaks
\begin{align*}
    \chibar(F) &\leq \EXSUB{\mu \sim \mathcal{Q}}{\sum_{v \in \mathcal{N}(D)} |v| \Delta^D_\mu(v) R^D_\mu(v)}\\
    &= \EXSUB{\mu \sim \mathcal{Q}}{\sum_{|v| < \sqrt{n}} |v| \Delta^D_\mu(v) R^D_\mu(v) + \sum_{\gamma \in \Gamma}\sum_{w \in \mathcal{N}(D_\gamma)} |\gamma \circ w| \Delta^D_\mu(\gamma \circ w) R^D_\mu(\gamma \circ w)}\\
    &\leq \EXSUB{\mu \sim \mathcal{Q}}{\sum_{|v| < \sqrt{n}} \sqrt{n} \Delta^D_\mu(v) R^D_\mu(v) + \sum_{\gamma \in \Gamma}\sum_{w \in \mathcal{N}(D_\gamma)} (\sqrt{n} + |w|) \Delta^D_\mu(\gamma \circ w) R^D_\mu(\gamma \circ w)}\\
    &\leq \EXSUB{\mu \sim \mathcal{Q}}{\sum_{v \in \mathcal{N}(D)} \sqrt{n} \Delta^D_\mu(v) R^D_\mu(v) + \sum_{\gamma \in \Gamma}\sum_{w \in \mathcal{N}(D_\gamma)} |w| \Delta^D_\mu(\gamma \circ w) R^D_\mu(\gamma \circ w)}\\
    &\leq \sqrt{n} + \sum_{\gamma \in \Gamma} \EXSUB{\mu \sim \mathcal{Q}}{\sum_{w \in \mathcal{N}(D_\gamma)} |w| \Delta^D_\mu(\gamma \circ w) R^D_\mu(\gamma \circ w)}\\
    &= \sqrt{n} + \sum_{\gamma \in \Gamma} \EXSUB{\mu \sim \mathcal{Q}}{\sum_{w \in \mathcal{N}(D_\gamma)} |w| \Delta^{D_\gamma}_{\mu|_\gamma}(w) R^{D_\gamma}_{\mu|_\gamma}(w) R^D_\mu(\gamma)}\\
    &= \sqrt{n} + \sum_{\gamma \in \Gamma}\sum_{\nu \in \support{\mathcal{Q}}} \mathcal{Q}[\nu]R^D_\nu(\gamma) \sum_{w \in \mathcal{N}(D_\gamma)} |w| \Delta^{D_\gamma}_{\nu|_\gamma}(w) R^{D_\gamma}_{\nu|_\gamma}(w)\\
    &= \sqrt{n} + \sum_{\gamma \in \Gamma} z_\gamma \sum_{\nu \in \support{\mathcal{Q}}}  \mathcal{Q}_\gamma[\nu|_\gamma] \sum_{w \in \mathcal{N}(D_\gamma)} |w| \Delta^{D_\gamma}_{\nu|_\gamma}(w) R^{D_\gamma}_{\nu|_\gamma}(w)\\
    &= \sqrt{n} + \sum_{\gamma \in \Gamma} z_\gamma \EXSUB{\mu \sim \mathcal{Q}_\gamma}{ \sum_{w \in \mathcal{N}(D_\gamma)} |w| \Delta^{D_\gamma}_{\mu}(w) R^{D_\gamma}_{\mu}(w)}
\end{align*}
Observe that for any $\gamma \in \Gamma$, $f|_\gamma \in \{\GapMaj, \neg\GapMaj\}$, hence following our choice of $D_\gamma$, we have:
$$\EXSUB{\mu \sim \mathcal{Q}_\gamma}{ \sum\nolimits_{w \in \mathcal{N}(D_\gamma)} |w| \Delta^{D_\gamma}_{\mu}(w) R^{D_\gamma}_{\mu}(w)} = \min_{D' \in \D(f|_\gamma)} \EXSUB{\mu \sim \mathcal{Q}_\gamma}{ \sum\nolimits_{w \in \mathcal{N}(D)} |w| \Delta^{D'}_{\mu}(w) R^{D'}_{\mu}(w)} \leq \chibar(f|_\gamma)$$
Finally, note that $\LR(\GapMaj) \leq O(n^{1/2})$, as witnessed by the algorithm that makes one random query and returns the result. Combining this observation together with \cref{thm:optimal} and the fact that $\chibar$ is inner-optimal yields $\chibar(f|_\gamma) \leq O(n^{1/2})$ for every $\gamma \in \Gamma$ so that:
\begin{align*}
    \chibar(F) &\leq \sqrt{n} + \sum\nolimits_{\gamma \in \Gamma} z_\gamma \EXSUB{\mu \sim \mathcal{Q}_\gamma}{\sum\nolimits_{w \in \mathcal{N}(D_\gamma)} |w| \Delta^{D_\gamma}_{\mu}(w) R^{D_\gamma}_{\mu}(w)}\\
    &\leq \sqrt{n} + O(\sqrt{n})\cdot \sum\nolimits_{\gamma \in \Gamma} z_\gamma\\
    &= \sqrt{n} + O(\sqrt{n})\cdot \sum\nolimits_{\gamma \in \Gamma} \EXSUB{\mu \sim \mathcal{Q}}{R^D_\mu(\gamma)}\\
    &\leq O(\sqrt{n}) \tag*{($\Gamma$ partitions the input space) \qedhere}
\end{align*}
\end{proof}

\part{Small-Bias Minimax} \label{part:two}

\section{Overview}\label{sec:ulr_overview}
In this \cref{part:two}, we prove the separation between $\LR$ and $\ULR$ in \cref{thm:lr-ulr}, restated below.
\LrUlr*

We start, in this section, by giving an outline of the proof. Without further delay, we define the separating function~$f$ below. For convenience, we use  $N=n^{O(1)}$ to denote the input length. It will be easy to show an upper bound $\LR(f)\leq O(n)$ below in \cref{lem:lr-ub}. The hard part is to show a lower bound $\ULR(f)\geq \Omega(n^{5/4})$, which will occupy us for \miniSecs.

\subsection{Choice of hard function} \label{sec:choice-function}

\begin{definition}[Separating function] \label{def:f}
We define a partial function $f\colon\{0,1\}^N\to\{0,1,*\}$ using parameters~$n,B$ and $b_X\leq o(b_Y)\leq o(1)$. The input string is of length $N\coloneqq Bn$ and we think of it as being composed of $B$ blocks, each of size $n$. The ``type'' of a block~$(x_1, x_2, \dots, x_n)$ is determined by $x_1\in\{0,1\}$:
\begin{itemize}
    \item If $x_1 = 0$, we say the block is \emph{easy} and the value of the block is $x_2$.
    \item If $x_1 = 1$, we say the block is \emph{hard} and the value of the block is $\Xor_{n-1}(x_2,\ldots,x_n)$.
\end{itemize}
The value of $f$ is the majority value of the blocks. To make $f$ partial, we promise that the number of hard block is either $0$ or $4Bb_Y$. Moreover:
\begin{itemize}
    \item If there are $0$ hard blocks, we guarantee there are at least $B/2 + Bb_X$ blocks with the same value.
    \item If there are $4Bb_Y$ hard blocks, we guarantee there are at least $B/2 + Bb_Y$ blocks with the same value.
\end{itemize}
We will set $b_X \coloneqq n^{-1}$, $b_Y \coloneqq n^{-3/4}$, $B \coloneqq 8n^{9/2}$, so that effectively $N = 8n^{11/2}$.
\end{definition}

\begin{lemma} \label{lem:lr-ub}
$\LR(f) \leq O(n)$
\end{lemma}
\begin{proof}
Let $R$ be the randomised tree that picks one block at random and outputs its value (i.e., easy blocks take two queries, hard blocks take $n$). Consider any input~$x \in \B^{Bn}$. Suppose first that the number of hard blocks in $x$ is 0. In this case, the number of queries $R$ makes is $2$ and the bias is at least $b_X$ so that the cost/bias ratio is $O(n)$. On the other hand, if the number of hard blocks is $4Bb_Y$, the bias is at least $b_Y$ while the expected query cost is $\cost(R, x) = n\cdot (4Bb_Y)/B  + 2\cdot (1-(4Bb_Y)/B) \leq 5 b_Y n$.
\end{proof}

\subsection{Choice of hard input distributions}

Our goal is now to prove a lower bound $\ULR(f)\geq\Omega(n^{5/4})$. To this end, we use the following minimax characterisation, which can be derived using the minimax theorem for ratios of bilinear functions~\cite{BenDavid2020minimax}
\begin{equation}\label{equation:ULR_minimax}
\ULR(f) ~\coloneqq~ \min_R\max_{x,y} \frac{\cost(R,x)}{\bias(R,y)}
~=~ \max_{\mu,\nu} \min_D \frac{\cost(D,\mu)}{\bias(D,\nu)}.
\end{equation}
We are thus faced with the task of finding a pair of hard distributions $\mu$ and $\nu$. To do so, we first define distributions $X$ and $Y$, over inputs with no hard blocks and with $4Bb_Y$ hard blocks, respectively. Naturally, these distributions combine the hard distribution for $\GapMaj$ and $\Xor$. Crucially, we define $Y$ as hiding the correct value of $f$ inside the hard blocks---this way, an algorithm that never bothers to solve hard blocks would see only block values \emph{negatively} biased against the correct value of $f$. Finally, we will define $\mu$ and $\nu$ as appropriate mixtures of $X$ and $Y$.

Specifically, we define $X^0$ as the result of setting all blocks to easy and $B/2 + Bb_X$ of them have value $0$. The process of picking the $B/2 + Bb_X$ many $0$-blocks is made at uniformly at random (without replacement). Observe that creating an easy $0$-block simply amounts to setting the underlying variables to~$0^n$. The distribution $X^1$ is defined analogously but with $B/2 + Bb_X$ many $1$-blocks. Note that $f(X^0) = 0$ and $f(X^1) = 1$. The definition of $Y^0$ is more interesting. We first select the location of $4Bb_Y$ hard blocks whose values are set to 0 using the hard distribution for $\Xor_{n-1}$ (uniform over $\Xor_{n-1}^{-1}(0)$). The remainder of the blocks are easy and $B/2 - 3Bb_Y$ are set to $0$ while $B/2 - Bb_Y$ are set to $1$. Observe that the values of the easy blocks are indeed negatively biased toward the right answer: the probability of getting the correct value by sampling one easy block is $(B/2 - 3Bb_Y)/(B - 4Bb_y) \leq 1/2 - b_Y$. The distribution $Y^1$ is defined similarly. We now let
\begin{align*}
X &\coloneqq\textstyle \frac{1}{2}X^0 + \frac{1}{2}X^1,\\
Y &\coloneqq\textstyle \frac{1}{2}Y^0 + \frac{1}{2}Y^1,\\
\mu &\coloneqq\textstyle \frac{1}{2}X + \frac{1}{2}Y,\\
\nu &\coloneqq (1-\lambda) X + \lambda Y,
\end{align*}
where we will set $\lambda \coloneqq \Theta(b_X/b_Y) = \Theta(n^{-1/4})$ with a carefully chosen implicit constant (see \cref{sec:ber-mix}).

\paragraph{Intuition.}
Here is the intuition behind our definition of the hard pair $\mu,\nu$. First, hard blocks are much more likely to be found in $\mu$ than in $\nu$. Hence, if an algorithm $D$ wants to keep $\cost(D,\mu)$ low, it should not solve hard blocks very often. On the other hand, consider the following algorithm $D$ that solves no hard blocks: Query a block, if it is easy, output its value; if it is hard, output a random guess. What is the bias of this algorithm? We have $\bias(D,X)=b_X$ and $\bias(D,Y)=-b_Y$ and thus the two biases cancel each other out: $\bias(D,\nu)=(1-\lambda)b_X - \lambda b_Y \approx 0$. The challenge for us will be to rule out every such decision tree that does not solve hard blocks. However, this analysis will become subtle. For example, if the algorithm witnesses a hard block (but does not solve it), it still learns that the input comes from $Y$ and it then knows that the rest of the blocks are negatively biased.

\subsection{Lower-bound plan}\label{section:ULR_lowerbound_plan}
Now that the hard distributions $\mu,\nu$ are chosen, our goal is to show for every deterministic tree $D$,
\begin{equation}\label{eq:to_prove_plan}
\frac{\cost(D,\mu)}{\bias(D,\nu)} \geq \Omega(n^{5/4}).
\end{equation}
We now outline our lower-bound strategy that we will carry out in the upcoming \cref{sec:two-simplifications,sec:two-hardness,sec:distinguish,sec:omitted-proofs}.

\begin{enumerate}[label=(\arabic*),itemsep=1em,leftmargin=3.5em]
\item[(\S\ref{sec:no-hard-blocks})] {\bf Simplification I.}
We start by making two simplifications.
First, we rule out any decision tree whose strategy is to solve hard blocks (querying all $n$ bits of a block) with noticeable probability. Such trees have too high a cost relative to $\mu$ (which includes hard blocks with probability $1/2$). We will henceforth assume that a given tree $D$ never solves a hard block. Moreover, we are left with a challenge of proving an $\LR$-style cost/bias trade-off relative to $\nu$ only. (We forget $\mu$ from now on.)

\item[(\S\ref{sec:streaming})] {\bf Simplification II.}
Second, we simplify the analysis of decision trees by switching to a \emph{infinite-stream} model of computation. Instead of the $N$-bit input distributions such as $X = \frac{1}{2}X^0 + \frac{1}{2}X^1$ we model them as \emph{infinite streams} $\smash{\tilde{X} = \frac{1}{2}\tilde{X}^0 + \frac{1}{2}\tilde{X}^1}$ that record the values of the blocks, with the original biases, but with more independence. For example, $\tilde{X}^1$ is an infinite sequence of iid $\Bern(1/2+b_X)$. Moreover, each of $\tilde{X},\tilde{Y},\tilde{\nu}$ is a mixture of at most 4 streams of iid symbols.

\item[(\S\ref{sec:two-hardness})] {\bf Two basic hardness results.}
At this point, it remains to prove a cost/bias trade-off for stream~$\tilde{\nu}$. In preparation for this, we establish some very basic query lower bounds that are well-known in the worst-case setting, but which have not been yet proved in the expected-cost setting. For example, every algorithm trying to distinguish between an iid stream of $\Bern(1/2+b)$ and an iid stream of $\Bern(1/2+b)$ must have bias at most $O(b\sqrt{\cost})$. Because the cost is measured in expectation, these basic facts turn out to be somewhat tricky to prove.

\item[(\S\ref{sec:distinguish})] {\bf Lower bound for \boldmath$\nu$-stream.}
We now have the tools to analyse the cost/bias trade-off for $\tilde{\nu}$. The analysis here is rather intricate, as there are several decision tree strategies that we need to defeat. Some technical calculations are relegated to \cref{sec:omitted-proofs} and \cref{app}.
\end{enumerate}

\section{Two simplifications} \label{sec:two-simplifications}
Having chosen two hard pair of distribution $\mu,\nu$ to be plugged in the minimax characterisation~\cref{equation:ULR_minimax} for $\ULR$, it remains to be shown that $\qbar(D, \mu)/\bias_f(D, \nu) \geq \Omega(n^{5/4})$ for all decision tree $D$ to prove \cref{thm:lr-ulr}. The purpose of this section is to make two simplifying steps (as sketched in \cref{section:ULR_lowerbound_plan}) and reduce the lower-bound task to showing
\begin{equation}\label{equation:ulr_lb_goal}
\frac{\qbar(D, \tilde{\nu})}{\TV(\tran(D, \tilde{\nu}^0), \tran(D, \tilde{\nu}^1))} \geq \Omega(n^{5/4})\quad\textup{for all deterministic decision trees $D$.}
\end{equation}
Here $D$ will be a ternary decision tree over $\T$ (where $\hb$ models a hard block that the algorithm did not solve) and $\tilde{\nu}$ an infinite stream over $\T$ representing $\nu$. The goal of this section is to define and justify precisely the reduction, leaving the proof of \cref{equation:ulr_lb_goal} for \cref{sec:distinguish}.

\subsection{Simplification I: No hard blocks} \label{sec:no-hard-blocks}
The first simplification step rules out any decision tree whose strategy is to solve blocks completely. In that regard, having $\lambda \ll 1$ is critical. Let us say that $D$ \textit{solves a hard block} on input $x$, denoted $S(D, x)$, if the leaf reached by $x$ on $D$ contains all the $n$ bits of some block with first bit equal to $1$. Observe that~$S(D, x)$ can only hold when $x$ actually contains a hard block, i.e., when it is coming from $Y$. Note that equating ``solving'' with ``querying the whole block'' is justified since the value $v$ of a hard block is hidden by the uniform distribution $\Xor_{n-1}^{-1}(v)$, offering zero bias until the very last query. Let $\nu[S]$ and $Y[S]$ be the probability that $D$ solves a hard block when the input is drawn from $\nu$, respectively $Y$. If $\nu[S] \geq \bias(t, \nu)/3$, the desired bound $\Omega(n^{5/4})$ in \cref{eq:to_prove_plan} is already attained because $\lambda = \Theta(b_X/b_Y)$ and
$$\cost(D, \mu) \geq \frac{1}{2} \cdot \cost(D, Y) \geq \frac{n}{2} \cdot Y[S] \geq \frac{n}{2\lambda} \cdot \nu[S] \geq \frac{n}{6\lambda} \cdot \bias(D, \nu) \geq \Omega(\bias(D, \nu) \cdot n^{5/4}).$$
Thus we may assume $\nu[S] \leq \bias(t, \nu)/3$ henceforth. We let $D'$ be the copy of $D$ that stops whenever it is about to solve a hard block, i.e., instead of querying the last remaining bit of a hard block, $D'$ stops and outputs the most likely answer. Note that $\qbar(D', \mu) \leq \qbar(D, \mu)$, but $D'$ still enjoys a constant fraction of the bias that $D$ has against $\nu$:
\begin{align*}
\bias(D', \nu) = 2\Pr_{x \sim \nu}[D'(x) = f(x)] - 1 &\geq 2\Pr_{x \sim \nu}[D'(x) = f(x) \text{ and } \neg S(D, x)] - 1\\
&= 2\Pr_{x \sim \nu}[\neg S(D, x)] \Pr_{x \sim \nu}[D(x) = f(x)] - 1\\
&\geq \bias(D, \nu)/3
\end{align*}
As $\qbar(D', \mu) \geq \qbar(D', \nu)/2$, we have reduced \cref{eq:to_prove_plan} to showing that $\qbar(D, \nu)/\bias(D, \nu) \geq \Omega(n^{5/4})$ for any decision tree $D$ that never solves hard blocks completely. \emph{Note that this is an $\LR$-style lower bound relative to a single hard distribution $\nu$ and against a restricted class of algorithms.}

\bigskip
\noindent
We may restrict the class of admissible algorithms even further. Observe first that a decision tree $D$ that does not solve any hard block can be simulated by a \emph{randomised} tree $R$ that never queries anything outside the first two variables $\{x_1, x_2\}$ of any given block. Indeed, if a block is easy, the values beyond the two first bits are fixed to $0^{n-2}$ and need not be queried. If a block is hard, then the marginal distribution of any proper subset of the variables $x_2,\ldots,x_n$ is uniform, and $R$ can simulate this distribution itself with internal randomness. We may now derandomise such an $R$ back into a deterministic tree: if $R$ has $\qbar(R, \nu)/\bias_f(R, \nu) \leq C$, then there must be a deterministic decision tree $D'$ in its support that also achieves $\qbar(D', \nu)/\bias_f(D', \nu) \leq C$.

After the above simplification, we only need to show that for any tree $D$ that is constrained to read at most variables $\{x_1, x_2\}$ within each block, $\qbar(D, \nu)/\bias(D, \nu) \geq \Omega(n^{5/4})$. At the cost of losing a constant factor in the bound, we may also assume that whenever $D$ queries $x_2$, it first queries $x_1$. Observe that such a $D$ only sees three types of events: easy 0-block, easy 1-block or hard block with undisclosed value. $D$ can thus be interpreted as a \emph{ternary} decision tree over the alphabet $\T$ (where $\hb$ means hard block with undisclosed value) trying to solve a ternary analogue $\overline{f}$ of $f$ against inputs sampled from the ternary analogue $\overline{\nu}$ of $\nu$. More precisely, $\overline{f}$ maps $\T^B\to\{0,1,*\}$ and $\overline{\nu}$ a distribution over $\T^B$ defined as before but now based on $\overline{X}$ and $\overline{Y}$:
\begin{enumerate}
\item $\overline{X}^0$: $B/2 + B b_X$ random locations are set to $0$, the rest are $1$.
\item $\overline{X}^1$: $B/2 + B b_X$ random locations are set to $1$, the rest are $0$.
\item $\overline{Y}^0$: $4Bb_Y$ locations are set to $\hb$, $B/2 - 3Bb_Y$ locations to $0$ and $B/2 - Bb_Y$ locations to $1$.
\item $\overline{Y}^1$: $4Bb_Y$ locations are set to $\hb$, $B/2 - 3Bb_Y$ locations to $1$ and $B/2 - Bb_Y$ locations to $0$.
\end{enumerate}
Following this discussion, to prove \cref{eq:to_prove_plan}, we only need to show:
\begin{equation}\label{eq:reduction_no_hb}
\frac{\qbar(D, \overline{\nu})}{\bias_{\overline{f}}(D, \overline{\nu})} \geq \Omega(n^{5/4}) \quad \textup{for all ternary decision tree $D$}
\end{equation}

\paragraph{Ternary decision tree.} We use an intuitive notion of ternary decision tree, where each query node has three children corresponding to each possible query response in $\T$. Akin to their binary analogues, we will see leaves of ternary tree as conjunction over literals with three possible states: positive, negative or \emph{witnessing}. This new witnessing literal type amounts to checking whether the variable equals $\hb$. Following this, we will say that a leaf $\ell$ is witnessing, in short $\hb \in \ell$, if it holds a witnessing literal.

\subsection{Simplification II: Streaming model} \label{sec:streaming}
Following \cref{sec:no-hard-blocks}, we are left with the task of showing \cref{eq:reduction_no_hb}. One of the main technical annoyances in working with $\overline{\nu}$ is that its entries are generated using a hypergeometric distribution, where, even in $\overline{X}^0$, the bits are not iid. To overcome this issue and simplify the calculations, we propose to replace $\overline{\nu}$ with a a multinomial variant $\tilde{\nu}$. This change requires attention on two counts. First, we need to ensure that $D$ is shallow enough and $B$ is large enough, thus asserting that the behavior of $D$ on $\tilde{\nu}$ is close to the behavior of $D$ on $\overline{\nu}$. Second and most importantly, observe that it is possible that $\widetilde{\nu}^0$ yields a 1-input (and $\widetilde{\nu}^1$ a 0-input). This makes the usual notion of bias moot, but we can still resort to the total-variation distance interpretation of the bias. Indeed, assuming without loss of generality that $D$ is optimally labelled for distinguishing $\overline{\nu}^0$ from $\overline{\nu}^1$:
$$\bias_{f'}(D, \overline{\nu}) = \TV(\tran(D, \overline{\nu}^0), \tran(D, \overline{\nu}^1)) = \frac{1}{2} \cdot \sum\nolimits_{\ell \in \L(D)} \left\vert \overline{\nu}^0[\ell] - \overline{\nu}^1[\ell] \right\vert$$
Thus, \cref{eq:reduction_no_hb} can actually be re-written swapping $\TV(\tran(D, \overline{\nu}^0), \tran(D, \overline{\nu}^1)$ in place of $\bias_{f'}(T, \overline{\nu})$ and this relaxed formulation opens the floor for the desired multinomial variant. $\tilde{\nu}$ is defined analogously to $\overline{\nu}$, by mixing appropriately the distributions $\tilde{X}^0$, $\tilde{X}^1$, $\tilde{Y}^0$ and $\tilde{Y}^1$:
\begin{itemize}
    \item $\tilde{X}^0$: $B$ iid $\Bern(1/2 - b_X)$ random variables.
    \item $\tilde{X}^1$: $B$ iid $\Bern(1/2 + b_X)$ random variables.
    \item $\tilde{Y}^0$: $B$ iid random variables taking value 0 with probability $1/2 - 3b_Y$, 1 with probability $1/2 - b_Y$ and $\hb$ with remaining probability $4b_Y$.
    \item $\tilde{Y}^1$: $B$ iid random variables taking value 0 with probability $1/2 - b_Y$, 1 with probability $1/2 - 3b_Y$ and $\hb$ with remaining probability $4b_Y$.
\end{itemize}

\begin{lemma}
If $\qbar(D, \tilde{\nu})/\TV(\tran(D, \tilde{\nu}^0), \tran(D, \tilde{\nu}^1) \geq \Omega(n^{5/4})$ for all $D$, then \cref{eq:reduction_no_hb} holds.
\end{lemma}
\begin{proof}
Fix some ternary decision tree $D$, and let us show that \cref{eq:reduction_no_hb} holds for $D$. We may assume without loss of generality that $D$ has depth bounded by $n^{5/4}$, as else $D$ would essentially already have $\LR$ ratio $\geq \Omega(n^{5/4})$ (see \cref{lemma:simplifying_depth} for details). The claim would also be vacuously true if $\bias_{f'}(D, \overline{\nu}) \leq n^{-5/4}$. Fix now any leaf $\ell \in \L(D)$ and let $q_0$, $q_1$ and $q_\hb$ be the number of negative, positive and witnessing literals in $\ell$ so that $|\ell| = q_0 + q_1 + q_\hb$. Recalling that $B = 8n^{9/2}$, we have $|\ell| \ll \sqrt{B}$, so that using \cref{lemma:hypergeometric_vs_binomial} for $X$ and \cref{lemma:hypergeometric_vs_multinomial} for $Y$:
$$\left\vert \tilde{X}^d[\ell] - \overline{X}^d[\ell]\right\vert \leq \frac{12|\ell|^2}{B} \cdot \tilde{X}^d[\ell] \quad\textup{and}\quad \left\vert \tilde{Y}^d[\ell] - \overline{Y}^d[\ell]\right\vert \leq \frac{2|\ell|^2}{Bb_Y}\cdot\tilde{Y}^d \quad \forall d \in \B$$
In short, $|\tilde{\nu}^d[\ell] - \overline{\nu}^d[\ell]| \leq n^{-5/4}\tilde{\nu}^d[\ell]/4$ and so $\cost(D, \overline{\nu}) \geq \cost(D, \tilde{\nu})/2$. Furthermore, both total variation distance are close:
\begin{align*}
\bias_{f'}(D, \overline{\nu}) = \sum\nolimits_{\ell \in \L(D)} \left\vert \overline{\nu}^0[\ell] - \overline{\nu}^1[\ell] \right\vert &\leq  \sum\nolimits_{\ell \in \L(D)} \left\vert \overline{\nu}^0[\ell] - \tilde{\nu}^0[\ell] \right\vert + \left\vert \tilde{\nu}^0[\ell] - \tilde{\nu}^1[\ell] \right\vert + \left\vert \overline{\nu}^1[\ell] - \tilde{\nu}^1[\ell] \right\vert\\
&\leq   \sum\nolimits_{\ell \in \L(D)} \left\vert \tilde{\nu}^0[\ell] - \tilde{\nu}^1[\ell] \right\vert + \sum\nolimits_{\ell \in \L(t)} n^{-5/4}\tilde{\nu}/2\\
&\leq n^{-5/4}/2 + \sum\nolimits_{\ell \in \L(D)} \left\vert \tilde{\nu}^0[\ell] - \tilde{\nu}^1[\ell] \right\vert\\
&\leq \bias_{f'}(D, \overline{\nu})/2 + \sum\nolimits_{\ell \in \L(D)} \left\vert \tilde{\nu}^0[\ell] - \tilde{\nu}^1[\ell] \right\vert
\end{align*}
Thus, if $\qbar(D, \tilde{\nu})/\TV(\tran(D, \tilde{\nu}^0), \tran(D, \tilde{\nu}^1) \geq \Omega(n^{5/4})$ holds for $D$, we have, as desired:
\begin{equation*}
\frac{\qbar(D, \overline{\nu})}{\bias_{f'}(D, \overline{\nu})} \geq \Omega(1) \cdot \frac{\qbar(D, \tilde{\nu})}{\TV(\tran(D, \tilde{\nu}^0), \tran(D, \tilde{\nu}^1))} \geq \Omega(n^{5/4}) \qedhere
\end{equation*}
\end{proof}
As an ultimate simplification, instead of multinomial distributions over $\T^B$, we will see $\tilde{X}^0$, $\tilde{X}^1$, $\tilde{Y}^0$ and $\tilde{Y}^1$ as \emph{infinite} iid streams and allow $D$ to be have unbounded length. Because of \cref{lemma:simplifying_depth}, this generalization adds no real power but this stream framework is generally nicer to work with. In particular, the parameter $B$ is not relevant anymore and as such we will only use $b_X$ and $b_Y$ for later sections. Following our reduction, proving $\ULR{f} \geq \Omega(n^{5/4})$ now amounts to proving the following:
\begin{equation}\label{equation:ulr_reduction_step2}
\frac{\qbar(D, \tilde{\nu})}{\TV(\tran(D, \tilde{\nu}^0), \tran(D, \tilde{\nu}^1)} \geq \Omega(n^{5/4})\quad\text{for all infinite ternary decision tree $D$}
\end{equation}

\subsubsection{Some thoughts on the $\LR$ streaming model}

Note that by the iid nature of random streams, the indices queried by decision trees do not matter. Actually, we could even force $D$ to query variable $x_1$ at level 1, $x_2$ at level 2 and so forth. This however doesn't prevent $D$ from adopting an adaptive strategy: $D$ can indeed decide \textit{when to stop} depending on past query answers. This allows for some wildly unbalanced decision trees that need to be ruled out, and as such it is a challenge to prove our results in the \textit{expected} query cost setting.

The lower bound \cref{equation:ulr_reduction_step2} can also be seen as a distribution testing hardness result. Consider the problem in which a secret $\varphi$ is sampled amongst $(\tilde{X}^0, \tilde{X}^1, \tilde{Y}^0, \tilde{Y}^1)$ with prior $(\lambda/2, \lambda/2, (1-\lambda)/2, (1-\lambda)/2)$ and one needs to decide whether $\varphi \in \{\tilde{X}^0, \tilde{Y}^0\}$ or $\varphi \in \{\tilde{X}^1, \tilde{Y}^1\}$ by making repeated queries to the stream $\varphi$. Then,~\cref{equation:ulr_reduction_step2} says that any decision tree $D$ accomplishing this task must have $\LR$ ratio $\geq \Omega(n^{5/4})$.

As a technicality, we will need to resort to $\LR$-streaming bounds for randomised decision tree. For this, we define the total variation distance on transcript distribution naturally with $\TV(\tran(R, P^0), \tran(R, P^1)) = \mathbb{E}_{D \sim R}[\TV(\tran(D, P^0), \tran(D, P^1))]$ (where $P^0$ and $P^1$ are two distributions). In particular, this is still consistent with the view that if $R$ is labelled optimally by $\{P^0, P^1\}$, then $\TV(\tran(R, P^0), \tran(R, P^1)) = \Pr_{x \sim P^0}[R(x) = P^0] - \Pr_{x \sim P^1}[T(x) = P^0]$

\section{Two basic hardness results} \label{sec:two-hardness}
Before tackling the proof of \cref{equation:ulr_reduction_step2} and ultimately showing $\ULR(f) \geq \Omega(n^{5/4})$, we first establish in this section a couple of basic results in the $\LR$-streaming model. Those self-contained results are a good starting point to get acquainted with $\LR$-style lower bounds and have the added benefit of being a key component of our main technical result \cref{theorem:main_technical}. 

\subsection{Source of hardness I: Bernoulli mixtures} \label{sec:ber-mix}
As a first step toward bounding the hardness of distinguishing $\tilde{\nu}^0$ from $\tilde{\nu}^1$ with a decision tree, we study an idealized version with no hard blocks. More precisely, for two parameters $b_X < o(b_Y) < o(1)$, we let $X^0$, $X^1$, $Z^0$ and $Z^1$ be iid random stream of $\Bern(1/2 - b_X)$, $\Bern(1/2 + b_X)$, $\Bern(1/2 + b_Y)$ and $\Bern(1/2 - b_Y)$ random variables and define the mixtures of random streams $M^0$ and $M^1$ with:
\begin{equation}\label{eq:setting_of_lambda}
\begin{array}{l}
M^0 \coloneqq  (1-\lambda)X^0 + \lambda Z^0\\
M^1 \coloneqq  (1-\lambda)X^1 + \lambda Z^1
\end{array} \qquad\text{where}\qquad \frac{1 - \lambda}{\lambda} = \frac{\ln\left(\frac{1+2b_Y}{1 - 2b_Y}\right)}{\ln\left(\frac{1+2b_X}{1 - 2b_X}\right)}
\end{equation}
Finally, we let $M \coloneqq (M^0 + M^1)/2$. We will show that $\TV(\tran(D, M^0), \tran(D, M^1)) \leq O(b_Xb_Y\qbar(D, M))$ for any deterministic $D$. The fine-tuning of $\lambda$ in \cref{eq:setting_of_lambda} will turn out to be a necessary technicality, but $\lambda = \Theta(b_X/b_Y)$ as \cref{lemma:approx_lambda} shows. Following the initial plan, distinguishing $M^0$ from $M^1$ can be interpreted as distinguishing $\tilde{\nu}^0$ from $\tilde{\nu}^1$ \emph{with no $\hb$}, in particular $Z$ carries overall \emph{negative bias}. This particular source of hardness will be used to analyse the bias brought by \textit{small} leaves, whose behavior for $\tilde{\nu}^0$ versus $\tilde{\nu}^1$ is close to $M^0$ versus $M^1$. As a secondary goal, the companion proof exemplifies one of the simplest way to obtain a $\LR$ lower bound, namely finding a \emph{hybrid} distribution---in this case the uniform distribution $U$---and apply a corruption bound.

\begin{theorem}\label{theorem:A0_vs_A1}
For any tree $D$, we have $\TV(\tran(D, M^0), \tran(D, M^1)) \leq  O(b_Xb_Y\qbar(D, M))$.
\end{theorem}
\begin{proof}
Following \cref{lemma:simplifying_depth}, we may assume without loss of generality that $D$ has depth bounded by $1/9b_Xb_Y$. Let $U$ be the random stream of $\Bern(1/2)$ variables and decompose the bias of $D$ as:
$$2\cdot \TV(\tran(D, M^0), \tran(D, M^1)) = \sum_{\ell \in \L(D)} \left\vert M^0[\ell] - M^1[\ell]\right\vert \leq \sum_{\ell \in \L(D)} \left\vert M^0[\ell] - U[\ell]\right\vert + \sum_{\ell \in \L(D)}  \left\vert M^1[\ell] - U[\ell]\right\vert$$
We focus on the first sum as the bound on the second follows by a symmetrical argument. The first sum can be interpreted as the bias held by $D$ in distinguishing $U$ from $M^0$. Letting $\L^U = \{\ell \in \L(D): U[\ell] \geq M^0[\ell]\}$:
\begin{align*}
\sum\nolimits_{\ell \in \L(D)} \left\vert M^0[\ell] - U[\ell]\right\vert &= 2 \sum\nolimits_{\ell \in \L^U} U[\ell] - M^0[\ell]\\
&\leq O(1) \cdot b_Xb_Y \sum\nolimits_{\ell \in \L^U} |\ell| \cdot U[\ell] \tag{by \cref{lemma:corruption_bound}}\\
&\leq O(1) \cdot b_Xb_Y \sum\nolimits_{\ell \in \L^U} |\ell| \cdot M^0[\ell] \tag{by \cref{lemma:corruption_bound} with $|\ell| \leq 1/9b_Xb_Y$}\\
&\leq O(1) \cdot b_Xb_Y \qbar(D, M) \tag*{\qedhere}
\end{align*}
\end{proof}

\subsection{Source of hardness II: Bernoulli with opposite bias}\label{sec:ber-opposite}
Our second hardness result tackles the problem of distinguishing a random stream $B^0$ of iid $\Bern(1/2 - b)$ variables from the symmetric random stream $B^1$ with variables sampled from $\Bern(1/2 + b)$, where $b \in o(1)$ is a parameter. This basic result will be employed several times in later section. For instances, witnessing leaves need to solve the $\tilde{Y}^0$ versus $\tilde{Y}^1$, which is essentially $B^0$ versus $B^1$ with $b \coloneqq b_Y$. In what follows, we let $B \coloneqq (B^0 + B^1)/2$.
\begin{theorem}\label{theorem:pi_vs_pis}
For any randomised decision tree $R$, $\TV(\tran(R, B^0), \tran(R, B^1)) \leq O(b \sqrt{\qbar(R, B)})$
\end{theorem}
Note that the theorem statement features a randomised decision tree instead of a deterministic one, a generalization needed for later use. There are essentially three ways to prove a result similar to the one of \cref{theorem:pi_vs_pis}. The first is to use a direct corruption bound, akin to \cref{theorem:A0_vs_A1} - but this would only give the weaker bound $\TV(\tran(R, B^0), \tran(R, B^1)) \leq O(b \qbar(R, B)))$. A second would be to leverage the machinery of later sections and especially \cref{lemma:Z0vsZ1_corruption} to get a bias bound \emph{per leaf}, ultimately yielding $\TV(\tran(R, B^0), \tran(R, B^1)) \leq O(b \sqrt{\qbar(R, B)} \textup{polylog}(\qbar(R, B)))$. Even though this is enough to cover the desired polynomial separation, \cref{theorem:pi_vs_pis} is optimal and we believe, interesting on its own. The crux is to re-use an idea brought by Sherstov~\cite{Sherstov2012} in the context of bounding the communication complexity of the gap-Hamming function. As a first step, we prove some hardness result in distinguishing $B$ from $U$.

\begin{lemma}\label{lemma:A_vs_U}
For any randomised decision tree $R$, $\TV(\tran(R, U), \tran(R, B)) \leq 4 b^2 \cost(R, U/2 + B/2)$.
\end{lemma}
\begin{proof}
We show this for a deterministic $D$ instead of $R$ as the randomised case follows by linearity of expectation. Let us begin by showing that for any $\ell \in \mathcal{L}(D)$, $B[\ell]/U[\ell] \geq 1 - 2|\ell|b^2$. Let $k \coloneqq |\ell|$ and supposing that $\ell$ has $k/2 + m$ positive literals (so $m \in [-k/2, k/2]$):
\begin{align*} 
\frac{B[\ell]}{U[\ell]} &= \frac{1}{2} \cdot \left(1 - 4b^2\right)^{k/2} \cdot \left[\left(\frac{1 + 2b}{1 -2b}\right)^m + \left(\frac{1 - 2b}{1 + 2b}\right)^m \right]
\end{align*}
The quantity within the square bracket is a function of $m$ and we find its minimum by setting its derivative equal to zero, yielding $m = 0$. Note that this shows that the most separating leaves have as many positive literals as negative ones. In any case, it holds that $B[\ell]/U[\ell] \geq (1 - 4b^2)^{k/2}$ so that by defining $\L^U \coloneqq \{\ell \in \L(D): U[\ell] \geq B[\ell]\}$ and using the approximation of \cref{lemma:approx_zoo}:
$$\TV(\tran(D, U), \tran(D, B))  = \sum_{\ell \in \L^U} U[\ell] - B[\ell] \leq \sum_{\ell \in \L^U} U[\ell] \cdot \left(1 - (1 - 4b^2)^{|\ell|/2} \right) \leq \sum_{\ell \in \L^U} U[\ell] \cdot 2b^2 |\ell|$$
We may finally conclude:
\begin{equation*}
\frac{\cost(D, U/2 + B/2)}{\TV(\tran(D, U), \tran(D, B))} \geq \frac{\cost(D, U)}{4b^2 \cdot \cost(D, U)} = 1/4b^2 \qedhere
\end{equation*}
\end{proof}

\begin{proof}[Proof of \cref{theorem:pi_vs_pis}]
As a first step, we prove the claim for deterministic $D$ and extend it to randomised ones at the end. Following an argument similar to the one of \cref{lemma:simplifying_depth}, we may assume that $D$ has depth bounded by $1/5b^2$. It will be convenient to see $D$ as having leaves labelled optimally by $\{B^0, B^1\}$. Let $\delta \coloneqq \TV(\tran(D, B^0), \tran(D, B^1)) = \Pr_{x \sim B^0}[D(x) = B^0] - \Pr_{x \sim B^1}[D(x) = D^0]$ be the bias of $D$ in $B^0$ against $B^1$ and note that we can mix $D$ with the trivial tree that always accept (or always reject) to get a randomised decision tree $R$ with centred acceptance probability:
$$\Pr\nolimits_{x \sim B^0}[R(x) = B^0] = 1/2 + \xi \quad\text{and}\quad \Pr\nolimits_{x \sim B^1}[R(x) = B^0] = 1/2 - \xi \quad\text{where}\quad \xi \geq \delta/6$$
The construction of $R$ is detailed in \cref{lemma:re_centering}, but $R$ has worst-case depth bounded by $1/5b^2$ too. We will use $R$ to build another randomised decision tree $R^\star$ which can distinguish $B$ from $U$, effectively reducing the hardness of $B^0$ versus $B^1$ to distinguishing $B$ from $U$. Define $\Rneg$ as a copy of $R$ with \emph{inverted} labels, i.e. leaves labelled with $B^0$ are now labelled with $B^1$ and reciprocally. The tree $R^\star$ we build will have $\TV(\tran(R^\star, U), \tran(R^\star, B)) \geq \Omega(\xi^2)$ and small cost. The construction of $R^\star$ depends on the value of $\alpha \coloneqq \Pr_{x \sim U}[R(x) = B^0] - 1/2$. If $\alpha \in [-\delta/7, \delta/7]$, we let $R^\star$ execute $R$ and $\Rneg$ in turn on independent bits and output $U$ if both run output $B^0$ and $B$ else. The independent runs can be performed by off-setting the query indices of $\Rneg$ by a large number, e.g. $\lceil 1/b^2 \rceil$. We have:
\begin{align*}
\Pr_{x \sim B}[R^\star(x) = U] &= \frac{1}{2} \cdot \Pr_{x \sim B^0}[R^\star(x) = U] + \frac{1}{2}\cdot\Pr_{x \sim B^1}[R^\star(x) = U]\\
&= \frac{1}{2} \cdot \Pr_{x, x' \sim B^0}[R(x) = \Rneg(x') = B^0] + \frac{1}{2} \cdot \Pr_{x, x' \sim B^1}[R(x') = \Rneg(x) = B^0] \\
&=  \left(\frac{1}{2} + \xi\right)\cdot\left(\frac{1}{2} - \xi \right)\\
&\leq \frac{1}{4} - \frac{\delta^2}{36}
\end{align*}
On the other hand, $\Pr_{x \sim U}[R^\star(x) = U] = (1/2 + \alpha)\cdot (1/2 - \alpha) \geq 1/4 - \delta^2/49$ and hence for this regime of $\alpha$, $R^\star$ achieves bias $\TV(\tran(R^\star, U), \tran(R^\star, B)) \geq 13\delta^2/1764$. Finally, if $\alpha \geq \delta/7$ we pick $R^\star \coloneqq R$ and since $R$ is centred, $\Pr_{x \sim B}[R(x) = U] = 1/2$. On the other hand, we have $\Pr_{x \sim U}[R(x) = U] \geq 1/2 + \xi/7$. The case $\alpha \leq \delta/7$ is analogous and requires picking $R^\star \coloneqq \Rneg$. In any case, we get a construction $R^\star$ with $\TV(\tran(R^\star, U), \tran(R^\star, A)) \geq \Omega(\delta^2)$ and worst-case depth $2/5b^2$, implying that for any $D'$ in the support of $R^\star$:
$$\qbar(D', B) = \sum_{\ell \in \L(D')} |\ell| \cdot B[\ell] \geq \sum_{\ell \in \L(D')} |\ell| \cdot U[\ell] \cdot \left(1 - 2|\ell| b^2\right) \geq \frac{1}{5}\sum_{\ell \in \mathcal{L}(D')} |\ell| \cdot U[\ell] = \frac{\qbar(D', U)}{5}$$
The first inequality holds because of the analysis in \cref{lemma:A_vs_U}. Applying linearity of expectation, we have that $\qbar(R^\star, U/2 + B/2) \leq 3\qbar(R^\star, B)$. Observe that the only non-trivial tree in the support of $R^\star$ is the initial $D$ so that $\cost(D, B) \geq \Omega(\cost(R^\star), U/2 + B/2)$ and hence using \cref{lemma:A_vs_U}:
$$\frac{\sqrt{\cost(D, B)}}{\delta} \geq \Omega\left( \frac{\cost(R^\star, U/2 + B/2)}{\TV(\tran(R^\star, U), \tran(R^\star, B))} \right)^{1/2} \geq \Omega(1/b)$$
To obtain the claim for randomised decision tree, we resort to Jensen's inequality:
\begin{align*}
\TV(\tran(R, B^0), \tran(R, B^1)) &= \EXSUB{D \sim R}{\TV(\tran(D, B^0), \tran(D, B^1))}\\
&\leq O(1) \cdot \EXSUB{D \sim R}{b \sqrt{\cost(D, B)}}\\
&\leq O(1) \cdot b \sqrt{\cost(R, B)} \qedhere
\end{align*}
\end{proof}

\section{Lower bound for \texorpdfstring{$\nu$}{nu}-stream} \label{sec:distinguish}
In this section, we finally prove that $\ULR(f) \geq \Omega(n^{5/4})$. Following \cref{sec:two-simplifications}, it suffices to prove \cref{equation:ulr_reduction_step2}. To keep things as general as possible, we will not work with $\tilde{\nu}$ directly but with an asymptotically equivalent version $\nu$ and drop the tilde notation which is too heavy, so that e.g. $Y$ is now a random stream over $\T$ instead of an hyper-geometric distribution over $\B^{Bn}$. All the random streams used for the remainder of the paper are summarised in \cref{table:distributions_summary}. Note that the stream $Y$ of \cref{table:distributions_summary} and the stream $\tilde{Y}$ of \cref{sec:two-simplifications} are indeed asymptotically equivalent, as the proof of \cref{theorem:ulr_lower_bound} shows. Let us now state our main technical result and show how to use it to get $\ULR(f) \geq \Omega(n^{5/4})$.

\begin{theorem}\label{theorem:main_technical}
For any $b_X < o(b_Y) < o(1)$, $\phb \in [3b_Y, 4b_Y]$ and deterministic decision tree $D$,
$$\frac{\cost(D, \nu)}{\TV(\tran(D, \nu^0), \tran(D, \nu^1))} \geq \Omega(1) \cdot \min\left\lbrace \frac{1}{b_X^{1/2}b_Y}, \frac{1}{b_Xb_Y^{1/3}} \right\rbrace$$
\end{theorem}

\begin{theorem}\label{theorem:ulr_lower_bound}
$\ULR(f) \geq \Omega(n^{5/4})$
\end{theorem}
\begin{proof}
Following \cref{sec:two-simplifications}, in order to prove $\ULR(f)\geq\Omega(n^{5/4})$, it is sufficient to prove \cref{equation:ulr_reduction_step2}. To this end, fix any deterministic ternary decision tree $D$ and let $\hat{b}_X$ and $\hat{b}_Y$ be the original parameters of \cref{sec:ulr_overview}, which were set to $n^{-1}$ and $n^{-3/4}$ respectively. By setting $b_X \coloneqq \hat{b}_X$, $b_Y \coloneqq \hat{b}_Y/(1-4\hat{b}_Y)$ and $\phb = 4\hat{b}_Y$, we have $\tilde{\nu} = \nu$, $b_Y = \Theta(\hat{b}_Y)$ and $\phb \in [3b_Y, 4b_Y]$ (for $n$ large enough), so that we may apply \cref{theorem:main_technical} directly:
$$\frac{\qbar(D, \tilde{\nu})}{\TV(\tran(D, \tilde{\nu}^0), \tran(D, \tilde{\nu}^1))} = \frac{\qbar(D, \nu)}{\TV(\tran(D, \nu^0), \tran(D, \nu^1))} \geq \Omega(1) \cdot \min\left\lbrace \frac{1}{b_X^{1/2}b_Y}, \frac{1}{b_Xb_Y^{1/3}} \right\rbrace$$
This is $\Omega(n^{5/4})$ for our initial setting of $\hat{b}_X$ and $\hat{b}_Y$.
\end{proof}

\renewcommand{\arraystretch}{1.2}
\begin{figure}[t]
\centering
\begin{minipage}{.65\textwidth}
\centering    
	\begin{tabular}{|r|l|l|l|}
		\hline
		Stream & $0$ & $1$ & $\hb$\\
		\hline
		\hline
		$X^0$ & $.5 + b_X$ & $.5 - b_X$ & 0\\
		\hline
		$X^1$ & $.5 - b_X$ & $.5 + b_X$ & 0\\
		\hline
		$Y^0$ & $(.5 - b_Y)(1 - \phb)$ & $(.5 + b_Y)(1 - \phb) $ & $\phb$\\
		\hline	
		$Y^1$ & $(.5 + b_Y)(1 - \phb)$ & $(.5- b_Y)(1 - \phb)$  & $\phb$\\
		\hline
		$Z^0$ & $.5 - b_Y$ & $.5 + b_Y$ & 0\\
		\hline
		$Z^1$ & $.5 + b_Y$ & $.5 - b_Y$ & 0\\
		\hline
		$U$ & $.5$ & $.5$ & 0\\
		\hline
	\end{tabular}
\end{minipage}%
\begin{minipage}{.35\textwidth}
	\centering
	\begin{align*}
		\nu^0 &\coloneqq  (1-\lambda)X^0 + \lambda Y^0\\
		\nu^1 &\coloneqq  (1-\lambda)X^1 + \lambda Y^1\\
		\nu &\coloneqq (\nu^0 + \nu^1)/2\\
		\\
		M^0 &\coloneqq  (1-\lambda)X^0 + \lambda Z^0\\
		M^1 &\coloneqq  (1-\lambda)X^1 + \lambda Z^1\\
		M &\coloneqq (M^0 + M^1)/2
	\end{align*}	    
\end{minipage}
\captionof{table}{A summary of the random streams used in this section. It should be read as e.g. the stream $Z^1$ has probability $1/2 - b_Y$ of producing a $1$. The mixture parameter $\lambda$ is set following \cref{eq:setting_of_lambda} with $b_X$ and $b_Y$ so that $\lambda = \Theta(b_X/b_Y)$.}
\label{table:distributions_summary}
\end{figure}

\subsection{Bias decomposition}
A decision tree trying to distinguish $\nu^0$ from $\nu^1$ may pick one of several different strategies or even a mixture of those. To encompass all types of strategy, we will split the bias contribution of each leaf depending on their length and witnessing qualities. For instance, leaves that witness (i.e. $\hb \in \ell$) have to solve the $Y^0$ versus $Y^1$ problem, whereas leaves that do not witness have the harder task of distinguishing $M^0$ from $M^1$. This effect however wears-off with an increasing depth: if some input reaches a very long non-witnessing leaf, then most likely the distribution was not $Y$ to start with and this leaf can thus focus on distinguishing $X^0$ from $X^1$. To formalize this idea, let us partition the leaves of a decision tree $D$ with $\L(D) = \Lwit(D) \cup \Lnotwit(D)$ where $\Lwit(D)$ contains all witnessing leaves and $\Lnotwit(D)$ the non-witnessing ones. For a pair of streams $P^0$ and $P^1$, we define the witnessing and non-witnessing bias with:
\begin{align*}
\bwit(D, P^0, P^1) \coloneqq \sum\nolimits_{\ell \in \Lwit(D)} \left\vert P^0[\ell] - P^1[\ell] \right\vert\\
\bnonwit(D, P^0, P^1) \coloneqq \sum\nolimits_{\ell \in \Lnotwit(D)} \left\vert P^0[\ell] - P^1[\ell] \right\vert
\end{align*}
In particular, $2 \cdot \TV(\tran(D, P^0), \tran(D, P^1)) = \bwit(D, P^0, P^1) + \bnonwit(D, P^0, P^1)$. This distinction between witnessing and non-witnessing leaves allows for a quick proof of \cref{theorem:main_technical}, provided that we have matching hardness result for the witnessing and non-witnessing trade-offs.
\begin{proof}[Proof of \cref{theorem:main_technical}]
Using \cref{theorem:witnessing_ratio} and \cref{theorem:nonwit_ratio} to bound the witnessing and non-witnessing trade-off,
\begin{align*}
\frac{\cost(D, \nu)}{\TV(\tran(D, \nu^0), \tran(D, \nu^1))} &\geq \min\left\lbrace \frac{\cost(D, \nu)}{\bwit(D, \nu^0, \nu^1)}, \,  \frac{\cost(D, \nu)}{\bnonwit(D, \nu^0, \nu^1)} \right\rbrace\\ &\geq \Omega(1) \cdot \min\left\lbrace \frac{1}{b_X^{1/2}b_Y}, \frac{1}{b_Xb_Y^{1/3}}\right\rbrace \qedhere
\end{align*}
\end{proof}
In the following sections, it will be convenient to use $\Delta(\ell)$ for the absolute difference between the number of positive and negative literal in the conjunction $\ell$. $\Delta(\ell)$ will be directly related to the distinguishing qualities of a leaf. For instance, if $\Delta(\ell) = 0$, then $\ell$ has no bias in distinguishing $X^0$ from $X^1$.

\subsection{Trade-off for witnessing leaves}

Since the distribution $X^0$ and $X^1$ never output $\hb$, witnessing leaves can only be reached if the distribution is $Y^0$ or $Y^1$. Hence, we can focus on bounding $\bwit(D, Y^0, Y^1)$ as $\bwit(D, \nu^0, \nu^1) = \lambda \cdot \bwit(D, Y^0, Y^1)$. The main idea is to bound $\bwit(D, Y^0, Y^1)$ using the hardness of distinguishing $B^0$ from $B^1$ with $b \coloneqq b_Y$, with care needed as $Y$ is ternary but $B$ is of binary nature. Note that under $Y$, we expect to see a $\hb$ after about $\Theta(1/\phb)$ queries and to simplify the argument, we will assume that any leaf witnesses a $\hb$ if its length is larger than $\Theta(1/\phb)$. This relaxed notion of witnessing can be understood as \emph{helping} the tree: if it has already made about $\Theta(1/\phb)$ queries, then we give it a $\hb$ \emph{for free}. To express this syntactically, let $\Lcut = \lceil 1/3\phb \rceil$ and define the set of stopping nodes $\mathcal{S}(D) \subseteq \L(D)$ as any node which witnesses for the first time or has not witnessed but is at depth $\Lcut$. Observe that the parent of any stopping node corresponds to a conjunction with no $\hb$ and that any witnessing leaf has a unique stopping node as ancestor. For each $s \in \mathcal{S}(D)$, let $D_s$ be the decision tree rooted at node $s$. With this notation in hand, we have:
\begin{equation}\label{equation:relaxed_witness}
\bwit(D, Y^0, Y^1) \leq \sum\nolimits_{s \in \mathcal{S}(D)} \sum\nolimits_{\ell \in \L(D_s)} \left\vert Y^0[s \circ \ell] - Y^1[s \circ  \ell] \right\vert
\end{equation}
The inequality instead of strict equality comes from our \textit{relaxed} notion of witnessing leaf, i.e. some large non-witnessing leaves are now also accounted for; but this is counterbalanced by the fact that those leaves tend to contribute greatly toward the expected cost. The connection between cost and bias will be made through $Y[\mathcal{S}(D)]$, the probability that a node from $\mathcal{S}(D)$ is reached by an input $x \sim Y$. Shallow trees will tend to have $Y[\mathcal{S}(D)]$ close to zero while tall trees will tend to have this probability close to 1. We now state and prove our bound for the witnessing trade-off.

\begin{theorem}\label{theorem:witnessing_ratio}
For any deterministic decision tree $D$,
$$\frac{\cost(D, \nu)}{\bwit(D, \nu^0, \nu^1)} \geq \Omega(1) \cdot \min\left\lbrace \frac{1}{b_X^{1/2}b_Y}, \frac{1}{b_Xb_Y^{1/3}}\right\rbrace$$
\end{theorem}
\begin{proof}
Using \cref{lemma:wit_bias_bound} and \cref{lemma:wit_cost_bound}, we have:
$$\frac{\cost(D, \nu)}{\bwit(D, \nu)} = \frac{\cost(D, \nu)}{\lambda \cdot \bwit(D, Y)}\geq \Omega(1) \cdot \frac{\max \left\lbrace\cost(D, X),\, \lambda \cost(D, Y),\, Y[\mathcal{S}(D)]/b_Y \right\rbrace}{\lambda \cdot \max \left\lbrace  b_Y^{4/3} \cost(t, X) ,\, b_Y \sqrt{Y[\mathcal{S}(D)] \cdot \cost(D, Y)}\right\rbrace}$$
If $b_Y^{4/3} \cost(D, X) \geq Y[\mathcal{S}(D)] \cdot \cost(D, Y)$, then we have:
$$\frac{\cost(D, \nu)}{\bwit(D, \nu)} \geq \Omega(1) \cdot \frac{\cost(D, X)}{\lambda b_Y^{4/3} \cost(D, X)} = \Omega\left(\frac{1}{b_X b_Y^{1/3}} \right)$$
The last inequality holds because $\lambda = \Theta(b_X/b_Y)$. If the other case holds, then we have:
\begin{equation*}
\frac{\cost(D, \nu)}{\bnonwit(D, \nu)} \geq \Omega(1) \cdot \max\left\lbrace\frac{\sqrt{\cost(D, Y)}}{b_Y \sqrt{Y[\mathcal{S}(D)]}} ,\,\frac{\sqrt{Y[\mathcal{S}(D)]}}{\lambda b_Y^2\sqrt{\cost(D, Y)}} \right\rbrace \geq \Omega\left(\frac{1}{b_X^{1/2}b_Y} \right) \qedhere
\end{equation*}
\end{proof}

\begin{lemma}\label{lemma:wit_bias_bound}
For any deterministic decision tree $D$:
$$\bwit(D, Y^0, Y^1) \leq O(1) \cdot \max \left\lbrace b_Y \sqrt{\cost(D, Y) \cdot Y[\mathcal{S}(D)]},\, b_Y^{4/3} \cost(D, X) \right\rbrace$$
\end{lemma}
\begin{proof}
Continuing on \cref{equation:relaxed_witness}, we further split the bias contribution by leaves that are stopping nodes and leaves which have an ancestor stopping node.
\begin{align*}
\bwit(D, Y^0, Y^1) &\leq \sum_{s \in \mathcal{S}(D)} \sum_{\ell \in \mathcal{L}(D_s)} \left\vert Y^0[s \circ \ell] - Y^1[s \circ  \ell] \right\vert \\
&= \sum_{s \in \mathcal{S}(D)} \sum_{\ell \in \mathcal{L}(D_s)} \left\vert Y^0[s] Y^0[\ell] - Y^1[s]Y^1[\ell] \right\vert \\
&= \sum_{s \in \mathcal{S}(D)} \sum_{\ell \in \mathcal{L}(D_s)} 2Y[s] \left\vert\left(\frac{1}{2} + \delta(s)\right) Y^0[\ell] - \left(\frac{1}{2} - \delta(s)\right) Y^1[\ell] \right\vert \quad \delta(s) \coloneqq \frac{Y^0[s] - Y^1[s]}{4Y[s]}\\
&\leq \sum_{s \in \mathcal{S}(D)} \sum_{\ell \in \mathcal{L}(D_s)} Y[s] \cdot \left\vert Y^0[\ell] - Y^1[\ell]\right\vert + 2Y[s] \cdot \left\vert \delta(s) \right\vert \cdot \left(Y^0[\ell] + Y^1[\ell]\right) \\
&= \sum_{s \in \mathcal{S}(D)} \sum_{\ell \in \mathcal{L}(D_s)} Y[s] \cdot \left\vert Y^0[\ell] - Y^1[\ell]\right\vert + \underbrace{\sum\nolimits_{s \in \mathcal{S}(D)}  \left\vert Y^0[s] - Y^1[s] \right\vert}_{\text{cash-out bias $b_C$}}
\end{align*}
\cref{lemma:cashout_bias} shows that $b_C \leq O(1) \cdot b_Y^{4/3} \cost(D, X)$ and we now argue that the first sum is bounded by $O(1) \cdot b_Y \sqrt{\cost(D, Y) \cdot Y[\mathcal{S}(D)]}$, thus finishing the claim. To do so, fix some $s \in \mathcal{S}(D)$ and let us assume that $D_s$ is optimally labeled with $\{Y^0, Y^1\}$ so that:
$$\sum\nolimits_{\ell \in \L(D_s)} \left\vert Y^0[\ell] - Y^1[\ell] \right\vert = 2 \cdot \left(\Pr\nolimits_{x \sim Y^0}[D_s(x) = Y^0] - \Pr\nolimits_{x \sim Y^1}[D_s(x) = Y^0] \right)$$
Now, $D_s$ can be transformed into a \emph{randomised} decision tree $R_s$ that solves $Z^0$ versus $Z^1$: $R_s$ runs $D_s$ as usual and swap a $\hb$ for query answers in $\B$ with probability $\phb$. $R_s$ further has $Y^0$ leaves relabeled by $Z^0$ and $Y^1$ leaves by $Z^1$, implying:
\begin{align*}
\sum\nolimits_{\ell \in \L(D_s)} \left\vert Y^0[\ell] - Y^1[\ell] \right\vert &= 2 \cdot \left( \Pr\nolimits_{x \sim Z^0}[R_s(x) = Z^0] - \Pr\nolimits_{x \sim Z^1}[R_s(x) = Z^0] \right)\\
&= 2 \cdot \TV(\tran(R_s, Z^0), \tran(R_s, Z^1))\\
&\leq O(1) \cdot b_Y \sqrt{\cost(R_s, Z)} \tag{by \cref{theorem:pi_vs_pis} with $b \coloneqq b_Y$}\\
&= O(1) \cdot b_Y \sqrt{\cost(D_s, Y)}
\end{align*}
Applying this bound to each $s \in \mathcal{S}(D)$, we get the desired bound on the left sum:
\begin{align*}
\sum_{s \in \mathcal{S}(D)} \sum_{\ell \in \mathcal{L}(D_s)} Y[s] \cdot \left\vert Y^0[\ell] - Y^1[\ell]\right\vert &\leq O(1) \cdot b_Y \sum_{s \in \mathcal{S}(D)} Y[s] \sqrt{\cost(D_s, Y)}\\
&\leq O(1) \cdot b_Y \sqrt{Y[\mathcal{S}(D)]} \cdot  \sqrt{\sum_{s \in \mathcal{S}(D)} Y[s] \cdot \cost(D_s, Y)}\\
&\leq O(1) \cdot b_Y \sqrt{Y[\mathcal{S}(D)]} \cdot  \sqrt{\cost(D, Y)}
\end{align*}
Where Cauchy-Schwarz inequality was used for the second inequality.
\end{proof}

\begin{lemma}\label{lemma:cashout_bias}
Following the notation of \cref{lemma:wit_bias_bound}, $b_C \leq O(1) \cdot b_Y^{4/3} \cost(D, X)$ for any deterministic decision tree $D$.
\end{lemma}
\begin{proof}
Without loss of generality, we may assume that $D$ has depth bounded by $\Lcut$ as there is no stopping node with depth greater than that. As a first step, we partition $\mathcal{S}(D) = \mathcal{S}^\hb \cup \Scut$ where $\mathcal{S}^\hb$ contains all the witnessing stopping nodes in $D$ and $\Scut$ the nodes that are stopping only because of their length being $\Lcut$. Note that any $s \in \mathcal{S}^\hb$ has parent node $p(s)$ representing a conjunction with no $\hb$. Therefore, we may substitute $Z$ to $Y$ as follows:
\begin{align*}
b_C &= \sum\nolimits_{s \in \Scut} \left\vert Y^0[s] - Y^1[s] \right\vert + \sum\nolimits_{s \in \mathcal{S}^\hb} \left\vert Y^0[s] - Y^1[s] \right\vert\\
&\leq \sum\nolimits_{s \in \Scut} \left\vert Z^0[s] - Z^1[s] \right\vert + \phb \sum\nolimits_{s \in \mathcal{S}^\hb} \left\vert Z^0[p(s)] - Z^1[p(s)]\right\vert
\end{align*}
The second sum bears a nice interpretation. Indeed, if $D'$ is a copy of $D$ that stops at depth $\Lcut - 1$ and has all witnessing paths removed then the second sum is equal to $\sum_{n \in \mathcal{N}(D')} \left\vert Z^0[n] - Z^1[n] \right\vert$, i.e. the bias $D'$ gets in distinguishing $Z^0$ from $Z^1$ while having the ability to \textit{cash-out} the current bias it has at each node. Now, if $D'$ happens to be close to the complete binary tree, then we may apply the hardness of distinguishing $Z^0$ from $Z^1$ at each level and get a global bound on the cash-out bias. This approach will however fail if the tree is largely unbalanced and we thus develop a more robust argument, which we re-use in later results. We first partition $\mathcal{S}^\hb = \bigcup \mathcal{S}^k$ where $\mathcal{S}^k \coloneqq \{s \in \mathcal{S}^\hb: |s| = k \}$ for $k \in [\Lcut]$. Fix now some $k \in [\Lcut]$ and observe that,
\begin{align*}
\sum_{s \in \mathcal{S}^k} \left\vert Z^0[p(s)] - Z^1[p(s)]\right\vert &\leq 8b_Y \sum_{s \in \mathcal{S}^k}\Delta(p(s))Z[p(s)] \tag{by \cref{lemma:Z0vsZ1_corruption}}\\
&\leq 16b_Yk^{2/3} \sum_{\Delta(s) \leq 2k^{2/3}} Z[p(s)] + 8b_Yk \sum_{\Delta(s) \geq 2k^{2/3}} Z[p(s)]\\
&\leq 16b_Yk^{2/3} \sum_{\Delta(s) \leq 2k^{2/3}} Z[p(s)] + 16b_Yke^{-k^{1/3}/48} \tag{by \cref{lemma:chernoff2_3}}\\
&\leq 16b_Yk^{2/3} \sum_{\ell \in \L(D): |\ell| \geq k} Z[\ell] + 16b_Yke^{-k^{1/3}/48}\\
&\leq 16b_Yk^{-1/3}\cost(D, Z) + 16b_Yke^{-k^{1/3}/48} \tag{Markov's inequality}\\
&\leq O(1) \cdot b_Yk^{-1/3}\cost(D, Z)
\end{align*}
We made the arbitrary choice to split between small and large $\Delta$ with cutoff parameter $2k^{2/3}$. Setting it to the limiting $k^{1/2}\textup{polylog}(n)$ would  have slightly improved the $\LR$ ratio against $\nu$ but this would ultimately yield $\lambda = \Theta(1/\textup{polylog}(n))$, thus \emph{worsening} the step in which we reduce to decision trees not solving hard blocks (see \cref{sec:two-simplifications}). Observing that the above chain of inequalities also holds for $\Scut$ with level $k = \Lcut$, we have:
\begin{align*}
b_C &\leq \sum_{s \in \Scut} \left\vert Z^0[s] - Z^1[s] \right\vert + \phb \sum_{s \in \mathcal{S}^*} \left\vert Z^0[p(s)] - Z^1[p(s)]\right\vert\\
&\leq O(1) \cdot b_Y\Lcutpower{-1/3}\cost(D, Z) + O(1) \cdot p_*b_Y\cost(D, Z) \sum\nolimits_{k \in [\Lcut]} k^{-1/3}\\
&\leq O(1) \cdot b_Y\Lcutpower{-1/3}\cost(D, Z) + O(1) \cdot p_*b_Y\Lcut^{2/3}\cost(D, Z) \\
&\leq O(1) \cdot b_Y^{4/3} \cost(t, Z) \tag{$\Lcut = \lceil 1/3p_* \rceil$ and $p_* = \Theta(b_Y)$}\\
&\leq O(1) \cdot b_Y^{4/3} \cost(t, X) &\tag*{(by  \cref{lemma:z_to_x_corruption_details}) \qedhere}
\end{align*}
\end{proof}

\begin{lemma}\label{lemma:wit_cost_bound}
For any deterministic decision tree $D$, $\cost(D, \nu) \geq \Omega(Y[\mathcal{S}(D)]/\phb)$.
\end{lemma}
\begin{proof}
Without loss of generality, we may assume that $D$ stops whenever it reaches a stopping node. Let us first argue that $\cost(D, \nu) \geq \Omega(\cost(D, Y))$. Fix $d \in \B$ and observe that sampling from $Y^d$ is the same as sampling from $Z^d$ while \textit{salting} the query answers by replacing them with a star with independent probability $\phb$ so that:
\begin{equation}\label{equation:salt}
\cost(D, Y^d) = \EXSUB{x \sim Z^d}{\EXSUB{\textup{salt}}{q(D, \textup{salt}(x))}} \leq \EXSUB{x \sim Z^d}{q(D,x)} = \cost(D, Z^d) \leq 78 \cost(D, X)
\end{equation}
The first inequality stems from the fact that flipping some answer with a star makes the leaf witness and $D$ thus directly stops. The last one is due by \cref{lemma:z_to_x_corruption_details}, recalling that $D$ has depth bounded by $\Lcut$. Using \cref{equation:salt} and the definition of $\nu$, we have $\cost(D, \nu) \geq \Omega(\cost(D, Y))$. Recall that there are two types of stopping nodes in $\mathcal{S}(D)$. The first type are nodes that witnesses for the first time and the second type are nodes that never witness but have reached depth $\Lcut$. We split $Y[\mathcal{S}(D)]$, according to both type:
$$p_1 \coloneqq \Pr\nolimits_{x \sim Y}[\textup{$D$ stops on \hb}] \quad\text{and}\quad p_2 \coloneqq \Pr\nolimits_{x \sim Y}[\textup{$D$ stops because of $\Lcut$}]$$
We may thus write $Y[\mathcal{S}[D]] = p_1 + p_2$, as a simple sanity check, note that it is possible to have $Y[\mathcal{S}(D)] \ll 1$, e.g. if $D$ has many small non-witnessing leaves. If $p_2 \geq p_1$, then $p_2 \geq Y[\mathcal{S}(D)]/2$ and so:
$$\cost(D, Y) \geq \sum\nolimits_{|\ell| = \Lcut} |\ell| Y[\ell] = p_2 \Lcut \geq \Omega(Y[\mathcal{S}(D)]/\phb)$$
Now, if $p_1 \geq Y[\mathcal{S}(D)]$, we let $D'$ be a decision tree that runs $D$ in turn until it witnesses a $\hb$ in the stream but for at most $\lceil 2/p_1 \rceil$ times. We ensure that the runs are independent by offsetting the query indices by a multiple of a large number (e.g. $10\Lcut$). With that number of runs, $D'$ has a constant probability of witnessing:
$$\Pr\nolimits_{x \sim Y}[\textup{$D'$ witnesses}] = 1 - (1- p_1)^{\lceil 2/p_1 \rceil} \geq 1 - e^{-2} \geq 3/4$$
Observe further that $D'$ never queries after witnessing and that $\cost(D, Y) \geq \Omega(\cost(D', Y) Y[\mathcal{S}(D)])$. Fix now $\L^1 = \{\ell \in \L(D'): \hb \in \ell \text{ and } |\ell| \leq \Lcut\}$ and $\L_2 = \{\ell \in \L(D'): \hb \in \ell \text{ and } |\ell| > \Lcut\}$ and note that $Y[\L^1] + Y[\L^2] \geq 3/4$. Because leaves of $D'$ can only have a $\hb$ as their last literal we have $Y[\ell] = (1-\phb)^{|\ell|}\phb Z[p(\ell)]$ where $p(\ell)$ is the parent node of $\ell$ and hence:
$$Y[\L^1] = \sum_{k = 1}^{\Lcut} \sum_{\substack{\ell \in \L_1:\\ |\ell| = k}} Y[\ell] \leq 2\phb \sum_{k = 1}^\Lcut (1-\phb)^{k-1} = 1 - (1 - \phb)^\Lcut \leq \phb\Lcut \leq 2/3$$
This shows that $Y[\L^2] \geq 1/12$ and thus:
\begin{equation*}
\cost(D, Y) \geq \Omega(1) \cdot Y[\mathcal{S}(D)] \cost(D', Y) \geq \Omega(1) \cdot Y[\mathcal{S}(D)] Y[\L^2] \Lcut \geq \Omega(Y[\mathcal{S}(D)/\phb]) \qedhere
\end{equation*}
\end{proof}

\subsection{Trade-off for non-witnessing leaves}
Non-witnessing leaves can be reached both by the $X$ and $Y$ distribution and as such have the harder task of distinguishing $\nu^0$ from $\nu^1$, unlike witnessing leaves that only need to solve $Y^0$ versus $Y^1$. As previously noted, this effect wears off following the depth of the tree: if the unknown stream reaches a long leaf with no $\hb$, then most likely the stream is part of $\{X^0, X^1\}$ and not $\{Y^0, Y^1\}$. To account for this, we will break the analysis into small and large leaves with cutoff parameter $\Lcut = \lceil 1/b_Y \rceil$. For a decision tree $D$, let $\Lsmall(D) = \{\ell \in \Lnotwit(D): |\ell| \leq \Lcut\}$ and $\Llarge(D) = \{\ell \in \Lnotwit(D): |\ell| > \Lcut\}$. We will argue that the bias brought by $\Lsmall(D)$ is capped by the hardness of the $M^0$ versus $M^1$ problem (see \cref{sec:ber-mix}) and that the bias brought by $\Llarge(D)$ can bounded using the $B^0$ versus $B^1$ problem with $b \coloneqq b_X$ (see \cref{sec:ber-opposite}).

\begin{theorem}\label{theorem:nonwit_ratio}
For any deterministic decision tree $D$, $\cost(D, \nu)/\bnonwit(D, \nu^0, \nu^1) \geq \Omega(1/b_Xb_Y^{1/3})$
\end{theorem}
\begin{proof}
Using the notation introduced above, we can split $\bnonwit(D, \nu^0, \nu^1)$ with:
\begin{align*}
\bnonwit(D, \nu^0, \nu^1) &= \sum_{\ell \in \Lsmall} \left\vert \nu^0[\ell] - \nu^1[\ell] \right\vert + \sum_{\ell \in \Llarge} \left\vert \nu^0[\ell] - \nu^1[\ell] \right\vert\\
&\leq \sum_{\ell \in \Lsmall} \left\vert \nu^0[\ell] - \nu^1[\ell] \right\vert + O(1) \cdot \sum_{\ell \in \Llarge} \left\vert X^0[\ell] - X^1[\ell] \right\vert + b_Y^3 \tag{by \cref{lemma:bias_spillover_details}}
\end{align*}
\cref{lemma:nonwit_large_bias} shows that the the second sum is bounded by $O(\cost(D, X)b_Xb_Y^{1/3})$ so that we only need to show that the first sum is also bounded by the same amount, which we do next. To analyse this sum, we may assume without loss of generality that $D$ is a binary decision tree of maximum depth $\Lcut$. Indeed, $\Lsmall$ contains small non-witnessing leaves only. This implies that for all $\ell \in \L(D)$, $Y[\ell] = (1 - \phb)^{|\ell|}Z[\ell]$ and so we may further break the sum into the contribution of $M^0$ versus $M^1$ and the one of $Z^0$ versus $Z^1$:
\begin{align*}
\left\vert \nu^0[\ell] - \nu^1[\ell] \right\vert &= \left\vert M^0[\ell] - \lambda Z^0[\ell] + \lambda Y^0[\ell] - M^1[\ell] + \lambda Z^1[\ell] - \lambda Y^1[\ell]\right\vert\\
&\leq \left\vert M^0[\ell] - M^1[\ell]\right\vert + \lambda \cdot \left\vert Y^0[\ell] - Y^1[\ell] - Z^0[\ell] + Z^1[\ell] \right\vert\\
&= \left\vert M^0[\ell] - M^1[\ell]\right\vert + \lambda \cdot (1 - (1-\phb)^{|\ell|} ) \cdot \left\vert  Z^0[\ell] - Z^1[\ell] \right\vert\\
&\leq \left\vert M^0[\ell] - M^1[\ell]\right\vert + \lambda \phb |\ell| \cdot \left\vert  Z^0[\ell] - Z^1[\ell] \right\vert \tag{by \cref{lemma:approx_zoo}}
\end{align*}
Using this insight, we can finally exploit the hardness of distinguishing $M^0$ from $M^1$:
\begin{align*}
\sum_{\ell \in \Lsmall} \left\vert \nu^0[\ell] - \nu^1[\ell] \right\vert &\leq \sum_{\ell \in \L(D)} \left\vert M^0[\ell] - M^1[\ell]\right\vert +  \lambda \phb \sum_{\ell \in \L(D)}  |\ell| \cdot \left\vert  Z^0[\ell] - Z^1[\ell] \right\vert\\
&= 2\cdot\TV(\tran(D, M^0), \tran(D, M^1)) + \lambda \phb \sum_{\ell \in \L(D)}  |\ell| \cdot \left\vert  Z^0[\ell] - Z^1[\ell] \right\vert\\
&\leq O(b_Xb_Y\cost(D, M)) + \lambda \phb \sum_{\ell \in \L(D)}  |\ell| \cdot \left\vert  Z^0[\ell] - Z^1[\ell] \right\vert \tag{by \cref{theorem:A0_vs_A1}}\\
&\leq O(b_Xb_Y\cost(D, X)) + \lambda \phb \underbrace{\sum\nolimits_{\ell \in \L(D)}  |\ell| \cdot \left\vert  Z^0[\ell] - Z^1[\ell] \right\vert}_{\textup{weighted bias $b_W$}} \tag{by \cref{lemma:z_to_x_corruption_details}}
\end{align*}
Since $\lambda = \Theta(b_X/b_Y)$ and $\phb = \Theta(b_Y)$, the only thing left to prove is that $b_W \leq O(b_Y^{1/3}\cost(D, X))$. We resort to an analysis similar to the one employed in \cref{lemma:cashout_bias}, except that this time we need to take into account the size of the leaf. To that end, let $\L^k \coloneqq \{\ell \in \L(D): |\ell| = k\}$ for $k \in [\Lcut]$ and let us bound the bias brought by level $k$:
\begin{align*}
\sum\nolimits_{\ell \in \L^k} k \cdot \left\vert Z^0[\ell] - Z^1[\ell]\right\vert &\leq 8b_Yk \sum\nolimits_{\ell \in \L^k} \Delta(\ell)Z[\ell] \tag{by \cref{lemma:Z0vsZ1_corruption}}\\
&\leq 16b_Yk^{5/3} \sum\nolimits_{\Delta(\ell) \leq 2k^{2/3}} Z[\ell] + 8b_Yk^2 \sum\nolimits_{\Delta(\ell) \geq 2k^{2/3}} Z[\ell] \\
&\leq 16b_Yk^{5/3} \sum\nolimits_{\Delta(\ell) \leq 2k^{2/3}} Z[\ell] + 16b_Yk^2e^{-k^{1/3}/48} \tag{by \cref{lemma:chernoff2_3}}
\end{align*}
Recalling that $D$ has depth bounded by $\Lcut = \lceil 1/b_Y \rceil$, we have $k^{5/3} \leq b_Y^{-2/3}k$ and hence:
\begin{align*}
b_W &= \sum\nolimits_{k \in [\Lcut]} \sum\nolimits_{\ell \in \L^k} k \cdot \left\vert Z^0[\ell] - Z^1[\ell]\right\vert\\
&\leq 16b_Y^{1/3} \sum\nolimits_{\ell \in \L(t)} |\ell| \cdot Z[\ell] + 16b_Y\sum\nolimits_{k \in [\Lcut]} k^2  e^{-k^{1/3}/48}\\
&\leq 16b_Y^{1/3} \cost(D, Z) + O(b_Y) \tag{by ratio test}\\
&\leq O(1) \cdot b_Y^{1/3}\cost(D, X) \tag*{(by \cref{lemma:z_to_x_corruption_details}) \qedhere}
\end{align*}
\end{proof}

\begin{lemma}\label{lemma:nonwit_large_bias}
For any deterministic decision tree $D$, $\sum_{\ell \in \Llarge(D)} \left\vert X^0[\ell] - X^1[\ell] \right\vert \leq O(1)\cdot b_Xb_Y^{1/3}\cost(D, X)$
\end{lemma}
\begin{proof}
We may assume without loss of generality that $D$ is over $\B$ only as any witnessing conjunction $\ell$ has $\left\vert X^0[\ell] - X^1[\ell]\right\vert = 0$. We let $\mathcal{C} \coloneqq \{c \in \mathcal{N}(t): |c| = \Lcut\}$ be the set of node at height $\Lcut$ and for each $c \in \mathcal{C}$, define $D_c$ to be the sub-tree of $D$ rooted at $c$. Re-cycling the decomposition used in the proof of \cref{lemma:wit_bias_bound}:
\[\sum\nolimits_{\ell \in \Llarge} \left\vert X^0[\ell] - X^1[\ell] \right\vert \leq \underbrace{\sum\nolimits_{c \in \mathcal{C}} X[c] \sum\nolimits_{\ell \in \mathcal{L}(D_c)} \left\vert X^0[\ell] - X^1[\ell]\right\vert}_{\text{large leaves bias $b_L$}} + \underbrace{\sum\nolimits_{c \in \mathcal{C}}  \left\vert X^0[c] - X^1[c] \right\vert}_{\text{cut-off bias $b_C$}}\]
We show next that $b_L, b_C \leq O(b_X b_Y^{1/3} \cost(D, X))$. To get a bound on $b_L$, observe that for any $c \in \mathcal{C}$, $\sum_{\ell \in \mathcal{L}(D_c)} \left\vert X^0[\ell] - X^1[\ell]\right\vert = 2 \cdot \TV(\tran(D_c, X^0), \tran(D_c, X^1))$ which can be bounded by $O(b_X \sqrt{\qbar(D_c, X)})$ using \cref{theorem:pi_vs_pis} with $b \coloneqq b_Y$ so that:
\begin{align*}
b_L &\leq O(1) \cdot b_X \sum_{c \in \mathcal{C}} X[c] \sqrt{\cost(D_c, X)}\\
&= O(1) \cdot b_X \sqrt{X[\mathcal{C}]} \cdot  \sqrt{\sum\nolimits_{c \in \mathcal{C}} X[c] \cdot \cost(D_c, X)} \tag{by Cauchy-Schwarz inequality}\\
&\leq O(1) \cdot b_Xb_Y^{1/2} \sqrt{\cost(D, X)} \cdot \sqrt{\sum\nolimits_{c \in \mathcal{C}} X[c] \cdot \cost(D_c, X)}\tag{by Markov's inequality and $\Lcut = \lceil 1/b_Y \rceil$}\\
&\leq O(1) \cdot b_Xb_Y^{1/2} \cost(D, X)
\end{align*}
We can finally bound $b_C$, using the fact that all $c \in \mathcal{C}$ have length $\Lcut = \lceil 1/b_Y \rceil$.
\begin{align*}
b_C &\leq O(1) \cdot b_X \sum\nolimits_{c \in \mathcal{C}} \Delta(c)X[c] \tag{by \cref{lemma:Z0vsZ1_corruption}}\\
&= O(1) \cdot b_X\Lcutpower{2/3} \sum\nolimits_{\Delta(c) \leq 2\Lcutpower{2/3}} X[c] + O(1) \cdot b_X\Lcut \sum\nolimits_{\Delta(c) \geq 2\Lcutpower{2/3}} X[c]\\
&= O(1) \cdot b_X\Lcutpower{2/3} \sum\nolimits_{\Delta(c) \leq 2\Lcutpower{2/3}} X[c] + O(1) \cdot b_X\Lcut e^{-\Lcut^{1/3}/48} &\tag{by \cref{lemma:chernoff2_3}}\\
&\leq O(1) \cdot 2b_X\Lcutpower{2/3} \sum\nolimits_{\Delta(c) \leq 2\Lcutpower{2/3}} X[c] + O(1) \cdot b_X\Lcut^{-1/3}\\
&\leq O(1) \cdot b_X\Lcut^{-1/3} \qbar(D, X) + O(1) \cdot b_X\Lcut^{-1/3} \tag{by Markov's inequality}\\
&\leq O(1) \cdot b_Xb_Y^{1/3}\cost(D, X)
\tag*{\qedhere}
\end{align*}
\end{proof}

\section{Technical lemmas} \label{sec:omitted-proofs}

\subsection{Some corruption bounds}

\begin{lemma}\label{lemma:corruption_bound}
For any non-witnessing conjunction $\ell$, $M^0[\ell]/U[\ell] \geq 1 - 8|\ell|b_Xb_Y$
\end{lemma}
\begin{proof}
Fix $k \coloneqq |\ell|$ and let $\ell$ have $k/2 + q$ positive variables for some $q \in [-k/2, k/2]$. As a function of $q$, the ratio $M^0[\ell]/U[\ell]$ can be expressed as:
$$r(q) \coloneqq (1-\lambda)\left(1 - 4b_X^2\right)^{k/2}\left(\frac{1- 2b_X}{1 + 2b_X}\right)^q + \lambda \left(1 - 4b_Y^2\right)^{k/2}\left(\frac{1 + 2b_Y}{1 - 2b_Y}\right)^q$$
Using Lemma \ref{lemma:location_min}, the minimizer of the ratio (extended to $\mathbb{R}$) is $q^\star \in [kb_Y / 7, kb_Y]$, so we may lower-bound the above with:
\begin{align*}
\min_{q \in [-k/2, k/2]} r(q) &\geq (1-\lambda)\left(1 - 4b_X^2\right)^{k/2}\left(\frac{1- 2b_X}{1 + 2b_X}\right)^{q^\star} + \lambda \left(1 - 4b_Y^2\right)^{k/2}\left(\frac{1 + 2b_Y}{1 - 2b_Y}\right)^{q^\star}\\
&\geq  (1-\lambda)\left(1 - 4b_X^2\right)^{k/2}\left(1-4b_X\right)^{kb_Y} + \lambda \left(1 - 4b_Y^2\right)^{k/2}\\
&\geq  (1-\lambda)\left(1 -2kb_X^2\right)\left(1-4kb_Xb_Y\right) + \lambda \left(1 - 2kb_Y^2\right)\\
&\geq  (1-\lambda)\left(1-6kb_Xb_Y\right) + \lambda \left(1 - kb_Y^2\right)\\
&\geq 1 - 6kb_Xb_Y - \lambda k b_Y^2
\end{align*}
Recalling that $\lambda \leq 2b_X/b_Y$, we get that $M^0[\ell]/U[\ell] \geq 1 - 8|\ell|b_Xb_Y$, as desired.
\end{proof}

\begin{lemma}[$Z$ vs. $X$]\label{lemma:z_to_x_corruption_details}
For any non-witnessing conjunction $\ell$ with $|\ell| \leq 1/b_Y$, $Z^d[\ell] \leq 78X[\ell]$ for $d\in\B$.
\end{lemma}
\begin{proof}
We prove the claim for $d = 0$, the other case being symmetric. Fix some conjunction $\ell$ and let $L \coloneqq \lceil1/b_Y \rceil$. We first demonstrate that $Z^0[\ell] \leq 26 U[\ell]$:
$$\frac{Z^0[\ell]}{U[\ell]} \leq (1 + 2b_Y)^L = \sum_{k = 0}^L \binom{L}{k} (2b_Y)^k \leq 1 + \sum_{k = 1}^L \left(\frac{2eL b_Y}{k} \right)^k \leq 1 + \sum_{k = 1}^\infty \left(\frac{2e}{k}\right)^k \leq 26$$
Where the first inequality is due to the worst-case consisting of a conjunction with $L$ positive literal and zero negative ones. We now show that $U[\ell] \leq 3X[\ell]$, thus finishing the claim. Observe that the conjunction maximizing the ratio $X[\ell]/U[\ell]$ has $L/2$ positive literal and $L/2$ negative one, hence:
\begin{equation*}
\frac{X[\ell]}{U[\ell]} \geq \frac{X^1[\ell]}{2U[\ell]} \overset{\textup{(b)}}{\geq} \frac{\left(1 + 2b_X\right)^{L/2}\left(1 - 2b_X\right)^{L/2}}{2} =  \frac{1 - 2b_X^2 L}{2} \geq \frac{1}{3} \qedhere
\end{equation*}
\end{proof}

\begin{lemma}[$Z^0$ vs. $Z^1$]\label{lemma:Z0vsZ1_corruption}
For any leaf $\ell$, $\left\vert Z^0[\ell] - Z^1[\ell] \right\vert \leq 8b_Y\Delta(\ell)Z[\ell]$
\end{lemma}
\begin{proof}
If $\ell$ is witnessing, the claim is trivially true so let us assume that $\hb \notin \ell$. By symmetry, we may assume that $\ell$ has more negative literals than positive ones so that $Z^1[\ell] \geq Z^0[\ell]$, thus:
$$\left\vert Z^0[\ell] - Z^1[\ell] \right\vert = Z^1[\ell] - Z^0[\ell] \leq 2Z[\ell] \cdot \left(1 - Z^0[\ell]/Z^1[\ell]\right)$$
Now, letting $q_0$, respectively $q_1$ be the number of negative, respectively positive, literals in $\ell$ and using the definition of $Z^0$ and $Z^1$, we have:
$$\frac{Z^0[\ell]}{Z^1[\ell]} = \frac{(1 + 2b_Y)^{q_0}(1 - 2b_Y)^{q_1}}{(1 + 2b_Y)^{q_1}(1-2b_Y)^{q_0}} = \left(\frac{1 - 2b_Y}{1 + 2b_Y} \right)^{\Delta(\ell)} \geq (1 - 4b_Y)^{\Delta(\ell)}$$
Combining both observations, we get:
\begin{equation*}
\left\vert Z^0[\ell] - Z^1[\ell] \right\vert \leq 2Z[\ell] \cdot \left( 1 - (1 - 4b_Y)^{\Delta(\ell)}\right) \leq 8b_Y\Delta(\ell)Z[\ell] \qedhere
\end{equation*}
\end{proof}

\subsection{Some bias transfers}

\begin{lemma}[$\nu$ to $X$ bias transfer]\label{lemma:bias_spillover_details}
Letting $\Llarge$ be defined as in Theorem \ref{theorem:nonwit_ratio}, it holds that:
$$\sum\nolimits_{\ell \in \Llarge} \left\vert\nu^0[\ell] - \nu^1[\ell]\right\vert \leq b_Y^2 + O(1) \cdot \sum\nolimits_{\ell \in \Llarge} \left\vert X^0[\ell] - X^1[\ell]\right\vert$$
\end{lemma}
\begin{proof}
Recall that $\Llarge$ only contains non-witnessing leaf of size at least $\Lcut = \lceil 1/b_Y \rceil$.
Using the definition of $\nu$ and the triangle inequality, we have:
$$\sum\nolimits_{\ell \in \Llarge} \left\vert\nu^0[\ell] - \nu^1[\ell]\right\vert \leq \sum\nolimits_{\ell \in \Llarge} \left\vert X^0[\ell] - X^1[\ell]\right\vert + \lambda \sum\nolimits_{\ell \in \Llarge} \left\vert Y^0[\ell] - Y^1[\ell]\right\vert$$
Hence, we need to focus on the second sum only. To bound it, we partition $\Llarge$ into balanced and unbalanced leaves:
$$\Llarge = \bigcup_{k = \Lcut}^\infty \mathcal{B}^k \cup \mathcal{U}^k\quad\text{where}\quad
\begin{array}{l}
\mathcal{B}^k \coloneqq \left\lbrace \ell \in \Llarge: |\ell| = k \quad\text{and}\quad \Delta(\ell) \leq |\ell|/2\right\rbrace\\
\mathcal{U}^k \coloneqq \left\lbrace\ell \in \Llarge: |\ell| = k \quad\text{and}\quad \Delta(\ell) > |\ell|/2 \right\rbrace
\end{array}$$
For leaves in $\mathcal{B}^k$, one can apply Lemma \ref{lemma:leafwise_bias_corruption} to get that $\lambda \left\vert Y^0[\ell] - Y^1[\ell]\right\vert \leq O(1) \cdot \left\vert X^0[\ell] - X^1[\ell]\right\vert$ so that we have:
\begin{align*}
\lambda \sum_{\ell \in \Llarge} \left\vert Y^0[\ell] - Y^1[\ell]\right\vert &\leq O(1) \cdot \sum_{\ell \in \Llarge} \left\vert X^0[\ell] - X^1[\ell]\right\vert + \lambda \sum_{k = \Lcut}^\infty \sum_{\ell \in \mathcal{U}^k} \left\vert Y^0[\ell] - Y^1[\ell] \right\vert\\
&\leq  O(1) \cdot \sum_{\ell \in \Llarge} \left\vert X^0[\ell] - X^1[\ell]\right\vert + \sum_{k = \Lcut}^\infty \sum_{\ell \in \mathcal{U}^k} Y[\ell]\\
&\leq  O(1) \cdot \sum_{\ell \in \Llarge} \left\vert X^0[\ell] - X^1[\ell]\right\vert + \sum_{k = \Lcut}^\infty e^{-k/100} \tag{*}\\
&\leq  O(1) \cdot \sum_{\ell \in \Llarge} \left\vert X^0[\ell] - X^1[\ell]\right\vert + 2e^{-\Lcut/100}\\
&\leq  O(1) \cdot \sum_{\ell \in \Llarge} \left\vert X^0[\ell] - X^1[\ell]\right\vert + b_Y^2 \tag{for $b_Y$ small enough}
\end{align*}
Where (*) is obtained by a slight modification of the proof of \cref{lemma:chernoff2_3}.
\end{proof}

\begin{lemma}\label{lemma:leafwise_bias_corruption}
For any non-witnessing leaf $\ell$ with $|\ell| \geq 1/b_Y$ and $\Delta(\ell) \leq |\ell|/2$,
$$\lambda \cdot \left\vert Y^0[\ell] - Y^1[\ell]\right\vert \leq O\left( \left\vert X^0[\ell] - X^1[\ell] \right\vert\right)$$
\end{lemma}
\begin{proof}
The rationale behind the statement is that the leaf $\ell$ maximizing the ratio between $\left\vert Y^0[\ell] - Y^1[\ell]\right\vert$ and $\left\vert X^0[\ell] - X^1[\ell] \right\vert$ has $\Delta(\ell)$ maximized (e.g. $\ell$ with $3|\ell|/4$ positive literals). This is however technically challenging to prove directly. Thus, we will split the proof in two cases: first with $\Delta(\ell) \leq 16/10b_Y$ and then with $\Delta(\ell) \geq 16/10b_Y$. For the second regime, we will actually be able to prove that the most-separating leaves have $\Delta(\ell) = |\ell|/2$. Fix now any non-witnessing leaf $\ell$, let $k \coloneqq |\ell|$ and $q \coloneqq \Delta(\ell)/2$. Since $\hb \notin \ell$, we have:
\begin{align*}
\left\vert Y^0[\ell] - Y^1[\ell]\right\vert &= \left(1/4 - b_Y^2 \right)^{k/2} \cdot \left(r_Y^q - r_Y^{-q} \right) \cdot \left(1 - p_*\right)^k &\text{where}\quad r_Y \coloneqq \frac{1 + 2b_Y}{1 - 2b_Y}\\
\left\vert X^0[\ell] - X^1[\ell]\right\vert &= \left(1/4 - b_X^2 \right)^{k/2} \cdot \left( r_X^q - r_X^{-q} \right) &\text{where}\quad r_X \coloneqq \frac{1 + 2b_X}{1 - 2b_X}
\end{align*}
As $b_X \leq b_Y$, we will ignore the terms $\left(1/4 - b_Y^2 \right)^{k/2}$ and $\left(1/4 - b_X^2 \right)^{k/2}$. For the first regime $q \leq 16/10b_Y$, we don't even need the \textit{dampening} term $(1-\phb)^k$ and we simply show that:
\begin{equation}\label{equation:anonymous_subclaim1}
r_Y^q - r_Y^{-q} \leq \frac{10000}{\lambda} \cdot \left(r_X^q - r_X^{-q} \right)
\end{equation}
Using the series representation of the exponential function and \cref{lemma:approx_zoo} to bound $\ln(r_Y)$, we have:
$$r_Y^q - r_Y^{-q} = 2\sum_{t\geq 0 \text{ odd}} \frac{\ln(r_Y)^t q^t}{t!} \leq e^{\ln(r_Y) q} - 1 \leq e^{6b_Yq} - 1$$
Now, using the fact that $\lambda \leq 3b_X/b_Y$ (see Lemma \ref{lemma:approx_lambda}), have:
$$\frac{10000}{\lambda} \cdot \left(r_X^q - r_X^{-q} \right) = \frac{20000}{\lambda} \sum_{t\geq 0 \text{ odd}} \frac{\ln(r_X)^t q^t}{t!} \geq  \frac{20000}{\lambda} \cdot \ln(r_X) q \geq \frac{80000}{3} \cdot b_Yq$$
Hence, \ref{equation:anonymous_subclaim1} holds if $b_Yq \leq 1.6$ so that the claim holds in the first regime. In the regime where $q \geq 16/10b_Y$, we will show that:
\begin{equation}\label{equation:anonynmous_subclaim2}
\lambda (1-\phb)^k\Phi(q) \leq 3 \quad\text{where}\quad \Phi(q) \coloneqq \frac{r_Y^q}{r_X^q - r_X^{-q}}
\end{equation}
Since $\Phi$ (as a real function over $q$) is increasing on the interval $[16/10b_Y, \infty)$ (see Lemma \ref{lemma:phi_increasing_second_regime}), the maximum of the left-hand side of equation \ref{equation:anonynmous_subclaim2} is attained at the boundary of the domain, i.e. for $q = k/4$. Therefore, in the second regime:
$$\lambda (1-\phb)^k \Phi(q) \leq \lambda (1-\phb)^k \Phi(k/4) =\frac{\lambda}{r_X^{k/4} - r_X^{-k/4}} \cdot \left[(1-\phb)^4 \cdot r_Y \right]^{k/4} \leq \frac{\lambda}{r_X^{k/4} - r_X^{-k/4}}$$
The last inequality holds because the quantity in the square bracket is $\leq 1$ (recall that $\phb \in [3b_Y, 4b_Y]$). Now, note that $\mu - \mu^{-1} \geq \lambda/3$ if $\mu \geq 1 + \lambda/3$ but since $k \geq 1/b_Y$ and $\lambda \leq 3b_X/b_Y$, we have:
$$\mu \coloneqq r_X^{k/4} \geq \left(\frac{1 + 2b_X}{1 - 2b_X} \right)^{k/4} \geq (1 + 4b_X)^{k/4} \geq 1 + kb_X \geq 1 + \lambda/3$$
So that \ref{equation:anonynmous_subclaim2} holds.
\end{proof}

\appendix
\addtocontents{toc}{\protect\setcounter{tocdepth}{1}}
\section{Appendix} \label{app}

\subsection{Hypergeometric vs.\ multinomial}

We let $H_3(N_a, N_b, N_c, k)$ be the hypergeometric distribution where one sample $k$ objects without replacement where there are $N_a$, $N_b$ and $N_c$ objects of type $a$, $b$ and $c$, respectively. This distribution is not independent and thus hard to work with. However, when $k$ is small enough, the hyper-geometric distribution becomes very close to the multinomial distribution $M(N_a/N, N_b/N, N_c/N, k)$ where $N = N_a + N_b + N_c$.

\begin{lemma}\label{lemma:hypergeometric_vs_multinomial}
If $q_a, q_b, q_c \in \mathbb{N}$ are such that $q_a + q_b + q_c = k$, $k \leq \sqrt{N}/2$, $q_a \leq N_a/2$, $q_b \leq N_b/2$ and $q_c \leq N_c/2$, then:
$$\left(1 - \frac{2q_a^2}{N_a} - \frac{2q_b^2}{N_b} - \frac{2q_c^2}{N_c} \right) \cdot \Pr_{M}[q_a, q_b, q_c] \leq \Pr_{H_3}[q_a, q_b, q_c] \leq \left(1 + \frac{4k^2}{N} \right) \cdot \Pr_{M}[q_a, q_b, q_c]$$
\end{lemma}
\begin{proof}
Using the definition of the hyper-geometric distribution, we have:
$$\Pr_{H_3}[q_a, q_b, q_c] = \frac{\binom{N_a}{q_a}\binom{N_b}{ q_b} \binom{N_c }{ q_c}}{\binom{N}{q_a + q_b + q_c}} = \frac{k!}{q_a! q_b! q_c!} \cdot \prod_{i = 0}^{q_a - 1} \frac{N_a - i}{N - i} \cdot \prod_{i = 0}^{q_b - 1} \frac{N_b - i}{N - q_a - i} \cdot \prod_{i = 0}^{q_c - 1} \frac{N^1 - i}{N - q_a - q_b - i}$$
Let us denote by $P$ the three products. We proceed by upper-bounding it:
$$P \leq \left(\frac{N_a}{N}\right)^{q_a} \cdot \left(\frac{N_b}{N - q_a}\right)^{q_b} \cdot \left(\frac{N_c}{N - q_a - q_b} \right)^{q_c} \leq \left(\frac{N_a}{N}\right)^{q_a} \cdot \left(\frac{N_b}{N}\right)^{q_b} \cdot \left(\frac{N_c}{N} \right)^{q_c} \cdot \left(\frac{1}{1 - k/N}\right)^k$$
Recall that $k \leq \sqrt{B}/2$, hence:
$$\left(\frac{1}{1 - k/N}\right)^k \leq \left(1 + \frac{2k}{N}\right)^k \leq e^{2k^2/N} \leq 1 + \frac{4k^2}{N}$$
The upper bound therefore follows. We give a lower bound to $P$ as follows:
\begin{align*}
P &\geq \prod_{i = 0}^{q_a - 1} \frac{N_a - i}{N} \cdot \prod_{i = 0}^{q_b - 1} \frac{N_b -  i}{N} \cdot \prod_{i = 0}^{q_c - 1} \frac{N_c - i}{N} \\
& = \left(\frac{N_a}{N}\right)^{q_a} \cdot \left(\frac{N_b}{N}\right)^{q_b} \cdot \left(\frac{N_c}{N} \right)^{q_c} \cdot \prod_{i = 0}^{q_a - 1} 1 - \frac{i}{N_a} \cdot \prod_{i = 0}^{q_b - 1} 1 - \frac{i}{N_c} \cdot \prod_{i = 0}^{q_c - 1} 1 - \frac{i}{N_c}\\
&\geq \left(\frac{N_a}{N}\right)^{q_a} \cdot \left(\frac{N_b}{N}\right)^{q_b} \cdot \left(\frac{N_c}{N} \right)^{q_c}  \cdot \left(1 - \frac{q_a}{N_a}\right)^{q_a} \cdot \left(1 - \frac{q_b}{N_b}\right)^{q_b} \cdot \left(1 - \frac{q_c}{N_c}\right)^{q_c}
\end{align*}
Recalling that $q_a \leq N_a/2$, $q_b \leq N_b /2$ and $q_c \leq N_c/2$, we get the desired lower bound:
\begin{align*}
\left(1 - \frac{q_a}{N_a}\right)^{q_a} \cdot \left(1 - \frac{q_b}{N_b}\right)^{q_b} \cdot \left(1 - \frac{q_c}{N_c}\right)^{q_c} &\geq \exp\left(-\frac{2q_a^2}{N_a} - 2\frac{q_b^2}{N_b} - \frac{2q_c^2}{N_c}\right)\\
&\geq 1 - \frac{2q_a^2}{N_a} - \frac{2q_b^2}{N_b} - \frac{2q_c^2}{N_c} \qedhere
\end{align*}
\end{proof}

The same holds in the case were there are two classes of objects. More specifically, we let $H_2(N_a, N_b, k)$ be the hyper-geometric distributions which amounts to sampling without replacement from a population with $N_a$ objects of type a and $N_b$ objects of type b and define $B(N_a/N, N_b/N, k)$ to be the classical binomial distribution (with replacement). Again, if $k$ is small enough then the distributions can be interchanged.

\begin{lemma}\label{lemma:hypergeometric_vs_binomial}
If $q_a, q_b \in \mathbb{N}$ are such that $q_a + q_b= k$, $k \leq \sqrt{N}/2$, $q_a \leq N_a/2$ and $q_b \leq N_b/2$, then:
$$\left(1 - \frac{2q_a^2}{N_a} - \frac{2q_b^2}{N_b}\right) \cdot \Pr_{M}[q_a, q_b] \leq \Pr_{H_2}[q_a, q_b] \leq \left(1 + \frac{4k^2}{N} \right) \cdot \Pr_{M}[q_a, q_b]$$
\end{lemma}
\begin{proof}
Similar to the proof of Lemma \ref{lemma:hypergeometric_vs_multinomial}.
\end{proof}

\subsection{Properties of some functions}

\begin{lemma}
The function $r(q)$ in the proof of \cref{lemma:corruption_bound} has minimizer $q^\star \in [kb_Y/7, kb_Y]$.
\label{lemma:location_min}
\end{lemma}
\begin{proof}
Setting $\partial r /\partial q$ equal to zero yields an equation for the minima:
$$(1-\lambda)\left(1 - 4b_X^2\right)^{k/2}\ln\left(\frac{1- 2b_X}{1 + 2b_X}\right)\left(\frac{1- 2b_X}{1 + 2b_X}\right)^qa + \lambda \left(1 - 4b_Y^2\right)^{k/2}\ln\left(\frac{1 + 2b_Y}{1 - 2b_Y}\right)\left(\frac{1 + 2b_Y}{1 - 2b_Y}\right)^q = 0$$
Shuffling around:
$$(1-\lambda)\left(1 - 4b_X^2\right)^{k/2}\ln\left(\frac{1+ 2b_X}{1 - 2b_X}\right)\left(\frac{1- 2b_X}{1 + 2b_X}\right)^q = \lambda \left(1 - 4b_Y^2\right)^{k/2}\ln\left(\frac{1 + 2b_Y}{1 - 2b_Y}\right)\left(\frac{1 + 2b_Y}{1 - 2b_Y}\right)^q$$
Shuffling around and using the precise definition of $\lambda$ (see \cref{eq:setting_of_lambda}):
$$\left(\frac{1 + 2b_X}{1 - 2b_X} \cdot \frac{1+2b_Y}{1 - 2b_Y}\right)^q = \frac{(1-\lambda)\left(1 - 4b_X^2\right)^{k/2}\ln\left(\frac{1+ 2b_X}{1 - 2b_X}\right)}{\lambda \left(1 - 4b_Y^2\right)^{k/2}\ln\left(\frac{1 + 2b_Y}{1 - 2b_Y}\right)} =  \left(\frac{1 - 4b_X^2}{1 - 4b_Y^2}\right)^{k/2}$$
This allows to isolate $q^\star$ and using the bounds of \cref{lemma:approx_zoo}, we have:
$$q^\star = \frac{k}{2} \cdot \frac{\ln(1 - 4b_X^2) - \ln(1 - 4b_Y^2)}{\ln\left(\frac{1 + 2b_X}{1 - 2b_X} \cdot \frac{1+2b_Y}{1 - 2b_Y}\right)} \implies \frac{k}{2} \cdot \frac{4b_Y^2 - 8b_X^2}{6b_X + 6b_Y} \leq q^\star \leq \frac{k}{2} \cdot \frac{8b_Y^2 - 4b_X^2}{4b_X + 4b_Y}$$
Finally, recall that $b_X \in o(b_Y) \in o(1)$ so that $q^\star \in [kb_Y/7, kb_Y]$
\end{proof}

\begin{lemma}\label{lemma:phi_increasing_second_regime}
The function $\Phi(q)$ of Lemma \ref{lemma:leafwise_bias_corruption} is increasing for $q \in [16/10b_Y, \infty)$.
\end{lemma}
\begin{proof}
The derivative of $\Phi$ is:
$$\frac{\partial \Phi}{ \partial q} = \frac{\ln(r_Y)r_Y^q\left(r_X^q - r_X^{-q} \right) - \ln(r_X)r_Y^q \left(r_X^q + r_X^{-q}\right)}{\left(r_X^q + r_X^{-q}\right)^2}$$
Thus, we only need to show that $\ln(r_Y) \left(r_X^q - r_X^{-q} \right) \geq \ln(r_X) \left(r_X^q + r_X^{-q}\right)$ for $q \in [16/10b_Y, \infty)$. Using the series representation of the exponential function and various bounds of \cref{lemma:approx_zoo}, we have:
$$\begin{array}{r@{}l@{}l} \ln(r_Y) \left(r_X^q - r_X^{-q}\right)\, &\geq \sum_{t \geq 0 \text{ odd}} c(t) q^t & \quad\text{where}\quad c(t) \coloneqq  8b_Y\ln(r_X)^t/t!\\
\ln(r_X) \left(r_X^q + r_X^{-q}\right)\, &\leq \sum_{t \geq 0 \text{ even}} d(t) q^t &\quad\text{where}\quad d(t) \coloneqq  12b_X\ln(r_X)^t/t! \end{array}$$
Observe that under the hypothesis that $q \geq 16/10b_Y$ it holds that $c(1) q^1/2 \geq d(0) q^0$. Thus, it only remains to show that for any $t \geq 1$:
$$\underbrace{c(t-1) q^{t-1}/2}_{o_1} + \underbrace{ c(t+1) q^{t+1}/2}_{o_2} \geq d(t) q^t$$
If $o_2 \geq d(t) q^t$, the claim is already good. If not, then we have: $t \geq b_Y \ln(r_X)  q/3b_X - 1$ so that:
$$\frac{o_1}{d(t)q^t} = \frac{t b_Y}{3b_X \ln(r_X) q} \geq \left(\frac{b_Y \ln(r_X) q}{3b_X} - 1 \right) \cdot \frac{b_Y}{3 b_X \ln(r_X) q} = \frac{b_Y^2}{9b_X^2} - \frac{b_Y}{3b_X\ln(r_X)q} \geq \frac{b_Y^2}{144 b_X^2}$$
Where the last inequality is due to the fact that $\ln(r_X) \geq 2b_X$ and $q \geq 16/10b_Y$. Since $b_X < o(b_Y)$, we have $o_1 \geq d(t) q^t$ and thus $\Phi$ is indeed increasing on $[16/10b_Y, \infty)$.
\end{proof}

\subsection{Properties of trees}
\begin{lemma}\label{lemma:simplifying_depth}
Let $P = (P^0 + P^1)/2$ be a distribution and $D$ a decision tree together with some $L \in \mathbb{N}$. If $D'$ is the version of $D$ that stops after $L$ queries, then:
$$\frac{\cost(D', P)}{\TV(\tran(D', P^0), \tran(D', P^1))} \geq L \implies \frac{\cost(D, P)}{\TV(\tran(D, P^0), \tran(D, P^1))} \geq \frac{L}{3}$$ 
\end{lemma}
\begin{proof}
Define $\Llarge(D) \coloneqq \{\ell \in \L(D): |\ell| \geq L\}$ and let $P[\Llarge(D)]$ be the probability that a leaf of $\Llarge$ is reached by $x \sim P$ in $D$. Observe that:
$$P[\Llarge(D)] \geq \TV(\tran(D, P^0), \tran(D, P^1))/3 \implies \cost(D, P) \geq L \cdot \TV(\tran(D, P^0), \tran(D, P^1))/3$$
Hence, we may assume for the remainder of the proof that $P[\Llarge(D)] \leq \TV(\tran(t, P^0), \tran(t, P^1))/3$. If $P[\Llarge(D)]$ is small, it must be that $D'$ holds a constant fraction of the bias of $D$. Indeed, letting $\L^0(D) \coloneqq \{\ell \in \L(D): P^0[\ell] \geq P^1[\ell]\}$:
\begin{align*}
\TV(\tran(D', P^0), \tran(D', P^1)) &= \sum\nolimits_{\ell \in \L^0(D')} P^0[\ell] - P^1[\ell]\\
&\geq \sum\nolimits_{\ell \in \L^0(D)} P^0[\ell] - P^1[\ell] - \sum\nolimits_{\ell \in \Llarge(D)} P^0[\ell] + P^1[\ell] \\
&= \TV(\tran(D, P^0), \tran(D, P^1)) - 2 \cdot P[\Llarge(D)]\\
&= \TV(\tran(D, P^0), \tran(D, P^1))/3
\end{align*}
Finally, as $D'$ is a truncated copy of $D$, it holds that $\qbar(D, P) \geq \qbar(D', P)$ and the claim follows.
\end{proof}

\begin{lemma}[Acceptance centring]\label{lemma:re_centering}
If $D$ is a deterministic decision tree over $\B^*$ labelled by $\{B^0, B^1\}$ with bias $\delta = \Pr_{x \sim B^0}[D(x) = B^0] - \Pr_{x \sim B^1}[D(x) = B^0]$, then there exists a randomised decision tree $R$ with $\qbar(R, B) \leq \qbar(D, B)$ and $\textup{depth}(R) \leq \textup{depth}(D)$ such that:
$$\Pr_{x \sim B^0}[R(x) = B^0] = \frac{1}{2} + \xi \quad\text{and}\quad \Pr_{x \sim B^1}[R(x) = B^0] = \frac{1}{2} - \xi \quad\text{where}\quad \xi \geq \delta/6$$
\end{lemma}
\begin{proof}
Let $p \coloneqq \Pr_{x \sim B^0}[t(x) = B^0]$ and suppose by symmetry that $p \leq 1/2$. For some $\alpha \in [0, 1]$ that we fix later, we define $R$ to query nothing and output $B^0$ with probability $\alpha$ and run $D$ with remaining probability $1 - \alpha$. As such, $\Pr_{x \sim B^0}[R(x) = B^0] = \alpha + (1-\alpha) \cdot p$ and $\Pr_{x \sim B^1}[R(x) = B^0] = \alpha + (1-\alpha) \cdot (p - \delta)$. Thus, by setting $\alpha \coloneqq 1 - 1/(2 - 2p + \delta)$, we have $\xi = \delta/(4 - 4p + 2\delta) \geq \delta/6$ and the desired acceptance probabilities.
\end{proof}

\subsection{Some inequalities}
\begin{lemma}\label{lemma:approx_lambda}
The mixture parameter $\lambda$ defined in \cref{eq:setting_of_lambda} satisfies $\lambda \in (b_X/b_Y) \cdot [2/3, 3]$.
\end{lemma}
\begin{proof}
Immediate by recalling that $b_X, b_Y \in o(1)$ and using inequalities \cref{equation:zoo_1} and \cref{equation:zoo_3} of \cref{lemma:approx_zoo}.
\end{proof}

\begin{lemma}\label{lemma:approx_zoo}
For any $x \in [0, 0.5]$, $y \in [0,1]$ and $k \in [0, \infty]$,
\begin{gather}
x \leq \frac{x}{1-x} \leq 2x \label{equation:zoo_1}\\
-2x \leq \ln(1-x) \leq -x \label{equation:zoo_2}\\
2x \leq \ln\left(\frac{1+x}{1-x}\right) \leq 3x \label{equation:zoo_3}\\
1 - (1 - y)^k \leq ky \label{equation:zoo_4}
\end{gather}
\end{lemma}
\begin{proof} 
Inequality \cref{equation:zoo_1} holds by inspection while \cref{equation:zoo_4} is proven in Lemma 3 of \cite{BenDavid2020minimax}. The upper bound of \cref{equation:zoo_2} is due to a truncation of the Taylor series whereas the lower bound comes from:
$$x - \ln(1-x) = \sum_{n = 2}^\infty \frac{x^n}{n} \leq x \cdot \sum_{n = 1}^\infty \frac{x^n}{n} = - x \ln(1-x) \leq -\ln\left(\frac{1}{2}\right)x \implies \ln(1-x) \geq -2x$$
The lower bound in \cref{equation:zoo_3} is again due to a truncation of the Taylor series while the upper bound is a combination of \cref{equation:zoo_2} and the identity $1+x \leq e^x$, i.e. $\ln((1 + x)/(1 -x)) = \ln(1+x) - \ln(1-x) \leq x + 2x$.
\end{proof}

\begin{lemma}\label{lemma:chernoff2_3}
If $D$ is a decision tree and $\mathcal{U}^k = \{\ell \in \L(D): |\ell| = k \textup{ and } \Delta(\ell) \geq 2k^{2/3}\}$ for all $k \in \mathbb{N}$, then:
$$\sum\nolimits_{\ell \in \mathcal{U}^k} Z[\ell] \leq 2e^{-k^{1/3}/48} \qquad \forall k \leq 1/64b_Y^3$$
\end{lemma}
\begin{proof}
Using the definition of $Z$, we can recast the sum as a probability:
$$\sum_{\ell \in \mathcal{U}^k} Z[\ell] \leq \sum_{\ell \in \mathcal{U}^k} 2 \cdot \max\left\lbrace Z^0[\ell], Z^1[\ell]\right\rbrace  \leq 2\sum_{q = k/2 + k^{2/3}}^k \binom{k}{q} \left(\frac{1}{2} + b_Y\right)^q \left(\frac{1}{2} - b_Y\right)^{k - q}$$
Now, the last quantity can be interpreted as the probability of having at least $k/2 + k^{2/3}$ successes in $k$ independent trials where the success probability is $1/2 + b_Y$. Therefore, we may leverage a standard Chernoff bound as follows:
$$\sum_{\ell \in \mathcal{U}^k} Z[\ell] \leq 2\Pr\left[\sum_{i \in [k]}\Bern(1/2 + b_Y) \geq (1+\delta)\mu\right] \quad \text{where } \mu = \frac{k}{2} + kb_Y \text{ and } \delta = \frac{k/2 + k^{2/3}}{\mu} -1$$
Note that under the hypothesis that $k \leq 1/64b_Y^3$, we have that $\delta \geq 0$ and $\delta^2\mu/3 \geq k^{1/3}/48$, thus:
\begin{equation*}
\sum_{\ell \in \mathcal{U}^k} Z[\ell] \leq 2\Pr\left[\sum_{i \in [k]}\Bern(1/2 + b_Y) \geq (1+\delta)\mu\right] \leq e^{-\delta^2\mu/3} \leq e^{-k^{1/3}/48} \qedhere
\end{equation*}
\end{proof}

\bigskip\bigskip

\subsection*{Acknowledgements}
We thank anonymous FOCS reviewers for their comments.

\bigskip

\DeclareUrlCommand{\Doi}{\urlstyle{sf}}
\renewcommand{\path}[1]{\small\Doi{#1}}
\renewcommand{\url}[1]{\href{#1}{\small\Doi{#1}}}
\bibliographystyle{alphaurl}
\bibliography{lr-refs}

\newcommand{\etalchar}[1]{$^{#1}$}
\begin{thebibliography}{BGKW20}

\bibitem[ABK16]{Aaronson2016}
Scott Aaronson, Shalev Ben{-}David, and Robin Kothari.
\newblock Separations in query complexity using cheat sheets.
\newblock In {\em Proceedings of the 48th Symposium on Theory of Computing
  (STOC)}, pages 863--876. ACM, 2016.
\newblock \href {https://doi.org/10.1145/2897518.2897644}
  {\path{doi:10.1145/2897518.2897644}}.

\bibitem[AGJ{\etalchar{+}}18]{Anshu2018}
Anurag Anshu, Dmitry Gavinsky, Rahul Jain, Srijita Kundu, Troy Lee, Priyanka
  Mukhopadhyay, Miklos Santha, and Swagato Sanyal.
\newblock A composition theorem for randomized query complexity.
\newblock In {\em Proceedings of the 37th Foundations of Software Technology
  and Theoretical Computer Science (FSTTCS)}, volume~93, pages 10:1--10:13,
  2018.
\newblock \href {https://doi.org/10.4230/LIPIcs.FSTTCS.2017.10}
  {\path{doi:10.4230/LIPIcs.FSTTCS.2017.10}}.

\bibitem[AKK16]{Ambainis2016}
Andris Ambainis, Martins Kokainis, and Robin Kothari.
\newblock Nearly optimal separations between communication (or query)
  complexity and partitions.
\newblock In {\em Proceedings of the 31st Computational Complexity Conference
  (CCC)}, pages 4:1--4:14. Schloss Dagstuhl, 2016.
\newblock \href {https://doi.org/10.4230/LIPIcs.CCC.2016.4}
  {\path{doi:10.4230/LIPIcs.CCC.2016.4}}.

\bibitem[BB19]{Blais2019}
Eric Blais and Joshua Brody.
\newblock Optimal separation and strong direct sum for randomized query
  complexity.
\newblock In {\em Proceedings of the 34th Computational Complexity Conference
  (CCC)}, pages 29:1--29:17. Schloss Dagstuhl, 2019.
\newblock \href {https://doi.org/10.4230/LIPIcs.CCC.2019.29}
  {\path{doi:10.4230/LIPIcs.CCC.2019.29}}.

\bibitem[BB20a]{BenDavid2020minimax}
Shalev Ben{-}David and Eric Blais.
\newblock A new minimax theorem for randomized algorithms.
\newblock In {\em Proceedings of the 61st Symposium on Foundations of Computer
  Science (FOCS)}. {IEEE}, nov 2020.
\newblock \href {https://doi.org/10.1109/focs46700.2020.00045}
  {\path{doi:10.1109/focs46700.2020.00045}}.

\bibitem[BB20b]{BenDavid2020comp}
Shalev Ben{-}David and Eric Blais.
\newblock A tight composition theorem for the randomized query complexity of
  partial functions.
\newblock In {\em Proceedings of the 61st Symposium on Foundations of Computer
  Science (FOCS)}. {IEEE}, nov 2020.
\newblock \href {https://doi.org/10.1109/focs46700.2020.00031}
  {\path{doi:10.1109/focs46700.2020.00031}}.

\bibitem[BDG{\etalchar{+}}20]{Bassilakis2020}
Andrew Bassilakis, Andrew Drucker, Mika G{\"o}{\"o}s, Lunjia Hu, Weiyun Ma, and
  Li-Yang Tan.
\newblock The power of many samples in query complexity.
\newblock In {\em Proceedings of the 47th International Colloquium on Automata,
  Languages, and Programming (ICALP)}, volume 168, pages 9:1--9:18. Schloss
  Dagstuhl, 2020.
\newblock \href {https://doi.org/10.4230/LIPIcs.ICALP.2020.9}
  {\path{doi:10.4230/LIPIcs.ICALP.2020.9}}.

\bibitem[BdW02]{Buhrman2002}
Harry Buhrman and Ronald de~Wolf.
\newblock Complexity measures and decision tree complexity: a survey.
\newblock {\em Theoretical Computer Science}, 288(1):21--43, 2002.
\newblock Complexity and Logic.
\newblock \href {https://doi.org/10.1016/S0304-3975(01)00144-X}
  {\path{doi:10.1016/S0304-3975(01)00144-X}}.

\bibitem[BGKW20]{BenDavid2020amp}
Shalev Ben{-}David, Mika G{\"o}{\"o}s, Robin Kothari, and Thomas Watson.
\newblock When is amplification necessary for composition in randomized query
  complexity?
\newblock In {\em Proceedings of the 24h International Conference on
  Randomization and Computation (RANDOM)}, volume 176, pages 28:1--28:16.
  Schloss Dagstuhl, 2020.
\newblock \href {https://doi.org/10.4230/LIPIcs.APPROX/RANDOM.2020.28}
  {\path{doi:10.4230/LIPIcs.APPROX/RANDOM.2020.28}}.

\bibitem[BK16]{Ben-David2016}
Shalev {Ben-David} and Robin Kothari.
\newblock Randomized query complexity of sabotaged and composed functions.
\newblock In {\em Proceedings of the 43rd International Colloquium on Automata,
  Languages, and Programming (ICALP)}, volume~55, pages 60:1--60:14, 2016.
\newblock \href {https://doi.org/10.4230/LIPIcs.ICALP.2016.60}
  {\path{doi:10.4230/LIPIcs.ICALP.2016.60}}.

\bibitem[BKLS20]{Brody2020}
Joshua Brody, Jae~Tak Kim, Peem Lerdputtipongporn, and Hariharan Srinivasulu.
\newblock A strong {XOR} lemma for randomized query complexity, 2020.
\newblock \href {http://arxiv.org/abs/2007.05580} {\path{arXiv:2007.05580}}.

\bibitem[DM21]{Dahiya2021}
Yogesh Dahiya and Meena Mahajan.
\newblock On (simple) decision tree rank.
\newblock In {\em Proceedings of the 41st oundations of Software Technology and
  Theoretical Computer Science (FSTTCS)}, volume 213, pages 15:1--15:16.
  Schloss Dagstuhl, 2021.
\newblock \href {https://doi.org/10.4230/LIPIcs.FSTTCS.2021.15}
  {\path{doi:10.4230/LIPIcs.FSTTCS.2021.15}}.

\bibitem[GJPW18]{Goos2018}
Mika G\"o\"os, T.~S. Jayram, Toniann Pitassi, and Thomas Watson.
\newblock Randomized communication versus partition number.
\newblock {\em ACM Transactions on Computation Theory}, 10(1), 2018.
\newblock \href {https://doi.org/10.1145/3170711} {\path{doi:10.1145/3170711}}.

\bibitem[GLSS19]{Gavinsky2019}
Dmitry Gavinsky, Troy Lee, Miklos Santha, and Swagato Sanyal.
\newblock A composition theorem for randomized query complexity via
  max-conflict complexity.
\newblock In {\em Proceedings of the 46th International Colloquium on Automata,
  Languages, and Programming (ICALP)}, volume 132, pages 64:1--64:13, 2019.
\newblock \href {https://doi.org/10.4230/LIPIcs.ICALP.2019.64}
  {\path{doi:10.4230/LIPIcs.ICALP.2019.64}}.

\bibitem[GM21]{Goos2021}
Mika G\"{o}\"{o}s and Gilbert Maystre.
\newblock A majority lemma for randomised query complexity.
\newblock In {\em Proceedings of the 36th Computational Complexity Conference
  (CCC)}, volume 200, pages 18:1--18:15. Schloss Dagstuhl, 2021.
\newblock \href {https://doi.org/10.4230/LIPIcs.CCC.2021.18}
  {\path{doi:10.4230/LIPIcs.CCC.2021.18}}.

\bibitem[GSS16]{Gilmer2016}
Justin Gilmer, Michael Saks, and Srikanth Srinivasan.
\newblock Composition limits and separating examples for some boolean function
  complexity measures.
\newblock {\em Combinatorica}, 36(3):265--311, 2016.
\newblock \href {https://doi.org/10.1007/s00493-014-3189-x}
  {\path{doi:10.1007/s00493-014-3189-x}}.

\bibitem[GTW21]{Girish2021}
Uma Girish, Avishay Tal, and Kewen Wu.
\newblock {Fourier Growth of Parity Decision Trees}.
\newblock In {\em Proceedings of the 36th Computational Complexity Conference
  (CCC)}, volume 200, pages 39:1--39:36. Schloss Dagstuhl, 2021.
\newblock \href {https://doi.org/10.4230/LIPIcs.CCC.2021.39}
  {\path{doi:10.4230/LIPIcs.CCC.2021.39}}.

\bibitem[JKS10]{Jain2010}
Rahul Jain, Hartmut Klauck, and Miklos Santha.
\newblock Optimal direct sum results for deterministic and randomized decision
  tree complexity.
\newblock {\em Information Processing Letters}, 110(20):893--897, 2010.
\newblock \href {https://doi.org/10.1016/j.ipl.2010.07.020}
  {\path{doi:10.1016/j.ipl.2010.07.020}}.

\bibitem[Li21]{Li2021}
Yaqiao Li.
\newblock Conflict complexity is lower bounded by block sensitivity.
\newblock {\em Theoretical Computer Science}, 856:169--172, feb 2021.
\newblock \href {https://doi.org/10.1016/j.tcs.2020.12.038}
  {\path{doi:10.1016/j.tcs.2020.12.038}}.

\bibitem[LMR{\etalchar{+}}11]{Lee2011}
Troy Lee, Rajat Mittal, Ben Reichardt, Robert Spalek, and Mario Szegedy.
\newblock Quantum query complexity of state conversion.
\newblock In {\em Proceedings of the 52nd Symposium on Foundations of Computer
  Science (FOCS)}, pages 344--353. IEEE, 2011.
\newblock \href {https://doi.org/10.1109/FOCS.2011.75}
  {\path{doi:10.1109/FOCS.2011.75}}.

\bibitem[Rei11]{Reichardt2011}
Ben Reichardt.
\newblock Reflections for quantum query algorithms.
\newblock In {\em Proceedings of the 22nd Symposium on Discrete Algorithms
  (SODA)}, pages 560--569, 2011.

\bibitem[Sav02]{Savicky2002}
Petr Savick{\'y}.
\newblock On determinism versus unambiquous nondeterminism for decision trees.
\newblock Technical Report TR02-009, Electronic Colloquium on Computational
  Complexity (ECCC), 2002.
\newblock URL: \url{http://eccc.hpi-web.de/report/2002/009/}.

\bibitem[She12]{Sherstov2012}
Alexander~A. Sherstov.
\newblock The communication complexity of gap hamming distance.
\newblock {\em Theory of Computing}, 8(8):197--208, 2012.
\newblock URL: \url{http://www.theoryofcomputing.org/articles/v008a008}, \href
  {https://doi.org/10.4086/toc.2012.v008a008}
  {\path{doi:10.4086/toc.2012.v008a008}}.

\bibitem[Tal13]{Tal2013}
Avishay Tal.
\newblock Properties and applications of boolean function composition.
\newblock In {\em Proceedings of the 4th Conference on Innovations in
  Theoretical Computer Science (ITCS)}, pages 441--454, 2013.
\newblock \href {https://doi.org/10.1145/2422436.2422485}
  {\path{doi:10.1145/2422436.2422485}}.

\bibitem[Ver98]{Vereshchagin1998}
Nikolai Vereshchagin.
\newblock Randomized boolean decision trees: Several remarks.
\newblock {\em Theoretical Computer Science}, 207(2):329--342, nov 1998.
\newblock \href {https://doi.org/10.1016/s0304-3975(98)00071-1}
  {\path{doi:10.1016/s0304-3975(98)00071-1}}.

\bibitem[Yao77]{Yao1977}
Andrew Yao.
\newblock Probabilistic computations: Toward a unified measure of complexity.
\newblock In {\em Proceedings of the 18th Symposium on Foundations of Computer
  Science (FOCS)}, pages 222--227, Oct 1977.
\newblock \href {https://doi.org/10.1109/SFCS.1977.24}
  {\path{doi:10.1109/SFCS.1977.24}}.

\end{thebibliography}

\end{document}